\documentclass[preprint,12pt]{elsarticle}

\usepackage{amssymb}
\usepackage{algorithm}
\usepackage{amsmath}
\usepackage{algpseudocode}
\usepackage{graphicx}
\usepackage{longtable}
\usepackage{geometry}
\usepackage{enumitem}
\usepackage{amssymb}
\usepackage{booktabs}
\usepackage{adjustbox}
\usepackage{tabularx}
\usepackage{float}
\usepackage{hyperref}
\usepackage{enumerate}
\usepackage{longtable}
\usepackage{multirow}
\usepackage{subcaption}

\algrenewcommand\algorithmicindent{1em}

\journal{Arxiv}

\begin{document}

\begin{frontmatter}



\title{Digital Transformation in the Water Distribution System based on the Digital Twins Concept}


\author[inst1,inst2]{MohammadHossein Homaei\corref{cor1}}
\ead{mhomaein@alumnos.unex.es}
\cortext[cor1]{Corresponding author}

\author[inst1]{Agust\'in Javier Di Bartolo}
\author[inst1]{Mar \'Avila} 
\author[inst1]{\'Oscar~Mogoll\'on-Guti\'errez}
\author[inst1]{Andr\'es Caro}

\affiliation[inst1]{
    organization={Universidad de Extremadura},
    addressline={School of Technology, Av/ Universidad S/N}, 
    city={Caceres},
    postcode={10003}, 
    state={Extremadura},
    country={Spain}
}

\affiliation[inst2]{
    organization={Ambling Company},
    addressline={Av. Martín Palomino, 90},
    city={Cáceres},
    postcode={10600},
    country={Spain}
}

\begin{abstract}
Digital Twins have emerged as a disruptive technology with great potential; they can enhance WDS by offering real-time monitoring, predictive maintenance, and optimization capabilities. This paper describes the development of a state-of-the-art DT platform for WDS, introducing advanced technologies such as the Internet of Things, Artificial Intelligence, and Machine Learning models. This paper provides insight into the architecture of the proposed platform-CAUCCES-that, informed by both historical and meteorological data, effectively deploys AI/ML models like LSTM networks, Prophet, LightGBM, and XGBoost in trying to predict water consumption patterns. Furthermore, we delve into how optimization in the maintenance of WDS can be achieved by formulating a Constraint Programming problem for scheduling, hence minimizing the operational cost efficiently with reduced environmental impacts. It also focuses on cybersecurity and protection to ensure the integrity and reliability of the DT platform. In this view, the system will contribute to improvements in decision-making capabilities, operational efficiency, and system reliability, with reassurance being drawn from the important role it can play toward sustainable management of water resources.

\end{abstract}


\begin{graphicalabstract}
\includegraphics[width=\textwidth]{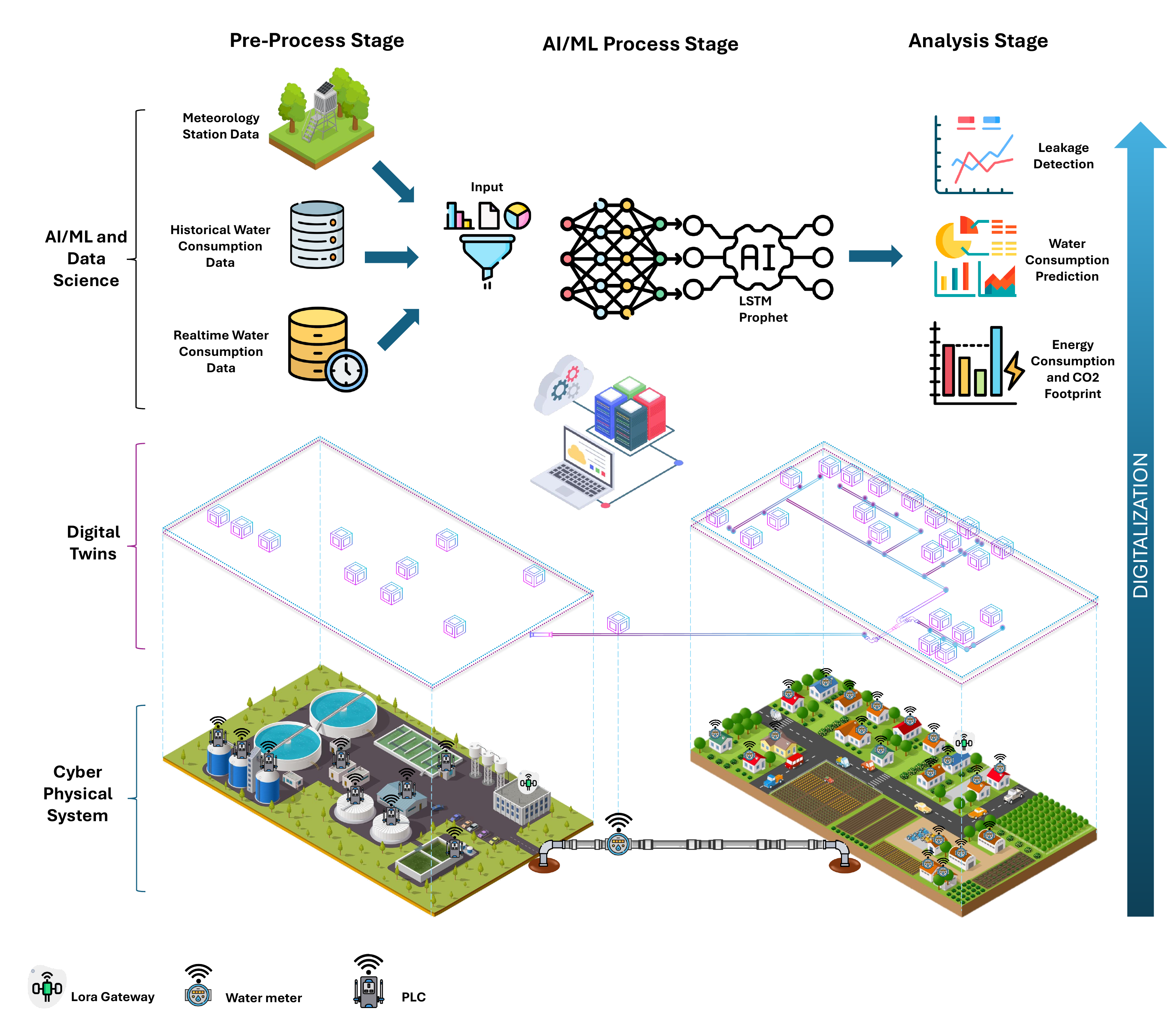}
\end{graphicalabstract}


\newpage

\begin{highlights}
\item Developed an integrated \textbf{Digital Twin (DT) platform} for water distribution systems, leveraging \textbf{IoT, AI/ML models, and cybersecurity} to optimize operations, reduce environmental impacts, and ensure data security.

\item Implemented advanced \textbf{AI/ML models} (LSTM, Prophet, LightGBM, XGBoost) with feature engineering and hyperparameter tuning, achieving a \textbf{Mean Absolute Error (MAE) of 5.76} and \textbf{Mean Absolute Percentage Error (MAPE) of 18.61\%} in 6-month water consumption forecasting.

\item Formulated a \textbf{Constraint Programming (CP)-based scheduling model} to address NP-hard challenges in task routing, preemptions, and dependencies, achieving \textbf{14\% reduction in completion time}, \textbf{25\% reduction in delays}, and \textbf{17\% lower CO$_2$} emissions.

\item Enhanced water system resilience by integrating \textbf{cybersecurity measures aligned with ISO 27001}, including \textbf{AES-128 encryption}, secure data transmission protocols, and \textbf{real-time threat detection} using Wazuh and Zabbix.

\item Demonstrated significant improvements in \textbf{operational efficiency}, \textbf{resource utilization}, and \textbf{forecasting accuracy}, setting a benchmark for \textbf{sustainable and secure digital transformation} in water distribution networks.

\item Provided a robust foundation for future \textbf{real-time optimization, scalability, and environmental sustainability} in water management, contributing to global \textbf{sustainable development goals (SDGs)}.
\end{highlights}


\end{frontmatter}

\newpage
\section{Introduction}\label{sec:Introduction}
Inefficient energy consumption and elevated expenses represent merely two of the challenges linked to the water supply system, an essential component of infrastructure that delivers water to final consumers, such as enterprises, the populace, and agricultural sectors \cite{Hu2021}. The establishment of facilities for the transportation of water is not a contemporary achievement of civilization. Throughout this era, developments in science and engineering have enabled water to be transported over long distances through the use of valves and pipelines \cite{DAYIOLU2021}. Aging, urbanization, population growth, climate change, and other factors challenge an already strained WDS \cite{Bauer2021}. The water supply and bent technology vicious cycle should be broken up simultaneously, reconstructing water networks designed to elevate efficiency, reliability, and promotion in this context even risks \cite{Beji2022}.

This type of digital transformation helps solve these systemic issues. Water utilities can track and control their systems remotely with IoT technology, AI, and DTs. Predictive maintenance, the ability to look after systems and devices in real-time, and simulations of systems are made possible to the benefit of system efficiency and, at the same time, minimize operational costs \cite{Khanna2020, Agostinelli2021}. However, despite these advances in digitalization related to WDS, there are still a number of significant challenges to overcome: data structuring, model precision, and interfacing with legacy systems \cite{Ciliberti2021}. Ensuring long-term viability and operational efficiency in WDS requires technological innovation and the adoption of sustainability-driven business models. Similar to how manufacturing firms are restructuring their business models to achieve sustainable value creation and delivery \cite{Jagani2023}, the water sector must integrate environmental, social, and governance principles into its digital transformation strategies.

The major focus on DTs technology is that it is presently one of the newsmaking, adopted adjunct technologies that will affect the operation of business organizations. They are defined as virtual models of discrete physical systems and run in real time with data and simulations important for the augmentation of performance of a system \cite{Jiang2021}. In water distribution, DTs can enhance system design, predict the needs around maintenance, conduct simulations concerning this system's operational capability, and many others \cite{Zekri2022}.
Unlike the above proposition, bottlenecks explicit of deploying DTs in this particular industry are insecurity, data protection, and the need for a competent workforce, among others \cite{Pylianidis2021}.

In this article, we leverage the application of DTs within the water industry and extend it to urban and rural water distribution and transmission networks to propose a novel platform model.

\subsection{Objective and Main Contributions of the Paper}

A new DT-based system for water supply systems is presented in this paper, in as much detail as possible. The proposed platform merges DTs with IoT technology to forecast and visualize water uses, thus enhancing the decision capabilities of whoever has authority over the management of the water utilities. Accordingly, the integration implied that the platform was supposed to incorporate the WDS imperatives for improved operational efficiency, regulatory observance, and good infrastructure. This paper will further discuss the characteristics of the platform to be created, its probable advantages, and different ways of overcoming the bottlenecks in the process of digitalization of WDS.

The main contributions of the paper are the following:

\begin{itemize}
    \item The paper proposes an innovative platform based on DTs applied to a WDS, which would be transferred/tested in an SME from the sector in a rural environment. Even if the platform is focused on the water sector and WDS, it will easily be replicable for other sectors or other commercial interests.
    \item TIt is composed of the infrastructures of AI and ML and IoT for the realization of a digital transformation of traditional business models into digital business models. This inclusion in DT, together with IoT devices that allow the acquisition of data used in the predictions, would be of huge interest due to the possible application in similar contexts for prediction purposes.
    \item The platform integrates with advanced Artificial Intelligence and Machine Learning models in water consumption prediction, hence providing more accurate demand forecasting and best resource management. Such predictive capability would help to achieve better decision-making in a water distribution system.
    \item It allows for maintenance scheduling, operator management of water distribution networks, realizes Digital Twins, and CP-based scheduling for optimization of maintenance activities. This ensures that operational efficiency is optimized, the occurrence of downtime is reduced, and service quality is improved through reliable management of the water distribution system.    \item In fact, the cybersecurity threats that, as a consequence of the paradigm shift implied by the use of DT models connected to IoT devices, have been studied and considered. Another contribution of the paper deals with identifying the cybersecurity strategies that are implemented in accordance with the norm compliance ISO 27001. The implemented security strategies will doubtless be of great interest in similar projects.
    \item It has been tested in a real practical environment what was foreseen with the platform, functionalities, and possibilities that it provides. The proposed DT model, the use of an AI tool, and IoT device identification- including the identification of cybersecurity strategies easily replicable in similar settings and projects, constitute one of the basic contributions of this work.
\end{itemize}

\begin{table}[h!]
\caption{List of Abbreviations} \label{tab:abbreviations}
\scriptsize
\centering
\begin{tabular}{ll}
\hline
\textbf{Abbreviation} & \textbf{Definition} \\
\hline
AEMET & Agencia Estatal de Meteorología \\
AES & Advanced Encryption Standard \\
AI & Artificial Intelligence \\
AMI & Advanced Metering Infrastructure \\
AMR & Automated Meter Reading \\
API & Application Programming Interface \\
CADF & Combined Anomaly Detection Framework \\
CAUCCES & Control de Agua Urbana, Cantidad y Calidad Excelentes y Sostenibles \\
CP & Constraint Programming \\
CPS & Cyber-Physical Systems \\
DL & Deep Learning \\
DMA & District Metered Area \\
DSS & Decision Support System \\
DT & Digital Twin \\
DWSs & Digital Water Services \\
ERD & Energy Recovery Device \\
EUI & Extended Unique Identifier \\
FDT & Functional Design Technology \\
GCN & Graph Convolutional Network \\
GDPR & General Data Protection Regulation \\
GIS & Geographic Information Systems \\
GNN & Graph Neural Network \\
HIP & Hydrologic Information Portal\\
HMI & Human-Machine Interface \\
HPP & High Pressure Pump \\
ICT & Information and Communication Technology \\
ISO & International Organization for Standardization \\
IoT & Internet of Things \\
LoRa & Long Range \\
LOS & Levels of Service \\
LSTM & Long-Short-Term Memory\\
ML & Machine Learning \\
MAE & Mean Absolute Error \\
MAPE & Mean absolute percentage error \\
MSE& Mean Squared Error \\
NASA & National Aeronautics and Space Administration \\
PAT & Pumps as Turbines \\
PCC & Pearson Correlation Coefficients \\
PLC & Programmable Logic Controller \\
PRV & Pressure Reduction Valves \\
QGIS & Quantum Geographic Information System \\
RO & Reverse Osmosis \\
SCADA & Supervisory Control And Data Acquisition \\
SME & Small and Medium-sized Enterprises \\
SSL & Secure Sockets Layer \\
SWaT & Secure Water Treatment \\
SWG & Smart Water Grid \\
TLS & Transport Layer Security \\
VPN & Virtual Private Network \\
WDN & Water Distribution Network \\
WDS & Water Distribution System \\
WRRF & Water Resource Recovery Facility \\
\hline
\end{tabular}

\end{table}

\subsection{Article Structure}

The rest of the paper is organized as follows: In Sec~\ref{sec:background}, \textit{Background and Motivation} reviews the current status, challenges within the WDS, and the need to adopt such Digital Transformation Technologies like DTs. Section~\ref{sec:related}, \textit{Literature's Review} discusses the existing literature on DT and presents the lacunars of the current research, setting the stage for our contributions. Section~\ref{sec:proposed}, \textit{Proposed DTs Platform} outlines the proposed platform and gives an overview of the system, functional capabilities, and methods to integrate with existing legacy systems. Section~\ref{sec:LSTM/Prophet/LightGBM}, \textit{Integrating AI/ML in Water Consumption Prediction} outlines the use case of the AI/ML method applied for water consumption prediction and leak detection based on the DTs platform.
In Section~\ref{sec:OptimisationWDS}, we solve one CP scheduling problem for task completion time, penalty reduction, CO\textsubscript{2} emission minimization, fuel consumption, and maximization of efficiency in a single-machine problem setting. Section~\ref{sec:Security measures}, \textit{Security Layer in DTs Platform}, presents some cybersecurity approaches in order to build up the resilience and sustainability of the DT platform. Finally, Section~\ref{sec:Conclusion}, \textit{Conclusion and Future Directions}, summarizes the contribution of this paper and provides some possible future directions for research in this domain.

\section{Background and Motivation}\label{sec:background}

\subsection{Digital Transforming}
Digital transformation in the water industry uses technologies like IoT, AI, ML, DL, and DT to make water supply and distribution more efficient, sustainable, and reliable \cite{Boyle2022, Haaker2021, Pivoto2021}. These tools can improve maintenance, optimize resources, and support better decision-making \cite{DAYIOLU2021,Feroz2021,Wjcicki2022}. Yet, as in other sectors experiencing similar changes, successfully adopting these advanced solutions in critical infrastructure can be challenging. National research and development initiatives have shown that strategic frameworks help integrate emerging technologies sustainably—and the same approach applies to DTs, AI, and IoT in water utilities \cite{JalaliSepehr2024}. When digital transformation aligns with such targeted R\&D efforts, it can promote long-term resilience, environmental responsibility, and more efficient resource management in the water sector.

\begin{itemize}
    \item Real-time monitoring: Digital transformation will enhance real-time monitoring of the WDS; this will help in the early detection and timely response in case of system failure and leaks. For example, sensors that detect a leak in the system and send real-time alerts to maintenance personnel who can act immediately and reduce water wastage \cite{BotnSanabria2022}.
    \item Predictive maintenance: The process of digital transformation facilitates predictive maintenance, which serves to avert system failures and diminish maintenance expenses. For instance, machine learning algorithms can be developed using system data to anticipate maintenance requirements, thereby enabling maintenance personnel to organize repairs prior to the occurrence of system malfunctions \cite{Futai2022}.
    \item Simulation capabilities: Digital conversion can provide simulation capabilities that can help optimize system design and operation.  For example, DTs can simulate system behavior under various operating conditions, allowing companies to optimize system design and identify potential problems before they occur \cite{Singh2021}.
    \item Data-driven decision making: Digital transformation can allow for better decision-making by availing data-driven insights as stated by \cite{arena2022novel}. For instance, system data can be analyzed for patterns and trends that may inform decisions related to system upgrades or improvements.
\end{itemize}

\subsection{Current Challenges}

Besides, digital transformation in WDS and more has its host of big challenges:

\begin{itemize}

    \item Integration with legacy systems: Integrating modern technology with the existing infrastructure is cumbersome and costly. In support of this process of digital transition, older WDS—which are very often incompatible with today's IoT devices and sensors—may involve extensive upgrades \cite{Nadkarni2020}.
    \item Data privacy and cybersecurity: Cybercriminals may be able to take advantage of more weaknesses as IoT devices and sensors proliferate, which might disrupt systems and expose data. For instance, a weakly secured sensor can give hackers access to compromised systems or pilfer confidential information \cite{Mendhurwar2019}.
    \item Proficient workforce: The management and upkeep of digital infrastructure require specialized knowledge. Utilities may face considerable challenges and must undertake substantial programmatic investments to equip their personnel with the necessary digital skills in regions lacking such expertise \cite{Broo2021}.
    \item Implementation costs: The initial costs of implementing digital transformation are often high, including investments in new technologies, infrastructure upgrades, and workforce training. These costs can be prohibitive for SMEs, limiting their ability to adopt advanced digital solutions and compete with larger entities \cite{Ko2021}.
    \item Escalating Cybersecurity Threats: As a consequence of digital transformation, a greater number of devices are being linked to the Internet, thereby augmenting the potential access points for cyberattacks. Assaults directed at critical infrastructure may lead to significant financial repercussions, harm to reputation, and even physical injury \cite{Homaei2024, Li2020}.
    \item Widening the digital divide: Digital transformation can exacerbate the digital gap by leaving behind people and communities who lack digital literacy or have restricted access to technology. This disparity impedes social and economic advancement and exacerbates existing inequalities \cite{AguileraCastillo2021}.
    \item Labour Displacement: The digital transformation of WDS should align with corporate social responsibility, addressing challenges like labor displacement caused by automation and AI. Organizations can mitigate job losses by fostering trust, equitable resource distribution, and social value through stakeholder engagement by introducing training and re-skilling programs to help workers transition to new roles. Integrating DT frameworks in water management advances technical efficiency, ethical stewardship, and workforce adaptability \cite{Callaway2024}.
    
    \item Privacy Issues: The significant collection and utilization of data have the potential to lead to privacy-related concerns. The implementation of digital technologies may be hindered by apprehensions regarding data security incidents or the handling of individual information, which could diminish public trust in these advancements \cite{Homaei2024, Homaei2022}.
    
    \end{itemize}

While digital transformation provides great dividends in terms of productivity and, therefore, innovative capacity in various industries, it needs to be pursued strategically and with awareness within WDS. In fact, critical challenges and risks threaten the successful digital transformation that service companies have to face. It is, therefore, especially relevant that a sound strategy, recognizing the peculiarities of WDS in terms of their complexity, criticality, and regulatory requirements, be developed. Through the adept management of these obstacles, electronic tools can facilitate the successful execution of digital transformation, thereby ensuring that its advantages are equitably shared among all stakeholders \cite{Shahi2020}. This holistic strategy will lay the groundwork for unlocking the complete potential of digital transformation in enhancing operational efficiency and promoting innovation throughout the industry.

\subsection{DTs in Water Industry}

This was born in the 1960s when most organizations, inclusive of NASA, applied physical twins of systems for the control of systems at remote locations, as was applied in the rescue of Apollo XIII \cite{ConejosFuertes2020, Haag2018, Ketzler2020}. These have evolved with time and debate to highly advanced virtual twins that simulate various ``what-if'' scenarios. This improves the city's ecosystem since the DT empowers planning and decision-making by highlighting the macro level of interaction that multiple components have, which enables these systems to work in a sustainable manner and with maximum efficiency.

The adoption of DTs across industries is novel, driving benefits from better product development to operations and decision-making. For example, electricity to produce more in manufacturing, to predict equipment failure, and to produce higher-quality products. The automobile industry uses them to simulate assembly lines, find out bottlenecks, and test new product designs \cite{Linda2022, YossefRavid2022, Bariah2022}. Healthcare - DTs model human organs or body parts to understand the conditions of patients better and plan surgeries. DTs in the construction industries are applied to fine-tune the designs of buildings and reduce energy consumption through simulations of wastes created by different materials, lighting, and ventilation systems. Transportation utilizes DTs for performance monitoring and predictive maintenance in view of safety and operational efficiency \cite{Homaei2024}.

 Henriksen et al. (2022) present the development of a novel type of DT called HIP DT for Adaptation to Climate Change, Water Management, and Disaster Risk Reduction. This paper delineates how the development and implementation of the national DK-model HIP have been done to realize real-time updating of simulations through the HIP portal. This approach focuses on how the betterment of Denmark's ability to cope with extreme weather events can be achieved by providing improved hydrological information. The system contributes to adaptive planning, ensures water security, reduces disaster vulnerability, and builds climate resilience by applying the latest hydrological modeling and machine learning techniques. Other important applications include real-time forecasting to help overcome the threats of flooding and drought, safeguarding water supplies, and supporting sustainable development projects \cite{Henriksen2022}.

Ibrahim Yousif's contribution \cite{Yousif2021} aims to leverage digital transformation in the water desalination process, improving smart facilities. The objective targeted here was to come up with DT models for two major items, namely a three-cylinder high-pressure pump energy recovery device-HPP-ERD and a three-stage RO membrane model. These DTs employ real-time data and IoT technologies to comprehensively monitor, detect faults, and predict them in the systems. Applying machine learning to higher-order signal processing will considerably increase the efficiency and reliability of the whole system. This research paper will present how digital transformation will result in an enhanced desalination process, reduced maintenance costs, and increased sustainability in the production of freshwater. 

Savic's work \cite{Savic2022}  presents a future of the water sector that is rather transformative, similar to what happened in the car and aerospace industries. Remote sensing, artificial intelligence detecting anomalies, and digital twins are identified as key enablers to enhance water management efficiency. This research puts a strong focus on embedding human capital into the process and using appropriate cybersecurity to manage risks related to automation. Lessons can be learned from the failures within other industries, such as those involving Tesla Autopilot and Boeing 737 MAX, and be applied for the advancement of digital transformation in the water industry, increasing its resilience and operational effectiveness.

\subsubsection{DTs in Water Field: Enhancing Efficiency and Sustainability}

DTs enjoy a wide range of applications from energy industries through increasing system efficiencies, reliability, and sustainability by adopting real-time monitoring, predictive maintenance, and data-driven decisions. Similarly, DTs hold huge promises of making paradigmatic changes in drinking WDNs management. The performances of the drinking water supply systems can be optimized by practicing various scenarios in the virtual replicas of the networks \cite{Zekri2022, Ramos2022}. Various most promising applications of DTs in drinking water WDNs are discussed in this section. 

One main use of DTs in WDN is enhancing hydraulic performance. By replicating how the network acts in various scenarios, operators can pinpoint possible problems like drops in pressure or water surges, allowing for knowledgeable choices regarding network layout and function. Moreover, DTs can identify areas at risk of pollution, enabling proactive measures to manage potential hazards \cite{ConejosFuertes2020, Bonilla2022, Henriksen2022}.

Besides, DTs improve the asset management of drinking WDNs. In other words, through virtual equivalent modeling of network assets, operators can test performance and predict faults before they occur. For example, the logical simulation of a pump or a valve operational behavior can allow operators to identify vulnerabilities and establish a plan for mitigating the potential damage, but also model the reaction of the network in case of power failures or natural disasters, among the quality of water.

Simulations of water flow in the network enable operators to identify possible locations of contamination and maintain good water quality accordingly \cite{Wei2022a, CardilloAlbarrn2021}. DTs also contribute to monitoring water consumption and detecting losses, reducing overall water waste and enhancing the overall efficiency of the network.

Hence, DTs are beneficial in planning emergency responses. Various contingency scenarios can be simulated to identify potential problems and develop effective contingency response strategies by operators. For example, simulating network behavior in the event of a power outage or natural disaster allows operators to identify the weak links and put in place strategies that reduce potential damage. In general, the output from DTs applied to WDNs can be summarized as optimization of hydraulic performance together with asset management, better monitoring of water quality, and emergency response planning. In such a way, the discussed technologies enable the operator to make data-driven decisions toward efficiency, reliability, and sustainability concerns in WDNs.

\section{Literature's Review}\label{sec:related} 
\subsection{Background Research} 

DTs in the water industry are improving the efficiency, reliability, and sustainability of water systems. In particular, WDS enables real-time monitoring and network behavior simulation. In this direction, the technology will enable predictive maintenance, better decision-making, and optimization of resources. DTs create a dynamic, virtual model of the WDN, bringing together real-time data, AI, and machine learning. This model helps identify potential issues, test scenarios, and understand the impacts of various decisions, reducing risks and improving service quality. This section reviews significant contributions to developing and applying DTs in WDS, highlighting advances and identifying ongoing challenges. Conducting a thorough and systematic literature review is paramount for identifying emerging trends, challenges, and solutions in applying DTs and AI/ML within WDS. Recent studies have emphasized the importance of refining literature review strategies to analyze big data trends and ensure research quality across different journal tiers \cite{MashhadiNejad2024}. Such methodological rigor helps build a robust theoretical foundation for understanding digital transformations in the water industry.

\subsubsection{Early Conceptual Foundations}

Early conceptual underpinning of DTs puts into consideration the problems regarding efficiency and sustainability in water treatments and distribution systems through sophisticated modeling combined with real-time data. Curl et al. (2019) focus on how DTs can provide in-depth and real-time representations of water treatment processes for predictive optimization and control. The approach has greatly optimized chemical dosing, energy use, and resource utilization to reduce operation costs and minimize environmental impact. For example, a case study at a North Carolina water treatment facility reported 10\% chemical savings with a 2\% improvement in water quality due to optimized coagulant dosing \cite{Curl2019}.

Conejos et al. (2020) further elaborate on developing and implementing DTs for managing drinking WDNs. They also refer to the main functionalities of the DTs: "Precise modelling of network behaviors, continuous data integration from systems like GIS, AMR, and SCADA are other advanced capabilities that include optimal network design, asset management, leak detection, and simulations of emergency responses." The case study performed on the WDN of Valencia-which supplies 1.6 million inhabitants-demonstrates remarkable added value regarding real-time monitoring, predictive maintenance, and efficiency enhancement of operations. This exemplifies the scalability and efficacy of decision trees in overseeing intricate water management systems, highlighting their potential as essential instruments for decision-making within contemporary water utility operations \cite{ConejosFuertes2020}.

Giudicianni et al. (2020) provide this basis by discussing the integration of advanced energy management and leakage control technologies within smart water grids. The study discusses the role of DTs, CPS, and blockchain in optimizing WDSs. Its core focus rests on the recovery of energy, and that is achieved by installing micro-turbines and PATs instead of traditional PRVs through the enhancement of energy efficiency together with leakage control. The study has also identified that the segmentation of water networks into DMAs is effective in improving monitoring and management. Giudicianni et al. trust that the integration of the concept of digital water will contribute significantly towards enhancing the sustainability and resilience of urban water systems to enable upcoming smart city development \cite{Giudicianni2020}.

\subsubsection{Expansion and Application in Various Water Systems}

The application of DTs in water systems has immensely increased, proving their flexibility and effectiveness for use across various environments. Valverde et al. (2021) extend the explanation of how DTs improve the performance and efficiency of the water infrastructure with respect to operational decision-making for sewer networks and water resource recovery facilities. The integration of real-time data and advanced models within the DTs allows for operational optimization and smarter handling of data in complex water systems. Case studies from Global Omnium in Valencia, Spain, Aarhus Vand in Denmark, and DC Water in the United States further illustrate that, in general, the technology of digital twins allows for real-time monitoring, integration, and holistic management of systems, underlining their flexibility in a variety of water management contexts \cite{valverde2021digital}.

Garrido-Baserba et al. (2020) relate the fourth revolution to the digitization impelled by big data and artificial intelligence in the water sector. These technologies enhance operation, maintenance, and sustainability when integrated with urban water infrastructure. AI and big data analytics enable real-time insights and predictive capabilities, changing how decisions are made, resources are recovered, and assets are managed. This digital transition will bring changes in the socio-economic perspective, it will influence new business models, and it will demand new research and also new workforce development to meet the future demands of the water sector \cite{GarridoBaserba2020}. 

Pedersen et al. 2021 investigate the Living and prototyping DTs for an urban water system that is focused on multi-purpose value creation via models and sensors. While a living DT performs real-time operational and control functionalities, in prototyping DT, it would be used for design and planning without any real-time data coupling. Implementation at VCS Denmark showed how the system management can be empowered with data links, simulation models, and enhanced analytics. This is emphasized through open data standards and intersectoral collaboration to maximize the impacts of DTs for efficiency, resilience, and sustainability in the management of urban water \cite{Pedersen2021}.

Hietala et al., 2021 present different forms of collaboration in the digital transformation of municipal wastewater management, focusing on inter-organizational cooperation between Finnish water utilities. They identify four main modes: autonomous, limited company, central service organization, and standardization. These collaboration modes enhance ICT development and deployment, providing benefits such as predictive maintenance, efficient resource allocation, and improved data integration. While autonomous development is prevalent, collaboration through limited companies and standardization offers significant advantages in managing digital transformation challenges and improving overall operational efficiency and sustainability \cite{Hietala2021}.

Van Rooij et al. (2021) proposed a DSS based on DTs to recover membranes in RO desalination plants. The study addresses bio-fouling caused by algal blooms, significantly impacting membrane efficiency. The DSS creates a DT of an RO vessel to evaluate maintenance strategies, including membrane cleaning, swapping, and replacement. Applied to the Carlsbad Desalination Plant, this approach optimizes maintenance schedules, reduces operational costs, and enhances the reliability and longevity of membrane systems, setting a new standard for membrane management in the desalination industry \cite{VanRooij2021}.

Udugama et al. 2021 explores the challenges and opportunities of applying Digital Twins to bio-manufacturing, highlighting potential gains in process efficiency and resource optimization. The following work details a five-step methodology for developing a full DT, starting with a basic steady-state process model and culminating in an advanced, validated model complete with bidirectional communications. A bench-scale ethanol fermentation serves as a test bed to demonstrate the improved monitoring and control functionalities of the developed framework. Key challenges include the need for high-fidelity models, real-time data integration, and operator interaction, with suggestions for future research to optimize biomanufacturing operations \cite{Udugama2021}.

Botin et al. (2021) study DT applications representing urban spaces and vehicles to create a Living Laboratory and demonstrate how DTs in urban environments can enable interaction in achieving the United Nations Sustainable Development Goals. The work describes a method for the development of DTs using a network of vehicle sensing devices which, by processing in real-time via edge computing, modelling software, and machine learning algorithms, resulting in DTs that shall be used to analyze the urban space evolution, mobility, and the interaction of vehicles for the valuation of insights into urban planning and improvements of infrastructure toward sustainable and resilient urban environments \cite{Botin2021digital}.

\subsubsection{Developing Advanced Methodologies and Frameworks}

The development of new methodologies and frameworks in WDN management has been strongly improved by combining DTs with legacy algorithms and innovative technologies. Ciliberti et al. (2021) propose a new digital transformation paradigm in the WDNetXL/WDNetGIS platform, focusing on life-cycle management and operational efficiency. The essential services, such as the Digital Water DMA Analyzer, optimize district metering area design for leakage reduction through pressure control, and the Digital Water Rehabilitation provides optimal pipe replacement plans. These digital services significantly improve the management and sustainability of WDNs by real-time data exploitation and advanced modeling techniques \cite{Ciliberti2021}.

Ramos et al. (2022) talk about the integration of DT technologies to enhance WDN efficiency, showing substantial water savings and improvement in system operations through optimization algorithms and real-time data collection using GIS. A case study in Lisbon shows the ability of DTs to bring about 28\% water savings through quick leak detection and optimal pressure control valve settings. This research puts forward the transformative effect of DTs on smart water management, emphasizing reduced water and energy losses and promoting sustainable operations \cite{Ramos2022}.

Pesantez et al.(2022) assess the impact of the COVID-19 pandemic on water infrastructure based on a DT framework by fusing AMI information with hydraulic modeling. The study highlights the dramatic changes in residential and nonresidential water use patterns due to social distancing measures, which are creating increased pressures and increased water age within the distribution system. This approach really depicts the potential of DTs in providing real-time insight and enhancing operational decision-making to improve infrastructure resilience during unforeseen events \cite{Pesantez2022}.

Bonilla et al. (2022) provide a state estimation methodology for WDSs using GCNs coupled with hydraulic models to constitute a DT. This framework estimates the pump speeds from available data on pressure and flow rates and makes a very accurate prediction of the system's hydraulic state. Validated in benchmark networks, this methodology shows high predictive accuracy and thus has great potential for DTs in improving WDS monitoring, management, and anomaly detection to improve overall operational efficiency and reliability \cite{Bonilla2022}.

Zekri et al. propose a smart water management framework using intelligent DTs and multi-agent systems in order to increase efficiency and distribution of water resources. This five-layer framework uses intelligent agents for data analytics, simulation of asset operation, and feedback from users in real-time. The framework is designed with much emphasis on autonomy and intelligence, using a reward-based mechanism to incentivize optimal water consumption—how DTs can improve asset management, leak detection, and system operations for sustainable use of water resources \cite{Zekri2022}.

Matheri et al. (2022) review the integration of DTs, AI, and data-driven optimization in wastewater treatment plants for sustainable circularity and intelligent operations. The study highlights how CPS can be deployed in real-time for monitoring, predictive maintenance, and enhancement of decision-making frameworks. In this regard, adopting circular bio-economy approaches will transform wastewater treatment into resource recovery facilities that support sustainable development goals. Implementations of these technologies under which considerable operational cost savings, significant improvements in system reliability, and compliance with better environmental standards are realized truly illustrate the transformative capability of DTs in wastewater management \cite{Matheri2022}.

\subsubsection{Recent Developments and Future Directions}

Recent progress in developing digital technologies for water systems has focused on improving real-time monitoring, operational efficiency, and sustainability by using new applications and advanced frameworks. Dodanwala et al. (2023) present a digital technology framework for LOS in relation to potable water infrastructure systems. This integrates real-time data acquisition with established service standard benchmarks so that the performance evaluation and management of the water infrastructures can be automated. This framework improves strategic decisions as well as operational efficiency in the development of a cyber equivalent for a physical infrastructure that is facing aging infrastructure and/or variable service conditions. It follows ISO 55000 guidelines based on dynamic data integration and automatic LOS assessments to ensure sustainability and resilience in the water delivery service \cite{Dodanwala2023}.

Ramos et al. (2023) confirm both applications of the SWG and DT technologies at the Gaula Water Distribution Network, Madeira, Portugal. Real-time monitoring, scenario analysis, and optimized pressure control were found to reduce significant water loss and improve the performance of the system. The DT model allowed for a potential reduction of 80\% in water losses, saving about EUR 165,000 every year. This study underlines the importance of using DT and SWG for effective modernization of water management practices toward sustainable water usage and increasing the efficiency of the network, in general \cite{Ramos2023smart}.

Overall, Grievson et al. (2022) describe the integration of digital solutions into water management: real-time optimization, total transparency, and predictive maintenance. A number of cases are presented that have underlined the transformative potential that digitalization can show in areas such as reduction of non-revenue water, improvement of the quality of water, and the management of infrastructure. Alongside opportunities, challenges about cybersecurity, data quality, and legacy systems call for collaboration in a multi-stakeholder perspective able to ensure digital transformation success. The research promotes creating future-proof frameworks that support innovative technologies, driving sustainable development and equitable water management practices \cite{Grievson022}.

Fu et al. (2022) introduced the development of DTs for biomanufacturing, considering a general framework for developing a full-fledged DT starting from a basic steady-state process model. For a case study on second-generation ethanol fermentation, this presented DT framework has been able to illustrate productivity enhancements of 20-33\%. The following critical success factors have been identified: modeling accuracy, human operator actions, and economic value proposition of the models. Notwithstanding these benefits, the study recognized that for such interaction to be effective, highly advanced digital infrastructure, together with carefully designed HMIs that will help enhance the accuracy and robustness of the DT system \cite{Fu2022}.

Pedersen et al.(2022) introduce a diagnostic framework for addressing uncertainties in integrated urban drainage models used in living DTs. The framework enhances iterative model improvements by classifying errors across urban drainage system components using hydro-logic and hydraulic signatures from water level sensors. Applied in Odense, Denmark, this approach reveals discrepancies in model inputs, structural attributes, and temporal variability, improving model accuracy and transparency. This diagnostic framework leads to more reliable and efficient urban water management \cite{Pedersen2022}.

Gino Ciliberti et al. (2023) present a transformative approach to the digitization of WDNs by developing standardized methods for creating DTs for WDNs. Advanced hydraulic modeling is combined with AI, machine learning, and deep learning techniques; thus, a conceptual framework for WDN digital transformation is presented. This framework integrates the whole DTs with advanced network analysis, developing DWSs as plugins for QGIS software: Enhancing WDN planning, management, and design, continuous improvement of their digital representation, and improving technical decision-making and overall efficiency \cite{GinoCiliberti2023}.

Additionally, Ramos et al. (2023) illustrated how DTs transform system operation and maintenance. DTs allow the detection of patterns in historical and real-time sensor data and thus inform predictive maintenance strategies, avoiding sudden failures and reducing costs to the very minimum. These technologies may be one of the primary drivers enabling the SDGs due to their reduced environmental footprint through better water management, pressure management, and resource preservation. There are still barriers, however, regarding scaling up with currently existing infrastructure in terms of data management issues, regulatory issues, and cybersecurity concerns. Further research is needed to improve forecasting water consumption and detecting leaks to fully leverage DT technology in enhancing WDSs \cite{Ramos2023}.

Torfs et al. (2024) introduce WRRF models transitioning into DT applications with a focus on how to overcome a lack of consensus in DT definition and application. The main differences from traditional simulation models toward DTs point to continuous and automated data links and dynamic updating. Integrating mechanical with data-driven models into hybrid frameworks enhances their predictive power, hence operational efficiency. Examples of success stories range from Changi Water Reclamation Plant to Kolding WRRF and have shown that decision improvements in operation and real-time optimizations have been possible by the implementation of DTs. Successful DT deployment in WRRFs needs a holistic approach that includes stakeholder buy-in and adequate data management \cite{Torfs2024}.

Menapace et al. (2024) present a proof-of-concept about optimal sensor placement using GNNs to support the development of DTs for WDSs. The paper presents a novel methodology which uses GNNs in the evaluation of pressure at the consumption nodes, guiding the optimal configuration of sensors with the goal of minimizing the estimation error. Applied to a synthetic case study, the approach demonstrated high accuracy in pressure estimation across various sensor configurations, highlighting the potential of GNNs in enhancing the accuracy and reliability of DTs in WDNs. This innovative method supports more effective monitoring and management of water systems, promoting improved operational efficiency and resource utilization \cite{Menapace2024}.

\pagebreak

\begin{scriptsize}
\begin{longtable}{ p{0.7cm} p{2cm} p{3cm} p{3cm} p{2cm} p{3cm}}
\caption{Comparison of DT Implementations in the Water Industry}
\label{table:comparative-analysis} \\
\hline
\textbf{No} & \textbf{Focus Area} & \textbf{Contributions/Goals} & \textbf{Methods} & \textbf{Case Studies/Apps} & \textbf{Outcomes} \\
\hline
\endfirsthead
\caption*{Comparison of DT Implementations in the Water Industry (Continued)} \\
\hline
\textbf{No} & \textbf{Focus Area} & \textbf{Contributions/Goals} & \textbf{Methods} & \textbf{Case Studies/Apps} & \textbf{Outcomes} \\
\hline
\endhead
\cite{Curl2019} 2019 & DTs in water treatment & Real-time optimization, sustainability & DTs, real-time data & North Carolina water treatment facility & 10\% reduction in chemical usage, 2\% water quality improvement \\
\hline
\cite{ConejosFuertes2020} 2020 & WDNs & Real-time monitoring, predictive maintenance & DTs, GIS, SCADA & Valencia, Spain & Improved operational efficiency, support for 1.6 million inhabitants \\
\hline
\cite{Giudicianni2020} 2020 & Smart water grids & Energy efficiency, leakage control & DTs, CPS, blockchain & Various implementations & Enhanced sustainability, resilience of urban water systems \\
\hline
\cite{Sun2020} 2020 & Urban water cycles & Real-time control, subsystem interoperability & Cyber-physical systems, model predictive control & Barcelona, Badalona & Reduced combined sewer overflows, optimized water usage \\
\hline
\cite{valverde2021digital} 2021 & Water infrastructure & Operational decision support & DTs, real-time data & Global Omnium, Aarhus Vand, DC water & Versatility and effectiveness in different contexts \\
\hline
\cite{Pedersen2021} 2021 & Urban water systems & Real-time operational and control functionalities & Living and prototyping DTs & VCS Denmark & Improved system management, efficiency, resilience \\
\hline
\cite{Hietala2021} 2021 & Wastewater Management & Inter-organisational collaboration & Autonomous, limited company, central service organisation, standardisation & Finnish water utilities & Enhanced ICT development, predictive maintenance \\
\hline
\cite{VanRooij2021} 2021 & Desalination plants & Membrane maintenance optimisation & DTs, decision support systems & Carlsbad desalination Plant & Reduced operational costs, enhanced membrane efficiency \\
\hline
\cite{Udugama2021} 2021 & Bio-manufacturing & Process efficiency, resource utilisation & DTs, real-time data integration & Ethanol fermentation process & Enhanced monitoring and control capabilities \\
\hline
\cite{Botin2021digital} 2021 & Urban spaces and vehicles & Sustainable urban environments & DTs, sensing devices, edge computing, machine learning & Various urban applications & Improved urban planning and infrastructure \\
\hline
\cite{Ciliberti2021} 2022 & WDNs & Asset management, pressure control & DTs, established algorithms & Lisbon, Portugal & 28\% water savings, reduced water and energy losses \\
\hline
\cite{Pesantez2022} 2022 & Water infrastructure & Impact assessment of COVID-19 & DTs, AMI data, hydraulic modelling & Mid-sized utility serving 60,000 people & Altered water demand patterns, improved operational decision-making \\
\hline
\cite{Bonilla2022} 2022 & WDSs & State estimation methodology & DTs, graph conventional networks & Two benchmark networks & High predictive accuracy, improved WDS monitoring \\
\hline
\cite{Zekri2022} 2022 & Smart water management & Water resource efficiency & DTs, multi-agent systems & Various implementations & Improved asset management, leak detection, system operation \\
\hline
\cite{Matheri2022} 2022 & Wastewater treatment plants & Process efficiency, resource recovery & DTs, AI, CPS & Various implementations & Reduced operational costs, enhanced system reliability \\
\hline
\cite{Dodanwala2023} 2023 & Potable water infrastructure & Real-time data collection, bench-marking & DTs, Levels of Service framework & Various implementations & Enhanced decision-making, operational efficiency \\
\hline
\cite{Ramos2023smart} 2023 & WDNs & Water loss reduction, system performance & Smart water grids, DTs & Gaula WDN, Madeira, Portugal & 80\% water loss reduction, EUR 165,000 annual savings \\
\hline
\cite{Grievson022} 2023 & Digital transformation & Digital solutions integration & DTs, real-time optimisation & Various case studies & Enhanced operational efficiency, resource management \\
\hline
\cite{Fu2022} 2023 & Biomanufacturing & DT framework & DTs, human-machine interfaces & Ethanol fermentation & 20-33\% productivity improvement \\
\hline
\cite{Ramos2023} 2023 & System operation and maintenance & Predictive maintenance, sustainability & DTs, historical and real-time data & Various implementations & Minimized maintenance costs, improved sustainability \\
\hline
\cite{GinoCiliberti2023} 2023 & WDNs & Digital transformation, AI integration & DTs, AI, machine learning, deep learning & QGIS software plugins & Improved planning, management, and design of WDNs \\
\hline
\cite{Torfs2024} 2024 & Water resource recovery & Real-time optimization, stakeholder involvement & DTs, hybrid frameworks & Changi water reclamation plant, Kolding WRRF & Significant operational efficiency improvements \\
\hline
\cite{Pedersen2022} 2024 & Urban drainage models & Diagnosing uncertainties & DTs, hydrologic and hydraulic signatures & Odense, Denmark & Improved model accuracy, enhanced urban water management \\
\hline
\cite{Menapace2024} 2024 & WDSs & Optimal sensor placement & DTs, graph neural networks & Synthetic case study & High accuracy in pressure estimation, improved monitoring \\
\hline
\end{longtable}
\end{scriptsize}

\subsection{Gaps in Research}

Despite all the advancements in applying DTs within the water industry, there still remain some major research gaps. A major gap is the integration and effective use of state-of-the-art AI and ML models with DTs for WDSs. While AI and ML hold prospects to enhance predictive analytics and operational efficiency, real-world applications are mostly curbed by the shortage of good-quality, granular time-series data needed to train strong models. The complexity of water consumption patterns and the requirement for much feature engineering make it difficult to implement models such as LSTM, Prophet, LightGBM, and XGBoost. It reflects the reality that AI models should be both accurate and generalizable within different contexts of the water industry.

Another critical gap is the optimization of maintenance operations within WDSs using DTs and advanced optimization algorithms, such as CP. How to integrate DTs with sophisticated scheduling models to address complex maintenance tasks with dynamic priorities and dependencies remains a less-explored avenue. This calls for setting up new research efforts toward developing scalable optimization methods for adaptation to the dynamic nature of WDSs that minimize operational costs and environmental impacts.

In addition, there is a need to develop scalable digital twin architectures that can handle the complexities involved in water distribution networks by considering diverse data sources, such as IoT sensor data, historical maintenance records, and environmental information. Current digital twin applications often focus on specific aspects, like pipeline health monitoring or water quality management, rather than working on an integrated view of all subsystems and external factors that influence water distribution. Such creation of integrated digital twin platforms would enable smooth interoperability between the different components involved, such as water quality assessment, leak detection, consumption prediction, and maintenance planning. That would need some research into data standardization methodologies and interoperability.

Successful DT adoption also needs a skilled workforce capable of managing these advanced systems. It is an imperative task to research programs for education and training tools to close the digital skill gap in the water industry. This will enable smoother transitions to digital platforms and realize maximum benefits from DT implementations.

Furthermore, digitalization in WDSs, especially regarding regulatory compliance and cybersecurity, is under-explored. With the increasing prevalence of digital systems, there is an acute need to ensure secure and resilient interconnected water distribution infrastructures from cyber threats. A great deal of research will be required to ensure DT platforms comply with relevant regulations, including data privacy policies like GDPR and ethical policies. Frameworks that properly balance operational efficiency against legal and moral imperatives will be critical to ensure both effective and responsible digital transformation of WDSs. The creation of the DT should, therefore, be a priority, enabling the discovery of those errors only when such a DT is created; hence, this paper proposes a DT platform for better and more precise water resource management in order to manage efficiently the support and maintenance of the water network.

\section{Proposed DTs Platform}\label{sec:proposed}

This manuscript introduces the CAUCCES platform, a DT-based system integrating AI models for water consumption forecasting, IoT infrastructures for gathering data in real-time, and cybersecurity measures compliant with the regulatory context. Successfully deployed in \href{http://www.ambling.es}{Ambling} \cite{ambling}, an SME in charge of the management of rural water supply services, the platform tackles challenges like water scarcity and demographic decline through showing how advanced sensing, communication, and information technologies may be beneficial in enhancing governance around water resources.

The platform provides an integrated framework for real-time monitoring and decision-making and digitization of the maintenance process for water distribution networks. It optimizes maintenance by using CP-based scheduling with DT technology to bring operational efficiency and service quality to a higher level, along with system reliability. Inspired by solutions for urban areas, CAUCCES tailors these technologies to the demands of rural needs, enabling functionalities such as remote meter readings, water quality monitoring, demand forecasting, and scenario-based system simulations.

Digital Twins, at the heart of the platform, are dynamic digital models of physical systems integrated in conformity with real-time sensor data to increase a system's efficiency and sustainability. This makes CAUCCES an enduring solution to the challenges of modern-day water distribution.

\begin{figure}[H]
\centering
\includegraphics[width=1.0\textwidth]{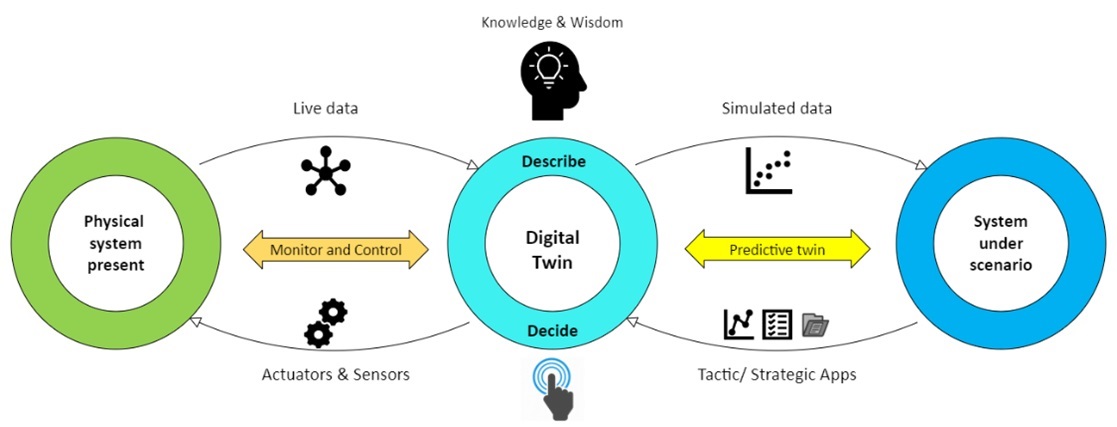}
\caption{\label{fig:DT}DTs in the WDS.}
\end{figure}

Figure \ref{fig:DT} classifies DT applications into operational and strategic. In an operational sense, DTs support monitoring and control; as such, they can also detect anomalies, optimizing any process—for instance, the automation of a water production facility. At a strategic level, "predictive twins" leverage simulated data to model potential scenarios to drive design changes in systems, train personnel, and simulate incidents proactively.

Models are usually used by water utilities for the prediction of system behavior under normal conditions. However, actual performance deviates due to real demand, maintenance, and aging infrastructure. By integrating these factors, a DT creates a dynamic, real-time model that continuously adapts to changes in order to improve daily operations and long-term planning. This is how DTs have the potential to transform the management of water from static models to dynamic, scenario-based tools that empower utilities to navigate the future with confidence.

\subsection{CAUCCES Architecture}

\begin{figure}[H]
    \centering
    \includegraphics[width=1\linewidth]{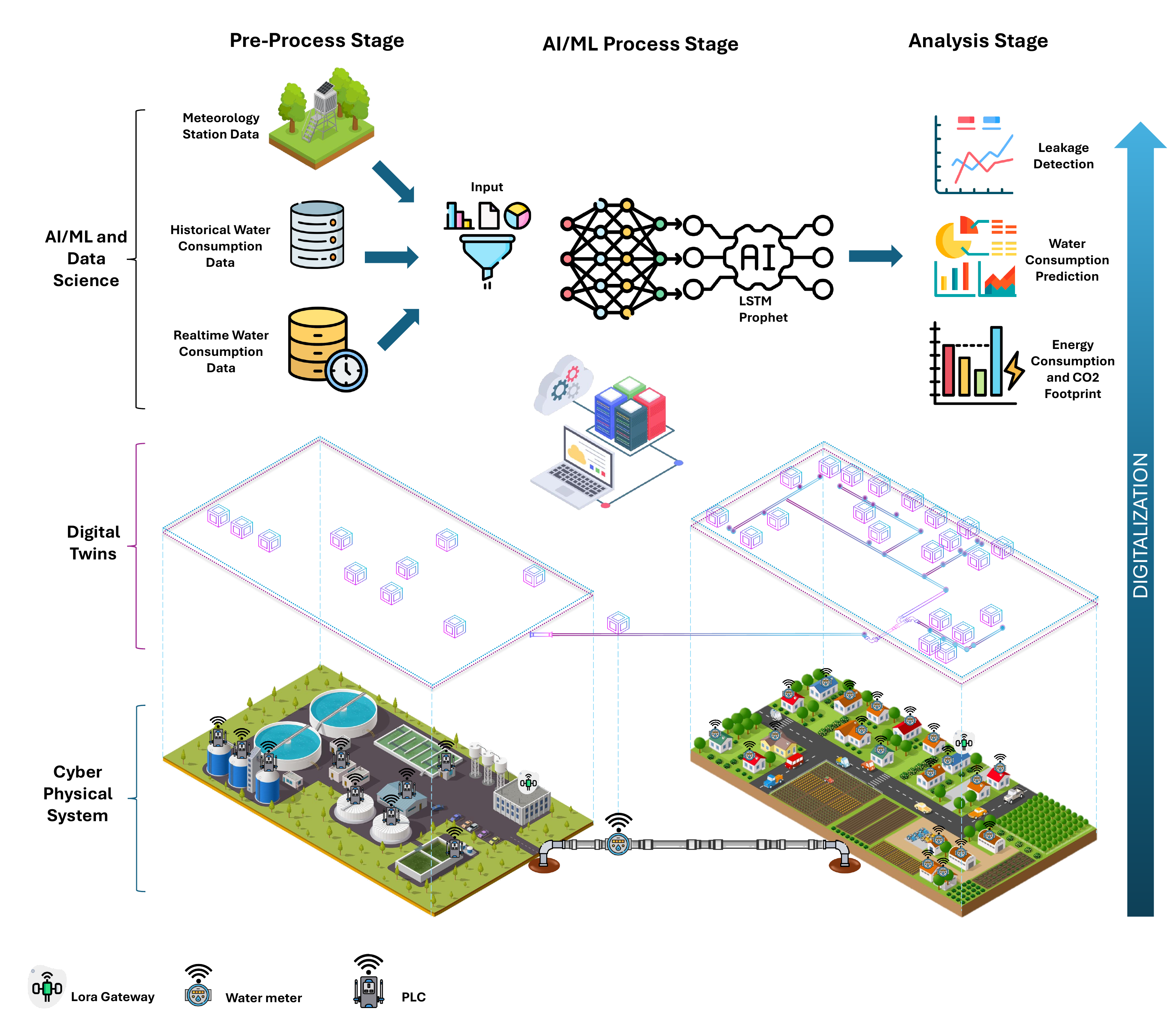}
    \caption{DTs Platform in the Water Distribution Networks}
    \label{fig:DT_Water}
\end{figure}

Figure \ref{fig:DT_Water} presents the architecture of the CAUCCES Smart Water Management System, which has been meticulously designed to integrate diverse data sources and cutting-edge technologies to enhance water management efficiency significantly. The architecture is structured into several interconnected layers, each tailored to perform specific functions contributing to the system's overall effectiveness and operational intelligence.

\subsubsection{Cyber Physical System}

The CPS represents the interconnected physical assets and digital system components that facilitate data generation, collection, and transmission. It comprises two key elements that work together to ensure the continuous flow of information from physical sensors to computational platforms:
\begin{itemize}

    \item \textbf{IoT Networks:}  These networks include IoT devices located throughout the WDS, including PLCs, meters, pumps, pipes, and gauges. Real-time data is collected at the various treatment, transmission, and distribution stages of water. The data collected becomes important in monitoring the performance of the system and determining inefficiencies or faults in the network. The deployment of IoT networks ensures extensive coverage of the entire water infrastructure with high-accuracy data capture.
    
    \item \textbf{Communication:}  The communication layer is important for the lossless and timely delivery of data gathered by IoT devices to downstream systems. In our architecture, data from IoT devices connected via the LoRaWAN protocol are aggregated and then securely transmitted to cloud platforms for further processing; thus, this ensures that the system has strong and scalable data handling, maintaining the integrity and timeliness of critical information across the network.

\end{itemize}

\subsubsection{DTs and System Integration Layer}

\begin{figure} [H]
    \centering
    \includegraphics[width=0.7\linewidth]{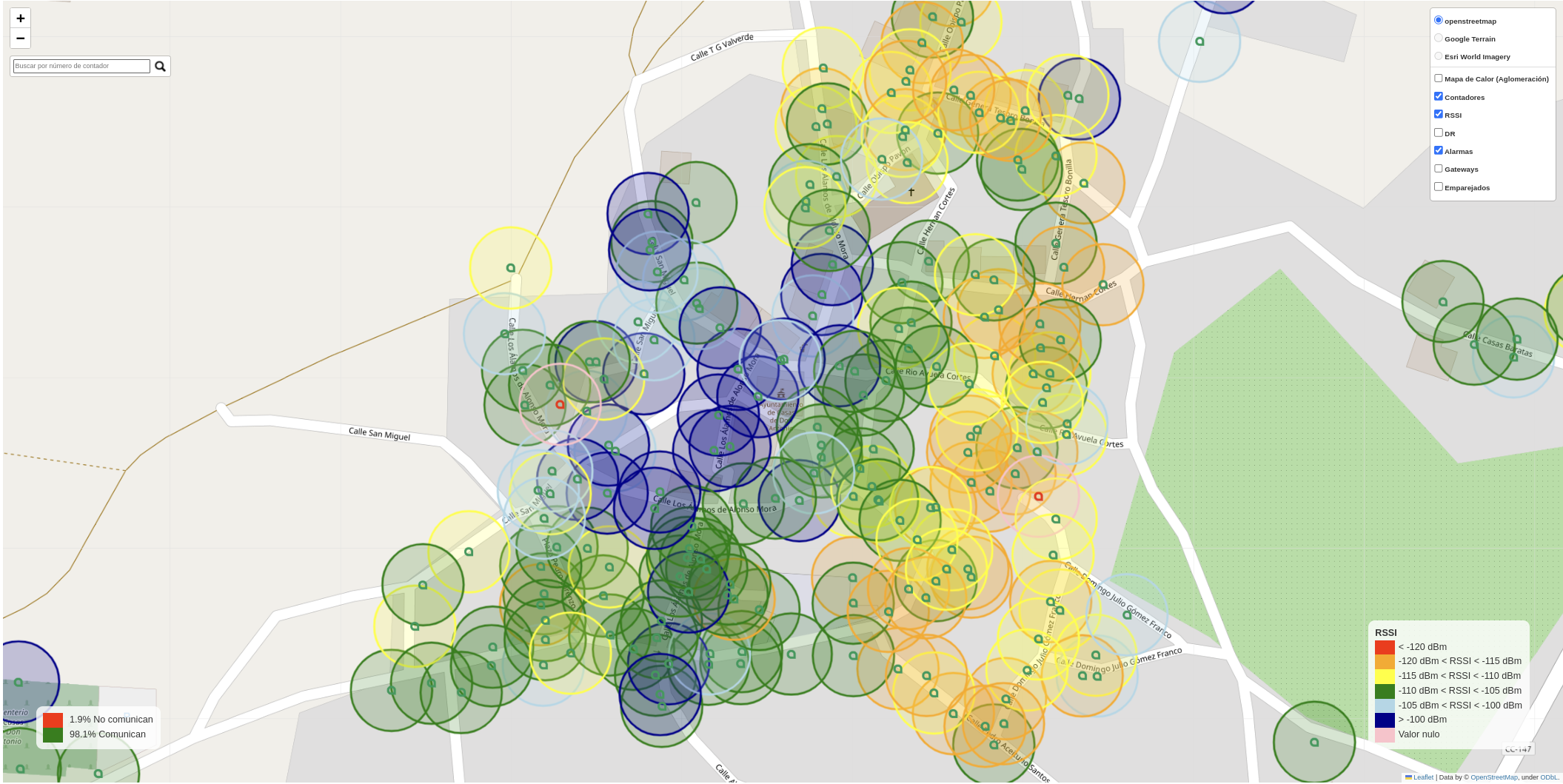}
    \caption{2D Map of the village and communication signals}
    \label{fig:2DMap}
\end{figure}

\begin{figure}[H]
    \centering
    \includegraphics[width=0.7\linewidth]{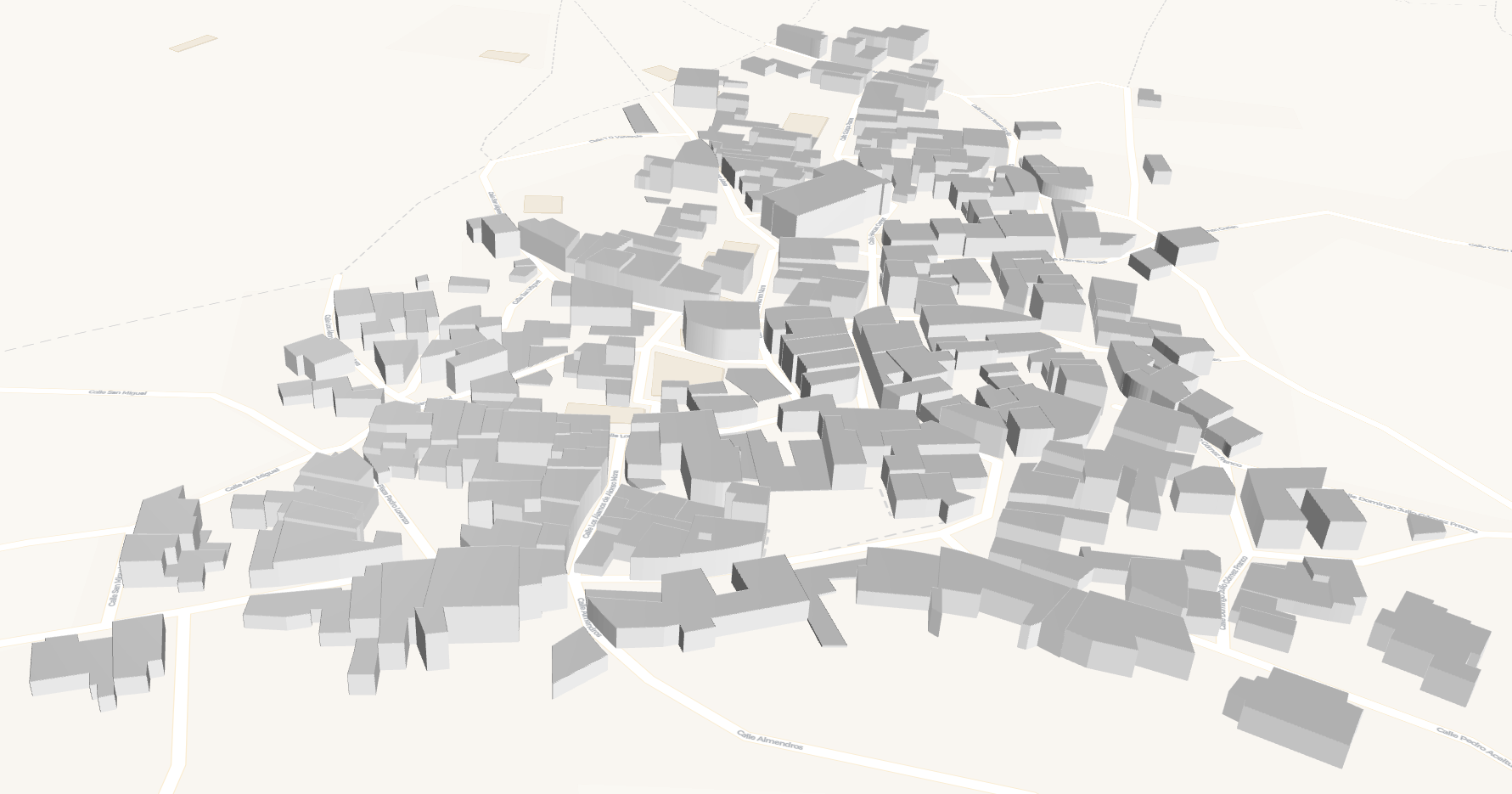}
    \caption{3D Map of the village}
    \label{fig:3DMap}
\end{figure}

At the heart of this architecture lies the DT component, which creates accurate virtual replicas of physical water systems. This core element encompasses various models and a computational engine that transforms real-world data into digital entities. It also enables seamless communication between the physical and virtual systems, allowing for higher-level analysis, optimization, and decision-making:

\begin{itemize}

    \item \textbf{API and Data Management Layer:} This layer is essential for data integration and management. It controls access, processing, and the exchange of data within the system in such a way as to provide a safe and orderly environment for the sharing of data through APIs. The platform provides access to the controlled data and functionality while allowing integration with other external systems; therefore, the collaboration between two or more platforms in operations becomes much more effective. The layer also ensures the system can handle large volumes of data efficiently, supporting scalability and compliance with regulations such as GDPR.

    \item \textbf{Visual Representation in GIS:} Although not a self-contained layer, Geographic Information Systems (GIS) are truly part of the system architecture. GIS provides visual insight by displaying data and results of analyses on geographic maps, which significantly enhances spatial analysis. The platform supports both two-dimensional and three-dimensional mapping capabilities, as illustrated in Figures \ref{fig:2DMap} and \ref{fig:3DMap}, respectively, which are pivotal in visualizing and interpreting the geographic information, fundamental for both operational and strategic decision-making in the water management domain.

\end{itemize}

\subsubsection{Processing and Analytics Layer}

The processing and analytics layer transforms raw data into actionable insights in the data analysis phase. This stage handles multiple data sources, categorized into real-time and historical data streams. The data typically undergo pre-processing, cleaning, and transformation on servers within this layer to ensure it is ready for optimization and predictive models. The critical components of this layer include:

\begin{itemize}
    \item Input Data: This layer gathers diverse types of data essential for analysis:
    \begin{itemize}
        \item Meteorological data is sourced via APIs from regional weather stations, providing crucial environmental context.
        \item Historical data consists of daily and even hourly water consumption records for the village, stored in dedicated databases for trend analysis and forecasting.
        \item Real-time data from meters, pumps, and PLC devices are continuously captured and stored in relevant databases, feeding into real-time monitoring and analysis models.
    \end{itemize}
    
    \item Artificial Intelligence Models and Forecasting: A key component of this layer is the AI/ML process stage, where artificial intelligence and machine learning models analyze historical and real-time data. To forecast water consumption, we employed a combination of advanced models to forecast water consumptions, including LSTM, Prophet, LightGBM, and XGBoost. These models enable the system to perform extensive optimization and computational tasks, predict future water demand, detect network leaks, and calculate energy usage and carbon dioxide emissions. Integrating time series data allows for highly accurate forecasting and efficient resource management \cite{khazrak2023}.
    
    \item Business Intelligence and Dashboard Tools: Processed data is presented visually through intuitive dashboards and detailed reports, providing users with real-time insights and historical data analysis. These tools empower decision-makers to make informed, data-driven choices, ensuring operational efficiency and long-term sustainability in water management.
   
\end{itemize}

\section{Integrating AI/ML in Water Consumption Prediction}\label{sec:LSTM/Prophet/LightGBM}

\subsection{Data Aggregation and Pre-Processing}

Water consumption data, originally recorded at 15-minute intervals, is aggregated into daily, weekly, and monthly totals to align with the temporal granularity of meteorological data. This alignment allows for practical correlation analysis at different timescales.

\subsection{Meteorological Variables}

\begin{figure}[H]
    \centering
    \includegraphics[width=0.96\linewidth]{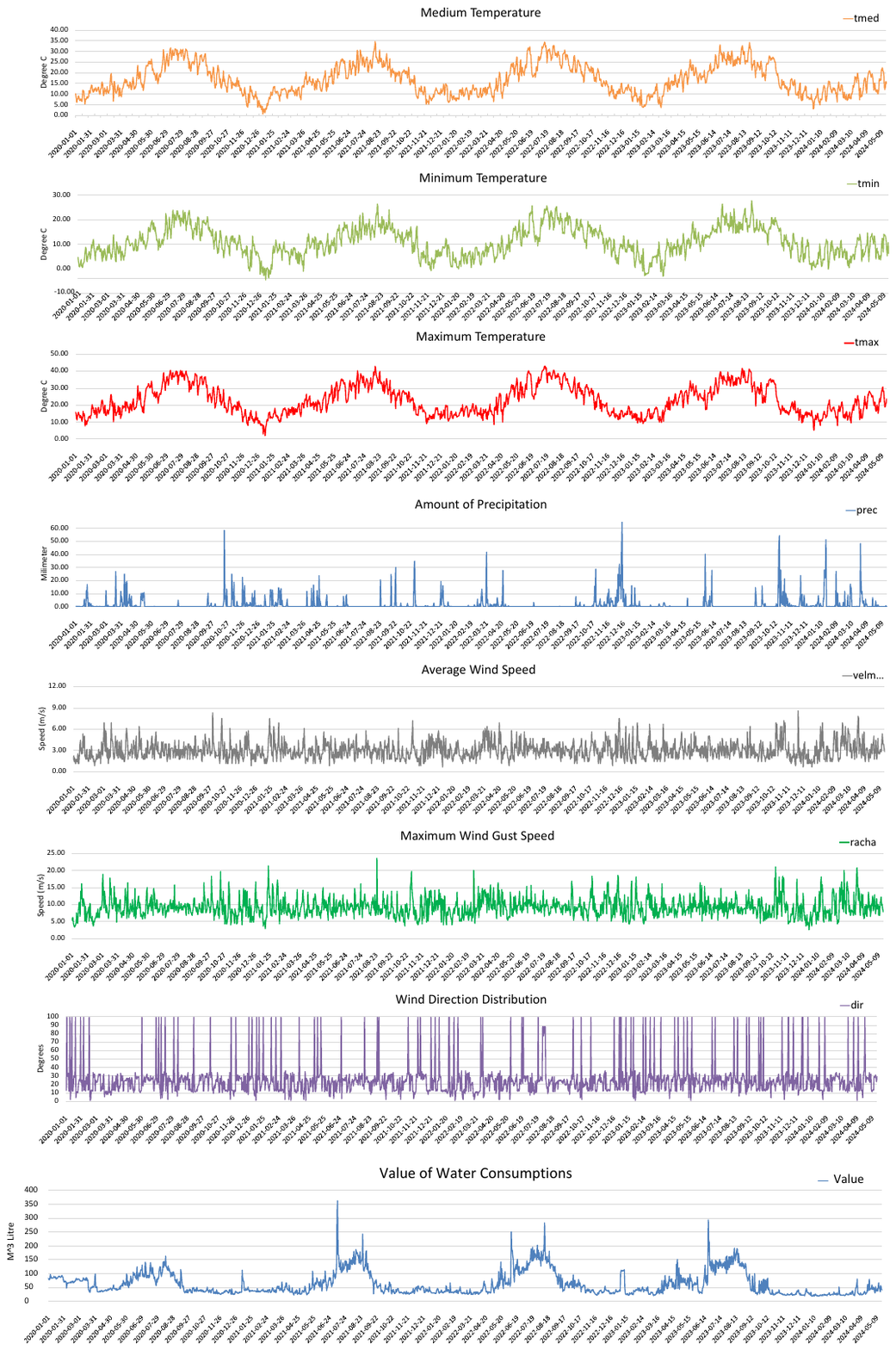}
    \caption{Meteorology and Water Consumption Data}
    \label{fig:Temperature}
\end{figure}

Data from the AEMET \cite{AEMET_OpenData} provides essential information about climatic conditions, augmenting the water consumption data and enabling the extraction of consumption patterns. These variables are listed in Table \ref{tab:meteorological_variables} and Figure \ref{fig:Temperature}.

\begin{table}[ht!]
\centering
\caption{Meteorological Variables from AEMET for Water Consumption Analysis}
\begin{scriptsize} 
\begin{tabular}{>{\arraybackslash}p{2.5cm} 
                >{\arraybackslash}p{12cm}}
\hline
\textbf{Variable} & \textbf{Description} \\
\hline
Date & The date when the data was recorded (year-month-day format). \\
Tmed & Average air temperature in °C, calculated from daily max and min temps. \\
Prec & Total precipitation in millimeters accumulated during the day. \\
Tmin & Minimum air temperature in degrees Celsius recorded during the day. \\
Hourtmin & The time (hh:mm format) when the minimum air temperature was recorded. \\
Tmax & Maximum air temperature in degrees Celsius recorded during the day. \\
Houratmax & The time (hh:mm format) when the maximum air temperature was recorded. \\
Dir & Average wind direction, derived from 10-minute instantaneous recordings, in degrees. \\
Velmedia & Average wind speed, derived from 10-minute instantaneous recordings, in m/s. \\
Maxvel & Maximum wind speed in meters per second recorded during the day. \\
Hourracha & The time (hh:mm format) when the maximum wind speed was recorded. \\
Sun & Duration of sunshine in hours recorded during the day. \\
PresMax & Maximum atmospheric pressure in hectopascals recorded during the day. \\
HouraPresMax & The time (hh:mm format) when the maximum atmospheric pressure was recorded. \\
PresMin & Minimum atmospheric pressure in hectopascals recorded during the day. \\
HourPresMin & The time (hh:mm format) when the minimum atmospheric pressure was recorded. \\
\hline
\end{tabular}
\end{scriptsize}
\label{tab:meteorological_variables}
\end{table}

\newpage
The following part evaluates the correlation between water consumption and meteorological parameters using Pearson's correlation coefficient ($R$), which ranges from $-1$ (perfect negative linear relationship) to $+1$ (perfect positive linear relationship). This analysis helps identify the most influential meteorological factors for inclusion in forecasting models.

Definition: Pearson’s Correlation Coefficient

\begin{equation}
R = \frac{n \sum{(x_i y_i)} - \sum{x_i} \sum{y_i}}{\sqrt{\left[n\sum{x_i^2} - \left(\sum{x_i}\right)^2\right] \left[n\sum{y_i^2} - \left(\sum{y_i}\right)^2\right]}}
\end{equation}

Where $n$ is the number of observations, $x_i$ and $y_i$ are the individual data points of variables $X$ and $Y$, and other terms are mathematical sums and products needed for the computation.

\begin{figure}[H]
    \centering
    \includegraphics[width=0.7\linewidth]{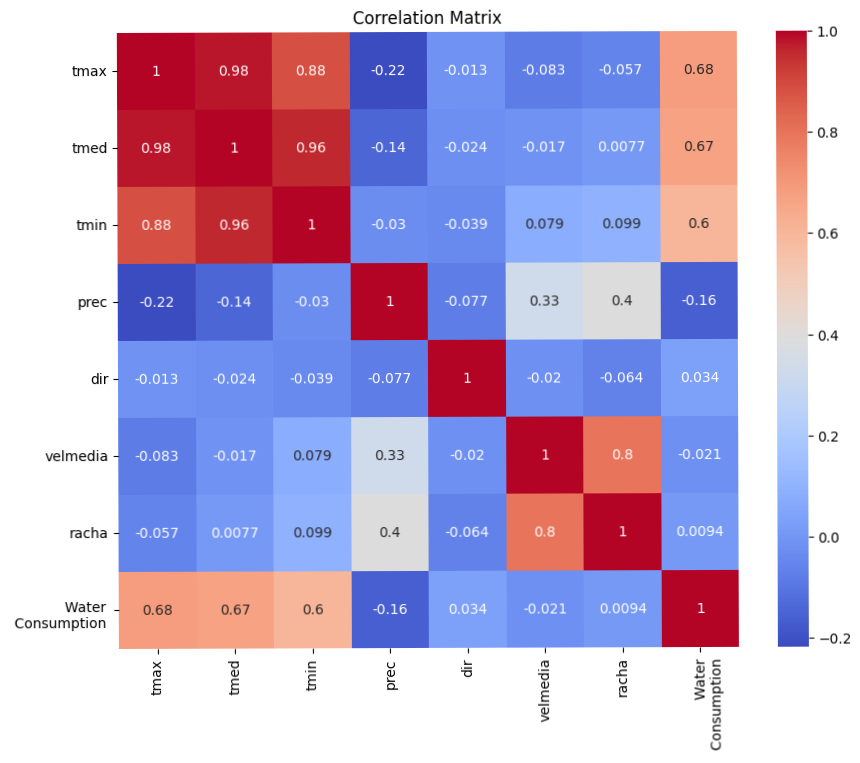}
    \caption{Correlation Matrix based on the parameters}
    \label{fig:correlation}
\end{figure}

Figure \ref{fig:correlation}, and Table \ref{tab:correlation_table} presents the Pearson correlation coefficients (R) for the relationships between various climatic factors and the meteorology station parameters, used as a proxy for water consumption. For the provided dataset, several climatic factors exhibit significant correlations with maximum daily temperature, suggesting that these variables may also impact water consumption if temperature is considered a surrogate for this measure. A correlation coefficient greater than 0.4 is considered significant, indicating a meaningful linear relationship between two variables. In this dataset, most of the climatic factors, such as maximum temperature (tmax), medium temperature (tmed), and minimum temperature (tmin), show strong positive correlations with the average temperature, indicating that changes in these factors will likely affect water consumption patterns.

\begin{table}[h]
\centering
\caption{PCC(R) for the Effects of Climatic Factors on Water Consumption}
\begin{scriptsize}
\begin{tabular}{>{\centering\arraybackslash}p{1.8cm} 
                >{\centering\arraybackslash}p{1.2cm} 
                >{\centering\arraybackslash}p{1.2cm} 
                >{\centering\arraybackslash}p{1.2cm} 
                >{\centering\arraybackslash}p{1.2cm} 
                >{\centering\arraybackslash}p{1.2cm} 
                >{\centering\arraybackslash}p{1.2cm} 
                >{\centering\arraybackslash}p{1.2cm}
                >{\centering\arraybackslash}p{1.4cm}}
\hline
\textbf{Climatic Factor} & \textbf{tmax} & \textbf{tmed} & \textbf{tmin} & \textbf{prec} & \textbf{dir} & \textbf{velmedia} & \textbf{racha} & \textbf{Water Cons} \\
\hline
\textbf{tmax}            & 1.000         & 0.980         & 0.884         & -0.217        & -0.013       & -0.083          & -0.057         & 0.683                 \\
\textbf{tmed}            & 0.980         & 1.000         & 0.959         & -0.144        & -0.024       & -0.017          & 0.008          & 0.669                 \\
\textbf{tmin}            & 0.884         & 0.959         & 1.000         & -0.030        & -0.039       & 0.079           & 0.099          & 0.604                 \\
\textbf{prec}            & -0.217        & -0.144        & -0.030        & 1.000         & -0.077       & 0.330           & 0.395          & -0.160                \\
\textbf{dir}             & -0.013        & -0.024        & -0.039        & -0.077        & 1.000        & -0.020          & -0.064         & 0.034                 \\
\textbf{velmedia}        & -0.083        & -0.017        & 0.079         & 0.330         & -0.020       & 1.000           & 0.800          & -0.021                \\
\textbf{racha}           & -0.057        & 0.008         & 0.099         & 0.395         & -0.064       & 0.800           & 1.000          & 0.009                 \\
\textbf{Water Cons}& 0.683         & 0.669         & 0.604         & -0.160        & 0.034        & -0.021          & 0.009          & 1.000                 \\
\hline
\end{tabular}
\end{scriptsize}
\label{tab:correlation_table}
\end{table}

However, some variables, such as wind speed (velmedia) and wind direction (dir), have weaker correlations, suggesting a lesser impact on water consumption when the temperature is used as a stand-in. Moreover, correlations among the variables highlight potential multicollinearity issues, which must be addressed when selecting independent predictor variables for any forecasting models to avoid invalid results. Based on the correlations, the average temperature (tmex), which has the highest correlation with maximum temperature (tmed), could be considered a significant input variable in a water consumption forecasting model using methods such as LSTM networks.

\subsection{LSTM Model}

The first model implemented on this platform for predicting water consumption is the LSTM network, which integrates climate variables along with real-time and historical water consumption data. By training on synthetic datasets derived from daily consumption records, LSTM captures long-term dependencies inherent in time series data. This makes it particularly effective in forecasting future water usage. The model combines historical water use patterns with real-time meteorological observations to deliver accurate and reliable predictions. Given its strength in analyzing sequential data, the LSTM model is crucial for strategic water management, enabling more informed decision-making and resource planning. Key Operations of LSTM Networks:

Forget Gate: The forget gate decides what information should be discarded from the cell state. It is computed as follows:
\begin{equation}
f_t = \sigma(W_f \cdot [h_{t-1}, x_t] + b_f) \label{eq:forget_gate}
\end{equation}

Input Gate:
The input gate updates the cell state by adding new information into the cell, calculated by:
\begin{equation}
i_t = \sigma(W_i \cdot [h_{t-1}, x_t] + b_i) \label{eq:input_gate}
\end{equation}

Candidate Layer:
This layer creates a vector of new candidate values that could be added to the state:
\begin{equation}
\tilde{C}_t = \tanh(W_C \cdot [h_{t-1}, x_t] + b_C) \label{eq:candidate_layer}
\end{equation}

Cell State Update:
The cell state is updated by combining the old state and the new candidate values, influenced by the forget and input gates:
\begin{equation}
C_t = f_t \cdot C_{t-1} + i_t \cdot \tilde{C}_t \label{eq:cell_state_update}
\end{equation}

Output Gate:
The output gate decides what the next hidden state should be, which contains information on previous inputs:
\begin{equation}
o_t = \sigma(W_o \cdot [h_{t-1}, x_t] + b_o) \label{eq:output_gate}
\end{equation}

Hidden State Output:
The hidden state contains the output information to be passed onto the next timestep:
\begin{equation}
h_t = o_t \cdot \tanh(C_t) \label{eq:hidden_state_output}
\end{equation}

Data preparation for LSTM involves applying Min-Max scaling for data normalization:

\begin{equation}
\label{eq:normalization}
x' = \frac{x - x_{\text{min}}}{x_{\text{max}} - x_{\text{min}}}
\end{equation}

The dataset is methodically partitioned into two segments: 80\% for training and 20\% for testing. This distribution allows the training set to predict future water consumption across specified periods effectively. Utilizing a structured approach, the LSTM network harnesses both historical and meteorological data to generate accurate forecasts of water consumption patterns. Below, we outline the step-by-step algorithm 1 for training the LSTM model tailored for predicting water consumption:

\begin{algorithm} \label{algorithm_LSTM}
\caption{LSTM for water consumption forecasting}
\scriptsize
\begin{algorithmic}[1]
\State \textbf{Initialize parameters:}
\State Define the number of LSTM units (neurons), learning rate, and epochs
\State Initialize weight matrices \(W_f\), \(W_i\), \(W_C\), \(W_o\), and bias vectors \(b_f\), \(b_i\), \(b_C\), \(b_o\)
\State \textbf{Preprocess input data:}
\State Normalize water consumption and meteorological data using Min-Max scaling
\State Divide the dataset into training, validation, and testing sets
\State Create sequences of input data \(X\) and target values \(Y\)
\State Reshape input data \(X\) to \((\text{num\_samples}, \text{sequence\_length}, \text{num\_features})\)
\State \textbf{Model Training:}
\State Initialize cell state \(C_0\) and hidden state \(h_0\) to zeros
\For{each epoch}
    \For{each batch in the training data}
        \For{each time step \(t\) in the input sequence}
            \State Compute Forget Gate: \(f_t = \sigma(W_f \cdot [h_{t-1}, x_t] + b_f)\)
            \State Compute Input Gate: \(i_t = \sigma(W_i \cdot [h_{t-1}, x_t] + b_i)\)
            \State Compute Candidate Cell State: \(\tilde{C}_t = \tanh(W_C \cdot [h_{t-1}, x_t] + b_C)\)
            \State Update Cell State: \(C_t = f_t \cdot C_{t-1} + i_t \cdot \tilde{C}_t\)
            \State Compute Output Gate: \(o_t = \sigma(W_o \cdot [h_{t-1}, x_t] + b_o)\)
            \State Update Hidden State: \(h_t = o_t \cdot \tanh(C_t)\)
        \EndFor
        \State Compute output (predicted water consumption): \(y_{\text{pred}_t} = \text{Dense}(h_t)\)
        \State Calculate batch loss: \(\text{MSE}(y_{\text{pred}_t}, y_{\text{true}_t})\)
        \State \textbf{Backpropagation through time (BPTT):}
        \State Calculate gradients of Loss w.r.t weights and biases
        \State Update \(W_f\), \(W_i\), \(W_C\), \(W_o\) and \(b_f\), \(b_i\), \(b_C\), \(b_o\) using an optimizer
    \EndFor
    \State Evaluate model on the validation set after each epoch
\EndFor
\State \textbf{Model Evaluation:}
\State Test the model on the testing set
\State Calculate and report performance metrics: RMSE and MAPE
\State \textbf{Model Deployment:}
\State Save the trained model for future use
\State Deploy the model for real-time or batch water consumption prediction
\end{algorithmic}
\end{algorithm}


\subsection{Prophet Model}

We employed the Prophet model to forecast daily water consumption from 2020 to 2024, integrating historical water usage data with maximum daily temperature as an external regressor. Prior analyses using the Pearson correlation coefficient indicated a significant positive correlation between water consumption and maximum temperature. This relationship suggests that higher temperatures lead to increased water usage due to activities like irrigation and cooling. By retaining missing values in the dataset, we utilized Prophet's capability to handle incomplete data internally, preserving the integrity and variability essential for robust forecasting. The model incorporated maximum temperature as an external regress to account for temperature-related variations in water demand. Prophet is particularly suitable for this application because it handles time series data with complex seasonal patterns and additional regressors. The model is formulated as:

\begin{equation}
y(t) = g(t) + s(t) + h(t) + \beta \cdot T_{\text{max}}(t) + \epsilon_t,
\end{equation}

where:
\begin{itemize}
    \item \( y(t) \) is the forecasted water consumption at time \( t \),
    \item \( g(t) \) represents the growth trend component (linear or logistic),
    \item \( s(t) \) denotes the seasonal component modeled using Fourier series to capture annual and weekly patterns,
    \item \( h(t) \) accounts for holiday effects using indicator functions,
    \item \( T_{\text{max}}(t) \) is the maximum temperature on day \( t \),
    \item \( \beta \) quantifies the impact of temperature on water consumption,
    \item \( \epsilon_t \) is the error term, assumed to be normally distributed with a mean of zero.
\end{itemize}

The inclusion of \( T_{\text{max}}(t) \) as an external regressor was implemented using Prophet's \texttt{add\_regressor} function, enabling the model to learn the relationship between temperature and water demand dynamically.

During the model fitting process, the Prophet model was trained on the historical dataset, optimizing the model parameters to minimize the discrepancy between the predicted and observed water consumption. Prophet employs maximum likelihood estimation (MLE) to estimate the parameters by optimizing the following objective function:

\begin{equation}
\min_{\theta} \sum_{t=1}^{n} \left( y(t) - \hat{y}(t; \theta) \right)^2,
\end{equation}

where:
\begin{itemize}
    \item \( \theta \) represents all model parameters, including \( \beta \), trend, and seasonal components,
    \item \( y(t) \) is the actual water consumption,
    \item \( \hat{y}(t; \theta) \) is the predicted water consumption based on parameters \( \theta \).
\end{itemize}

After completing the training and validation phases, the model generated forecasts for future water consumption over two periods: from January 1, 2023, to July 1, 2024 (18 months), and from January 1, 2024, to July 1, 2024 (6 months). These forecasts utilized patterns identified from historical data, including trends, seasonal variations, holiday effects, and the influence of maximum temperatures. The anticipated water consumption values were calculated using:

\begin{equation}
\hat{y}(t) = \hat{g}(t) + \hat{s}(t) + \hat{h}(t) + \hat{\beta} \cdot T_{\text{max}}(t),
\end{equation}

where \( \hat{g}(t) \), \( \hat{s}(t) \), \( \hat{h}(t) \), and \( \hat{\beta} \) are the estimated components from the fitting stage, and \( T_{\text{max}}(t) \) represents the projected maximum temperature values, either derived from historical trends or provided as external forecasts.

Algorithm 2 outlines the steps in applying the Prophet model for forecasting water consumption.

\begin{algorithm} \label{algorithm_Prophet}
\caption{Prophet for water consumption forecasting}
\scriptsize
\begin{algorithmic}[1]
\State \textbf{Initialize Parameters:}
\State Define forecasting horizon \( n \) days
\State Set growth model \( g(t) \) as 'linear' or 'logistic'
\State Specify relevant seasonalities (e.g., daily, weekly, yearly)
\State \textbf{Preprocess Input Data:}
\State Collect historical water consumption data \( y(t) \) and maximum temperature data \( T_{\text{max}}(t) \)
\State Prepare dataset with columns:
\State \quad Date (\( t \)): \text{'ds'}
\State \quad Water consumption (\( y(t) \)): \text{'y'}
\State \quad Maximum temperature (\( T_{\text{max}}(t) \)): \text{'T\_max'}
\State Handle missing values appropriately (Prophet can handle them internally)
\State \textbf{Model Configuration:}
\State Initialize Prophet model with specified growth and seasonalities:
\State \quad model \( = \) Prophet(growth\( = g(t) \), daily\_seasonality\( = \)True,
\State \quad \quad \quad \quad weekly\_seasonality\( = \)True, yearly\_seasonality\( = \)True)
\State Add maximum temperature as an external regressor:
\State \quad model.add\_regressor('T\_max')
\State \textbf{Model Fitting:}
\State Fit the Prophet model to the historical data:
\State \quad model.fit(data)
\State \textbf{Forecasting:}
\State Create a future dataframe for \( n \) days ahead:
\State \quad future \( = \) model.make\_future\_dataframe(periods\( = n \))
\State Obtain future maximum temperature \( T_{\text{max}}(t) \) for future dates
\State Add \( T_{\text{max}}(t) \) to the 'future' dataframe under 'T\_max'
\State Generate forecast:
\State \quad forecast \( = \) model.predict(future)
\State Extract predicted water consumption \( \hat{y}(t) \) from 'forecast' dataframe
\State \textbf{Model Deployment:}
\State Save the trained model for future use:
\State \quad model.save('prophet\_model.pkl')
\State Deploy the model for real-time or batch water consumption prediction
\end{algorithmic}
\end{algorithm}

\newpage

\subsection{LightGBM Model with Feature Engineering}

To enhance the forecasting accuracy of daily water consumption, we employed the LightGBM model, a gradient-boosting framework renowned for its efficiency and performance in handling large-scale data and complex features. This model integrates feature engineering techniques to capture temporal patterns and dependencies inherent in water consumption data.

LightGBM operates by constructing an ensemble of decision trees, where each subsequent tree focuses on correcting the errors of the previous ones. This iterative boosting process allows the model to capture nonlinear relationships and interactions between features, making it well-suited for time-series forecasting tasks. Key components of the LightGBM model are:
\begin{itemize}
    \item Objective Function: The model minimizes an objective function that combines a loss function with a regularization term to prevent overfitting:

\begin{equation}
\mathcal{L} = \sum_{i=1}^{n} l(y_i, \hat{y}_i) + \Omega(T), \label{eq:objective_function_LightGBM}
\end{equation}

where:
\begin{itemize}
    \item $y_i$ is the actual water consumption at time $i$,
    \item $\hat{y}_i$ is the predicted water consumption,
    \item $l(y_i, \hat{y}_i)$ is the loss function, typically MSE for regression tasks,
    \item $\Omega(T)$ is the regularization term for the complexity of the trees $T$.
\end{itemize}

\item Feature Engineering: Various features were created to enhance the model's predictive performance:

\begin{itemize}
    \item Lag Features: Previous water consumption and temperature values were included to capture temporal dependencies:
    \begin{equation}
    y_{\text{lag}_k} = y_{t - k}, \quad T_{\text{max, lag}_k} = T_{\text{max}, t - k}, \label{eq:lag_features}
    \end{equation}
    where $k$ is the lag period (e.g., 1 day, 7 days).

    \item Rolling Means: Moving averages were computed to smooth out short-term fluctuations:
    \begin{equation}
    \bar{y}_t = \frac{1}{w} \sum_{i=t - w + 1}^{t} y_i, \quad \overline{T}_{\text{max}, t} = \frac{1}{w} \sum_{i=t - w + 1}^{t} T_{\text{max}, i}, \label{eq:rolling_means}
    \end{equation}
    where $w$ is the window size (e.g., 7 days).

    \item Temporal Indicators: Features such as day of the week and weekend indicators were added to capture weekly seasonal patterns:
    \begin{equation}
    \text{IsWeekend}_t = 
    \begin{cases}
    1 & \text{if day of week} \geq 5, \\
    0 & \text{otherwise}.
    \end{cases} \label{eq:temporal_indicators}
    \end{equation}
\end{itemize}

\item Data Normalization: Feature scaling was performed using standardization to ensure that all features contribute equally to the model training:

\begin{equation}
x' = \frac{x - \mu}{\sigma}, \label{eq:standardization}
\end{equation}

where $x$ represents the original feature value, $\mu$ is the mean, and $\sigma$ is the standard deviation of the feature.
\end{itemize}

The LightGBM model effectively captures complex temporal and seasonal patterns in water consumption da by integrating these engineered features. This approach enhances the model's ability to provide accurate and reliable forecasts, which are crucial for efficient water resource management. In following the pseudocode of the LightGBM model shown in algorithm 3. 

\begin{algorithm} [H]
\label{algorithm_LightGBM}
\caption{LightGBM with Feature Engineering for Water Consumption Prediction}
\scriptsize
\begin{algorithmic}[1]
\State \textbf{Initialize Parameters:}
\State Set LightGBM hyperparameters: learning rate $\eta$, number of leaves $num\_leaves$, feature fraction $ff$, bagging fraction $bf$, bagging frequency $bfreq$, number of boosting rounds $num\_boost\_round$
\State \textbf{Data Preparation:}
\State Load dataset with dates $ds$, water consumption $y$, and maximum temperature $T_{\text{max}}$
\State Convert date strings to datetime objects
\State Remove missing values in $y$ and forward-fill missing values in $T_{\text{max}}$
\State \textbf{Feature Engineering:}
\State Create lag features:
\State \quad $y_{\text{lag1}} = y_{t-1}$, \quad $y_{\text{lag7}} = y_{t-7}$
\State \quad $T_{\text{max, lag1}} = T_{\text{max}, t-1}$, \quad $T_{\text{max, lag7}} = T_{\text{max}, t-7}$
\State Create rolling mean features with window size $w=7$:
\State \quad $\bar{y}_t = \dfrac{1}{w} \sum_{i=t-w+1}^{t} y_i$
\State \quad $\overline{T}_{\text{max}, t} = \dfrac{1}{w} \sum_{i=t-w+1}^{t} T_{\text{max}, i}$
\State Create temporal indicators:
\State \quad Day of week: $day\_of\_week_t = \text{DayOfWeek}(t)$
\State \quad Is weekend indicator:
\State \quad $\text{is\_weekend}_t = 
\begin{cases}
1 & \text{if } day\_of\_week_t \geq 5 \\
0 & \text{otherwise}
\end{cases}$
\State Remove any rows with missing values resulting from feature creation
\State \textbf{Feature Scaling:}
\State Standardize features using z-score normalization:
\State \quad For each feature $X$:
\State \quad \quad $X' = \dfrac{X - \mu_X}{\sigma_X}$
\State \textbf{Split Data:}
\State Define split date (e.g., January 1, 2024)
\State Split data into training set (dates before split date) and testing set (dates on or after split date)
\State Extract feature matrix $X$ and target vector $y$ for both sets
\State \textbf{Prepare Data for LightGBM:}
\State Create LightGBM datasets:
\State \quad Training data: $train\_data = \text{lgb.Dataset}(X_{\text{train}}, y_{\text{train}})$
\State \quad Validation data: $valid\_data = \text{lgb.Dataset}(X_{\text{test}}, y_{\text{test}}, \text{reference}=train\_data)$
\State \textbf{Model Training:}
\State Train the model using the training data:
\State \quad $lgbm\_model = \text{lgb.train}(\text{params}, train\_data, num\_boost\_round,$
\State \quad \quad $\text{valid\_sets}=[valid\_data], \text{early\_stopping\_rounds}=100)$
\State \textbf{Model Prediction:}
\State Predict on the testing set:
\State \quad $\hat{y} = lgbm\_model.predict(X_{\text{test}}, \text{num\_iteration}=lgbm\_model.best\_iteration)$
\State \textbf{Model Deployment:}
\State Save the trained model for future use
\State Deploy the model for real-time or batch water consumption prediction
\end{algorithmic}
\end{algorithm}

\subsection{XGBoost Model with Feature Engineering}

We employed the XGBoost model to enhance the accuracy of daily water consumption forecasting. XGBoost, known for its efficiency in regression tasks and regularization techniques to prevent overfitting, is well-suited for time-series forecasting. The dataset consists of daily water consumption and maximum temperature records from 2020 to 2024. Preprocessing steps included converting date strings to datetime objects, renaming columns for consistency, removing missing values in water consumption, and forward-filling missing temperature values to ensure data continuity.

\begin{algorithm}[H]

\label{algorithm_XGBoost}
\caption{XGBoost with Hyperparameter Tuning and Feature Engineering}
\scriptsize
\begin{algorithmic}[1]

\State \textbf{Initialize Parameters:}
\State Define hyperparameter search space for:
\State \quad Number of estimators $n_{\text{estimators}}$
\State \quad Maximum depth $\text{max\_depth}$
\State \quad Learning rate $\eta$
\State \quad Subsample ratio $\text{subsample}$
\State \quad Column subsample ratio $\text{colsample\_bytree}$
\State \quad Regularization parameters $\gamma$, $\alpha$, $\lambda$

\State \textbf{Data Preparation:}
\State Load dataset with dates $ds$, water consumption $y$, and maximum temperature $T_{\text{max}}$
\State Convert date strings to datetime objects
\State Remove missing values in $y$ and forward-fill missing values in $T_{\text{max}}$

\State \textbf{Feature Engineering:}
\State Create lag features: \quad // Use previous time steps as features
\State \quad $y_{\text{lag1}} = y_{t-1}$, \quad $y_{\text{lag7}} = y_{t-7}$
\State \quad $T_{\text{max, lag1}} = T_{\text{max}, t-1}$, \quad $T_{\text{max, lag7}} = T_{\text{max}, t-7}$

\State Create rolling mean features with window size $w=7$: \quad // Capture trends over time
\State \quad $\bar{y}_t = \dfrac{1}{w} \sum_{i=t-w+1}^{t} y_i$
\State \quad $\overline{T}_{\text{max}, t} = \dfrac{1}{w} \sum_{i=t-w+1}^{t} T_{\text{max}, i}$

\State Create temporal indicators: \quad // Add time-related patterns
\State \quad Day of week: $day\_of\_week_t = \text{DayOfWeek}(t)$
\State \quad Is weekend indicator:
\State \quad $\text{IsWeekend}_t = 
\begin{cases}
1 & \text{if } day\_of\_week_t \geq 5 \\
0 & \text{otherwise}
\end{cases}$

\State Remove any rows with missing values resulting from feature creation

\State \textbf{Feature Scaling:}
\State Standardize features using z-score normalization:
\State \quad $X' = \dfrac{X - \mu_X}{\sigma_X}$ \quad // $\mu_X$ is the mean and $\sigma_X$ is the standard deviation

\State \textbf{Split Data:}
\State Define split date (e.g., January 1, 2024)
\State Split data into training set (dates before split date) and testing set (dates on or after split date)
\State Extract feature matrix $X$ and target vector $y$ for both sets

\State \textbf{Hyperparameter Tuning:}
\State Define time series cross-validation strategy with $k$ folds
\State Initialize XGBoost regressor model
\State Perform randomized search over hyperparameter space using cross-validation to find optimal hyperparameters $\Theta^*$

\State \textbf{Model Training:}
\State Train the XGBoost model with optimal hyperparameters $\Theta^*$ on the training data

\State \textbf{Model Prediction:}
\State Predict on the testing set:
\State \quad $\hat{y}_t = \text{XGBoostModel.predict}(X_{\text{test}})$

\State \textbf{Objective Function:}
\State Minimize the objective function:
\begin{equation}
\mathcal{L}(\Theta) = \sum_{i=1}^{n} l(y_i, \hat{y}_i) + \sum_{k=1}^{K} \Omega(f_k)
\end{equation}
\State Where:
\State \quad $l(y_i, \hat{y}_i)$ is the loss function (e.g., MAE or MSE)
\State \quad $\Omega(f_k)$ is the regularization term for tree $f_k$
\State \quad $K$ is the number of trees in the ensemble

\State \textbf{Model Deployment:}
\State Save the trained model for future use
\State Deploy the model for real-time or batch water consumption prediction

\end{algorithmic}
\end{algorithm}

\newpage

\subsection{Model Evaluation}

The forecasting models were evaluated using three essential statistical metrics: Mean Absolute Error (MAE), Mean Absolute Percentage Error (MAPE), and Root Mean Squared Error (RMSE). MAE measures the average magnitude of the errors in a set of predictions, without considering their direction, providing a straightforward assessment of prediction accuracy in the same unit as the original data. MAPE quantifies the error as a percentage, making it especially useful for comparing performance across different scales or datasets. RMSE, on the other hand, penalizes larger errors more heavily by squaring them before averaging, offering insight into the variability of the errors and emphasizing significant deviations in the predictions.

\textbf{MAE} represents the average of the absolute differences between the predicted and actual values, providing a straightforward measure of the model's prediction accuracy:
\begin{equation}
\text{MAE} = \frac{1}{n} \sum_{i=1}^{n} \left| y_i - \hat{y}_i \right|,
\end{equation}

\textbf{MAPE} measures the average absolute percentage difference between the predicted and actual values, offering insight into the model's accuracy relative to the actual consumption:
\begin{equation}
\text{MAPE} = \frac{100\%}{n} \sum_{i=1}^{n} \left| \frac{y_i - \hat{y}_i}{y_i} \right|,
\end{equation}

\textbf{RMSE} is the square root of the Mean Squared Error and represents the standard deviation of the prediction errors, providing a measure of the average magnitude of these errors:
\begin{equation}
\text{RMSE} = \sqrt{\frac{1}{n} \sum_{i=1}^{n} (y_i - \hat{y}_i)^2},
\end{equation}

where \( n \) is the number of observations, \( y_i \) is the actual water consumption, and \( \hat{y}_i \) is the predicted water consumption.

By utilizing these evaluation metrics, we identified each model's strengths and areas for improvement, guiding further refinements to enhance forecasting accuracy. This comprehensive evaluation is crucial for developing reliable water demand forecasts essential for effective water resource management.

\begin{figure*}[htbp]
    \centering
    \begin{minipage}{0.48\textwidth}
        \centering
        \includegraphics[width=\textwidth]{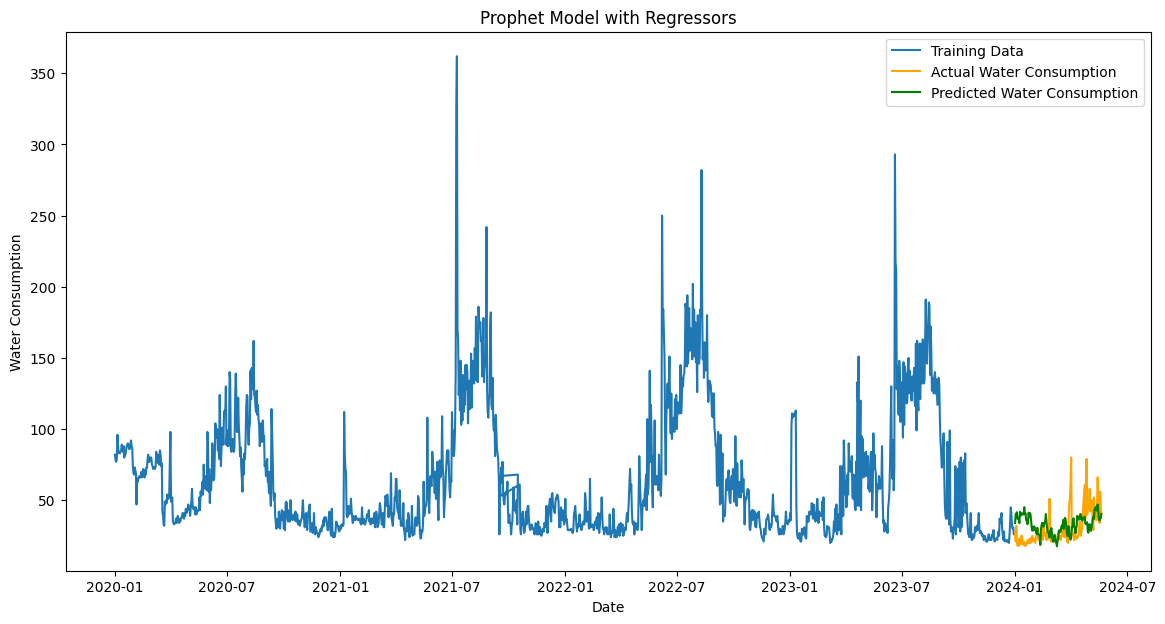}
        \caption*{(a) Six-month forecasting}
        \label{fig:PMR6M}
    \end{minipage}\hfill
    \begin{minipage}{0.48\textwidth}
        \centering
        \includegraphics[width=\textwidth]{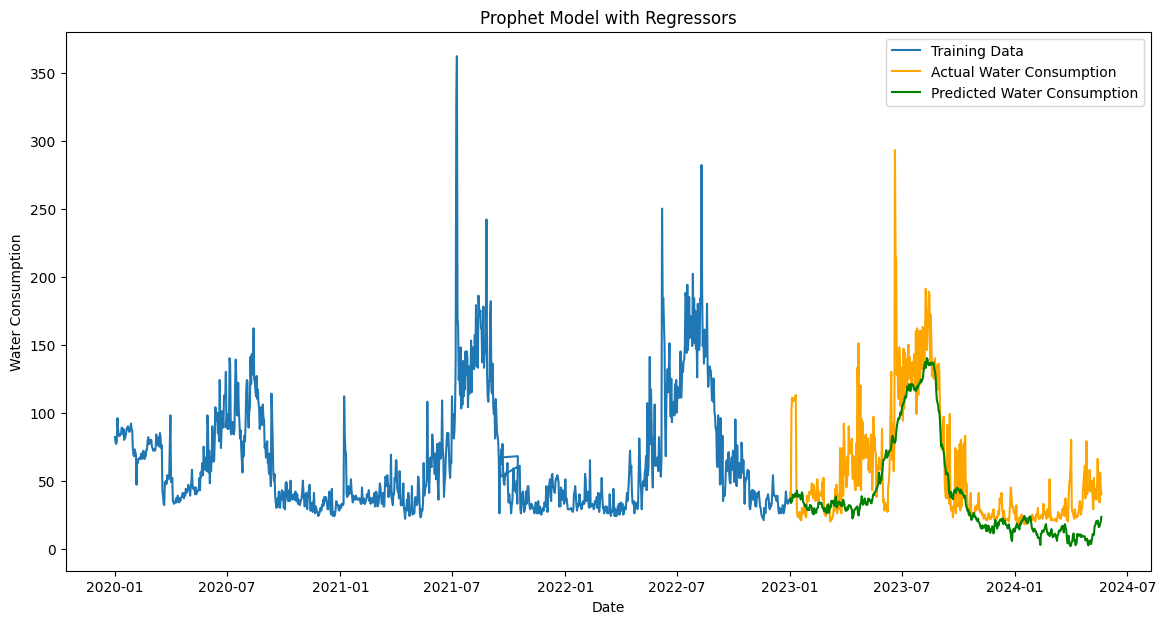}
        \caption*{(b) Eighteen-month forecasting}
        \label{fig:PMR18M}
    \end{minipage}
    \caption{Prophet Model with Regressors (six and eighteen months forecasting). Subfigure (a) shows the model's short-term performance, while subfigure (b) highlights long-term trends and limitations.}
    \label{PMR}
\end{figure*}

Results of water consumption forecasting by the Prophet model with six regressors are presented in Figure \ref{PMR}, and they show large shortcomings for all forecast periods. Looking at the 6-month forecasting, while the model picks up broad seasonal trends and cyclical changes, there are deviations at sharp peaks and troughs that manifest a lack of robustness in handling transient phenomena or unmodeled noise. The identified discrepancies suggest that the model's reliance on pre-determined regressors may not adequately capture the complex nature of water consumption dynamics, especially in the short term. Although the overall correlation with observed data appears to be acceptable, these problems raise some doubts about the reliability of the model for medium-term operational planning in environments characterized by frequent sudden changes. Looking ahead to the 18-month horizon, the shortcomings of the model become even more apparent. While some deeper seasonal patterns are still captured, there is a visible rise in cumulative error as the model struggles to mimic more complex variations. This kind of deterioration in performance could be due to a combination of static correlations between regressors and water use, along with a diminished importance of the chosen exogenous variables over time, and an inability to adapt to changing dynamics. Results: The findings indicated that the Prophet model with regressors performed reasonably well in short-term forecasting, though serious shortcomings occurred as this went further. Consequently, its adequacy for longer-term operational planning and strategic decision-making was highly questionable.

\begin{figure*}[htbp]
    \centering
    \begin{minipage}{0.48\textwidth}
        \centering
        \includegraphics[width=\textwidth]{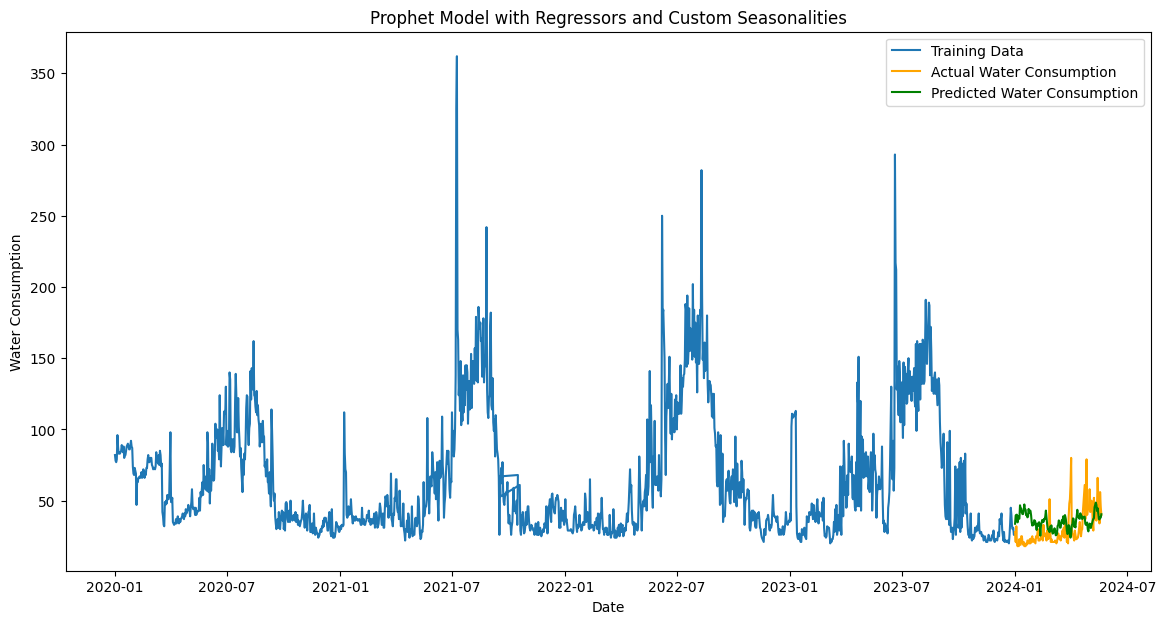}
        \caption*{(a) Six-month forecasting}
        \label{fig:PMRCS6M}
    \end{minipage}\hfill
    \begin{minipage}{0.48\textwidth}
        \centering
        \includegraphics[width=\textwidth]{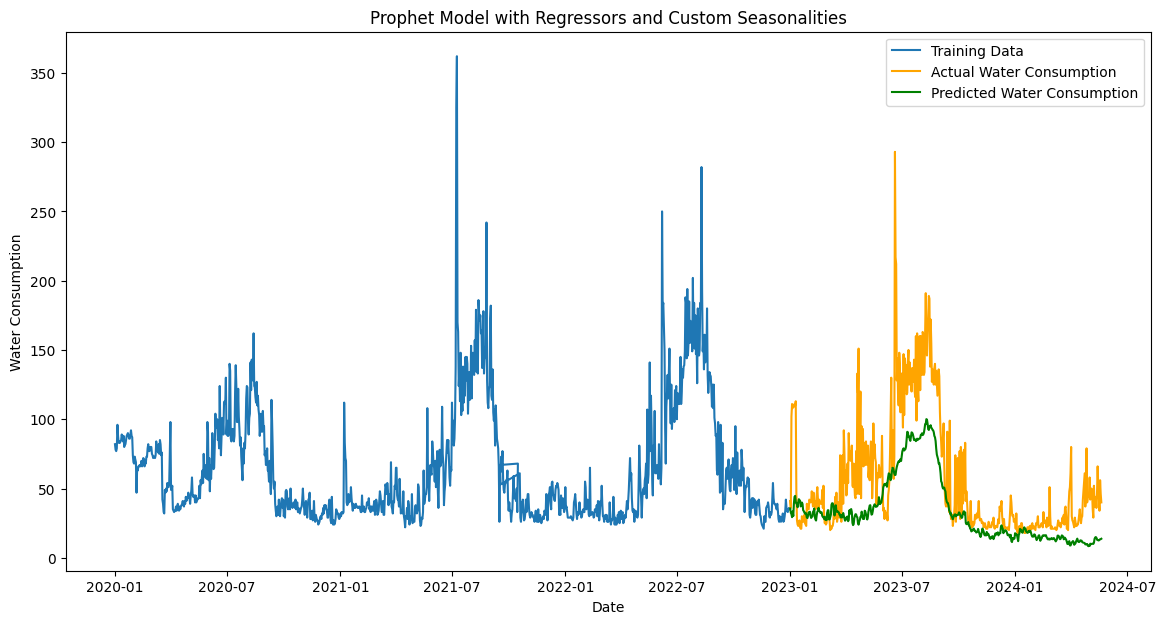}
        \caption*{(b) Eighteen-month forecasting}
        \label{fig:PMRCS18M}
    \end{minipage}
    \caption{Prophet Model with Regressors and Custom Seasonalities (six and eighteen months forecasting). Subfigure (a) shows improvements in short-term predictions, and subfigure (b) reflects better seasonal pattern detection.}
    \label{PMRCS}
\end{figure*}

The performance of the Prophet model with regressors and custom seasonalities, as shown in Figure \ref{PMRCS}, despite capturing broad trends correctly, reveals several critical deficiencies. For the 6-month forecast, the model does indicate adherence to general consumption patterns; however, large deviations from actual data, particularly for sudden changes or high peaks, suggest that its representation of short-term fluctuations is not adequate. The seasonalities specific to certain domains try to catch the periodic fluctuations in demand and operational cycles; still, this customized approach hardly makes up for the discrepancies with rapid changes or special events in water use. It implies that the tailored seasonalities and regressors alone may not reach a degree of accuracy satisfactory for reliable short-term forecasts. Over the course of the 18 months for which forecasting was conducted, the limitations of the model become more pronounced. While it is able to reproduce the general patterns of the seasons over the long term, results diverge substantially from the actual data at higher temporal resolutions. This increasing divergence shows that the reliance of the model on fixed seasonalities and rigid associations with regressors is not sufficient to handle the compounding uncertainties and shifting dynamics over time. While the tailored seasonalities improve slightly the representation of general trends compared to models using generalized seasonalities, this improvement is of little consequence and does not approach the large disparities in capturing real-world variability. It results in that, though the Prophet model with its seasonality tailored incorporates some usefulness in recognizing general trends, its predictive precision for both immediate and long-term time frames is yet far from satisfactory, which strongly limits its use in practical applications characterized by high dynamism and variability.

\begin{figure*}[htbp]
    \centering
    \begin{minipage}{0.48\textwidth}
        \centering
        \includegraphics[width=\textwidth]{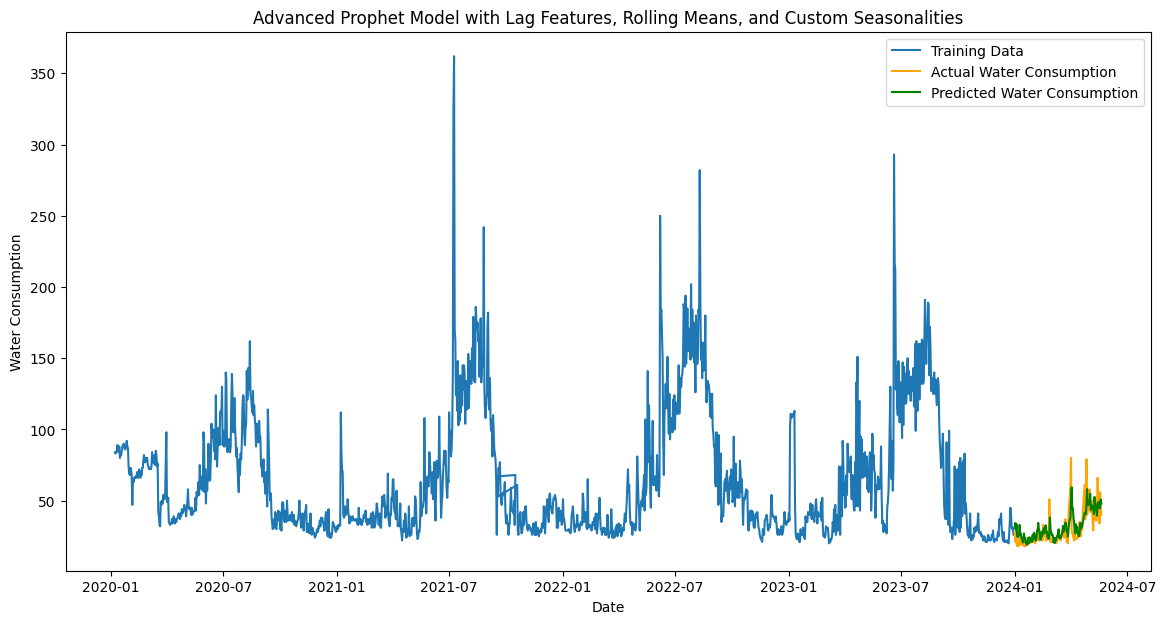}
        \caption*{(a) Six-month forecasting}
        \label{fig:APMLFRMCS6M}
    \end{minipage}\hfill
    \begin{minipage}{0.48\textwidth}
        \centering
        \includegraphics[width=\textwidth]{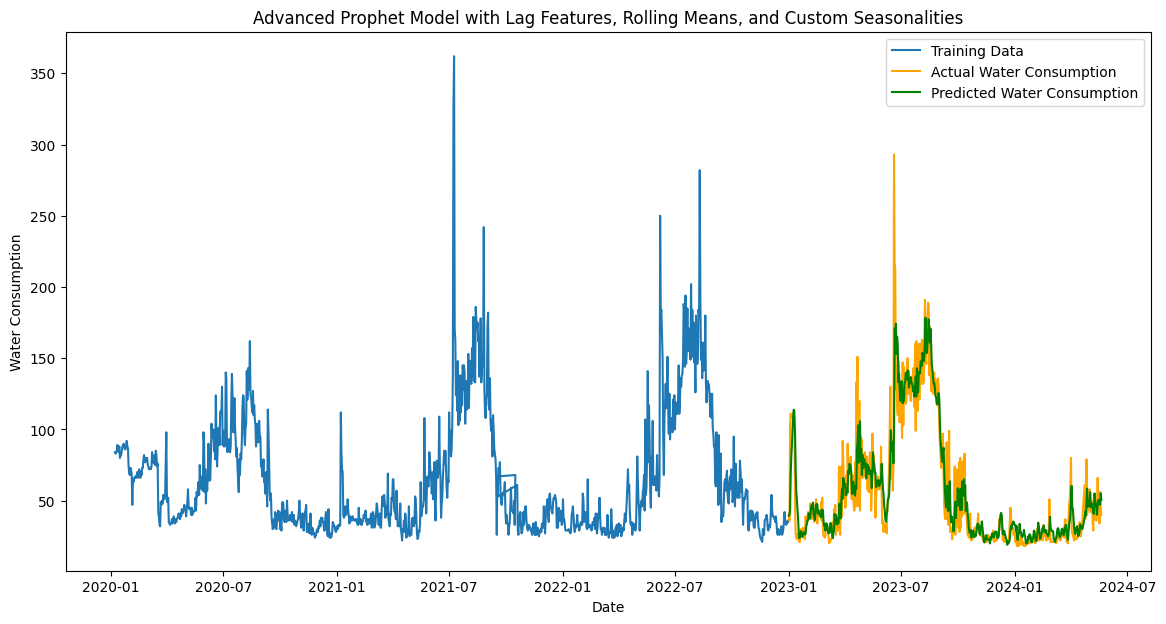}
        \caption*{(b) Eighteen-month forecasting}
        \label{fig:APMLFRMCS18M}
    \end{minipage}
    \caption{Advanced Prophet Model with Lag Features, Rolling Means, and Custom Seasonalities (six and eighteen months forecasting). Subfigure (a) demonstrates short-term improvements, while subfigure (b) focuses on long-term trends.}
    \label{APMLFRMCS}
\end{figure*}

The advanced variant of the Prophet model-which includes lag features, rolling means, and custom seasonalities showed in Figure \ref{APMLFRMCS} presents an improved performance in predictions by taking into consideration historical dependencies with softly varying trends. Looking at its 6-month-ahead forecast, the model fits well with the actual data of consumption and captures the short-term fluctuations and sudden changes in the trend. The lag features allow the model to capture temporal dependencies, and the rolling means create a smoothed representation of periodic variability that can work to improve the accuracy of the operational planning. Lastly, customized seasonalities further improve the model's ability to conform to domain-specific periodicities that are highly effective at managing dynamic shifts in water demand patterns. In the 18-month projection, the model maintains its ability to capture seasonal patterns, though it is not as good at capturing finer-scale fluctuations and anomalies.
These differences can be explained by less temporal relevance of lagged variables on longer forecasting periods and the inherently higher problem of modeling unexpected changes in consumption behavior.

\begin{figure*}[htbp]
    \centering
    \begin{minipage}{0.48\textwidth}
        \centering
        \includegraphics[width=\textwidth]{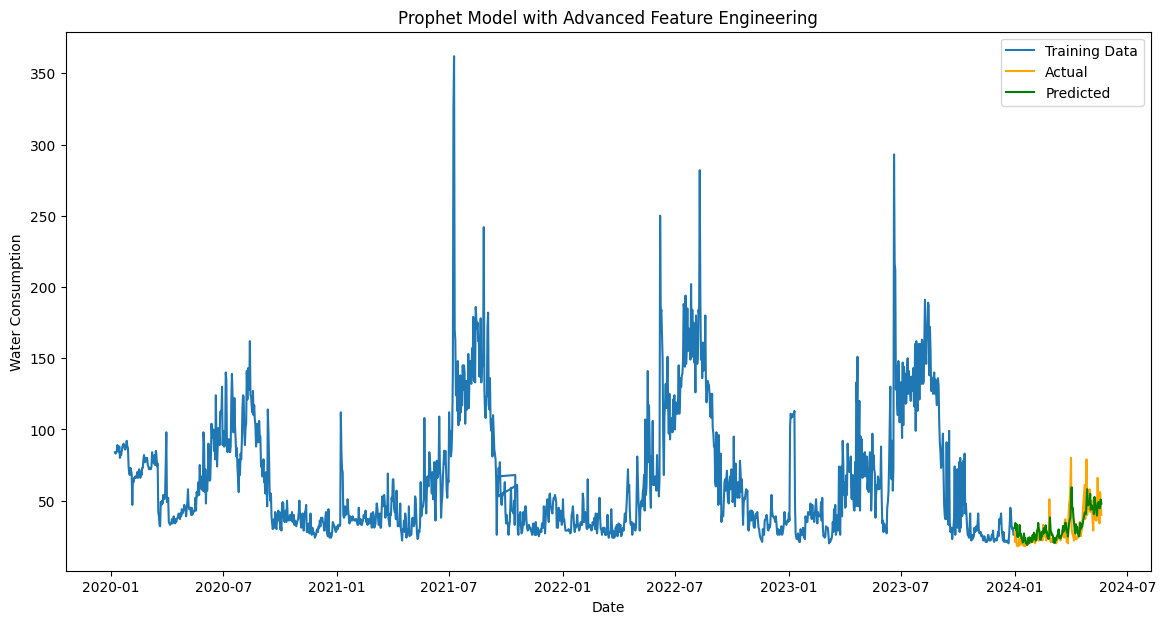}
        \caption*{(a) Six-month forecasting}
        \label{fig:PMAFE6M}
    \end{minipage}\hfill
    \begin{minipage}{0.48\textwidth}
        \centering
        \includegraphics[width=\textwidth]{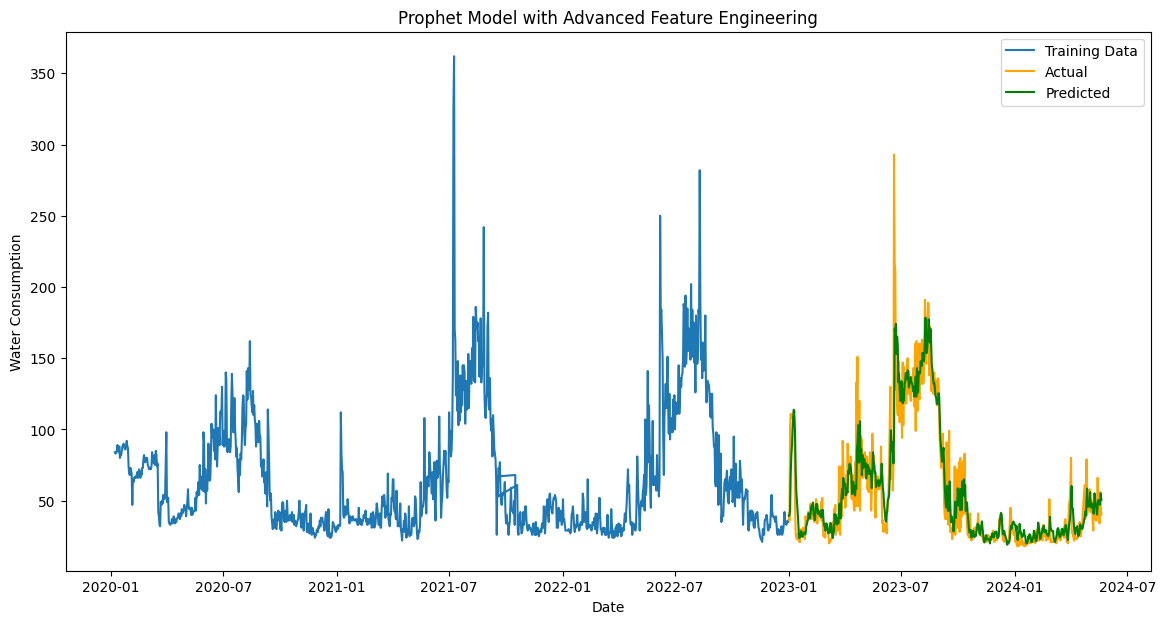}
        \caption*{(b) Eighteen-month forecasting}
        \label{fig:PMAFE18M}
    \end{minipage}
    \caption{Prophet Model with Advanced Feature Engineering (six and eighteen months forecasting). Subfigure (a) captures detailed seasonal trends, while subfigure (b) shows long-term prediction limitations.}
    \label{PMAFE}
\end{figure*}

The developed Prophet Model with Advanced Feature Engineering showed in Figure \ref{PMAFE} leverages an expansive set of lagged variables, rolling statistics, and domain-specific customizations that enhance forecasting precision. It can also be seen that the 6-month forecast exhibits a high degree of agreement with real water consumption trends. The lag features embed temporal dependencies within the model, while the rolling mean features reduce the observed short-term variability and can thus better capture the fundamental patterns. Complemented by supplementary contextual data in the day-of-week effects and indications of weekends, the model can consider such a cyclical shift in demand fluctuations. The incorporation of multiplicative seasonality, along with Fourier-based seasonal components for both weekly and monthly cycles, allows the model to respond effectively to the intricate seasonal patterns associated with water consumption. In the case of the 18-month forecast, the model maintains strong performance in identifying long-term seasonal trends; however, it exhibits certain discrepancies when forecasting abrupt transitions and anomalies. This can be attributed to the fact that the predictive power of lagged variables and their rolling means weaken with the extension of horizons. Moreover, there are possible external influences not represented by the engineered features. Still, the inclusion of country-specific holidays and domain-specific seasonality somewhat combats these weaknesses and allows the model to remain robust over this longer forecast period. 

\begin{figure*}[htbp]
    \centering
    \begin{minipage}{0.48\textwidth}
        \centering
        \includegraphics[width=\textwidth]{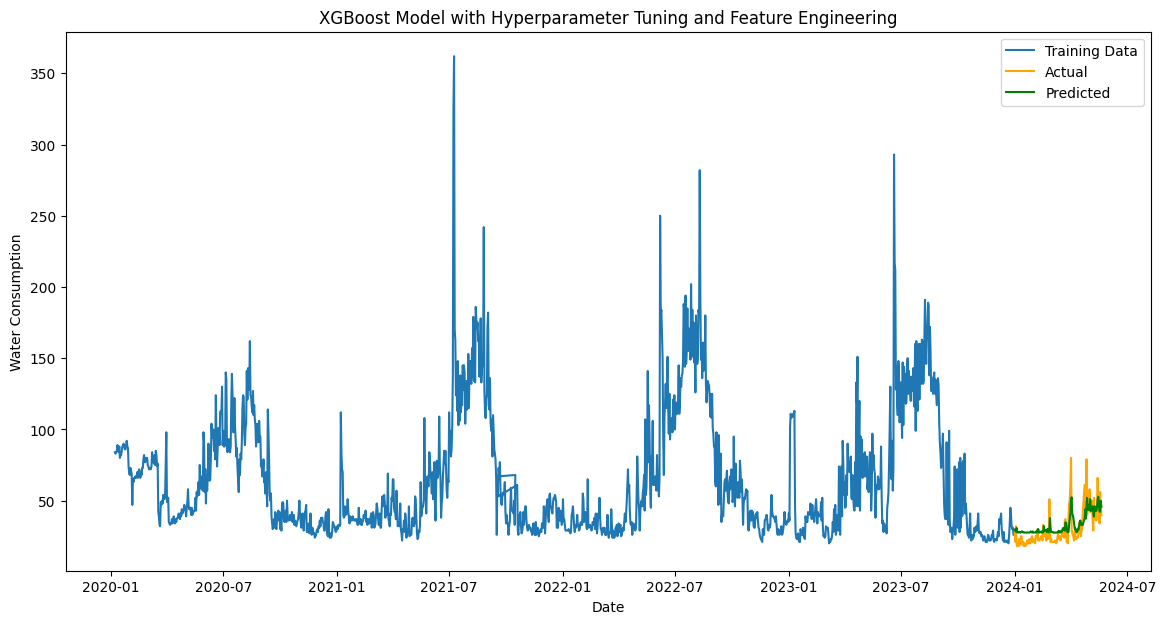}
        \caption*{(a) Six-month forecasting}
        \label{fig:XMHTFE6M}
    \end{minipage}\hfill
    \begin{minipage}{0.48\textwidth}
        \centering
        \includegraphics[width=\textwidth]{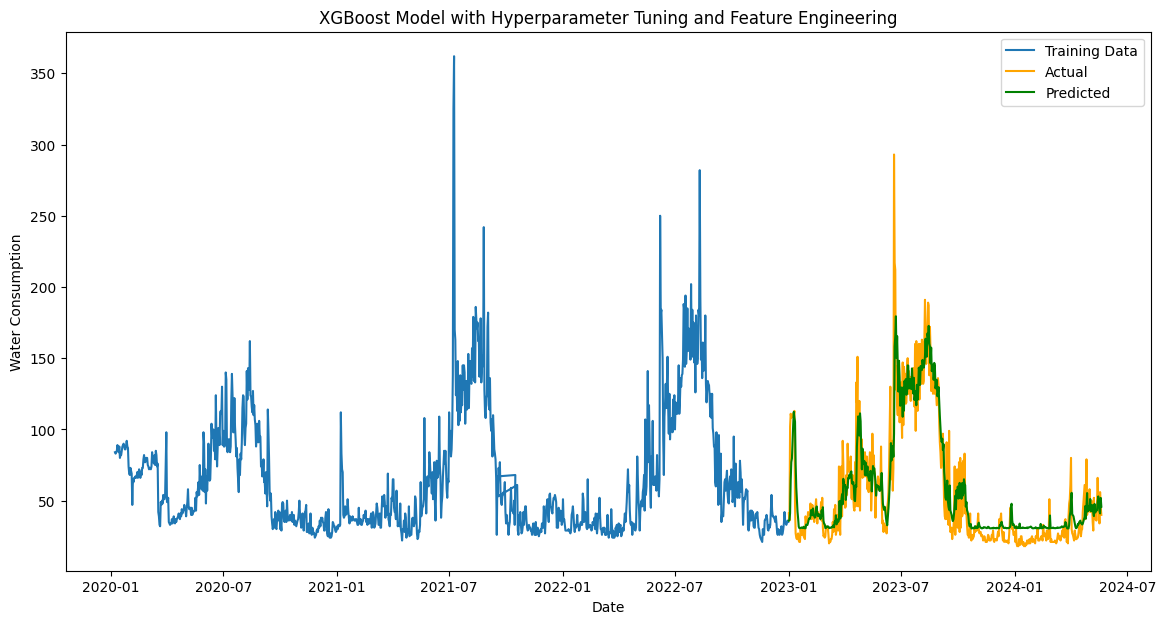}
        \caption*{(b) Eighteen-month}
        \label{fig:XMHTFE18M}
    \end{minipage}
    \caption{XGBoost Model with Hyperparameter Tuning and Feature Engineering (six and eighteen months forecasting). Subfigure (a) shows precise short-term forecasts, while subfigure (b) captures long-term seasonal trends.}
    \label{XMHTFE}
\end{figure*}

The predictions of XGBoost model with feature engineering plotted in Figure \ref{XMHTFE} exhibits strong performance in forecasting water consumption, benefiting from the same comprehensive set of engineered features as the advanced Prophet model. In the 6-month forecast, the model effectively captures short-term fluctuations and seasonality, closely aligning with actual water consumption data. The integration of lag features, rolling statistics, and temporal variables (day of the week, weekend indicators) allows XGBoost to leverage both temporal dependencies and cyclical demand patterns. Minor deviations during extreme consumption peaks suggest potential overfitting to specific patterns or the need for additional external variables. In the 18-month forecast, the model continues to demonstrate robustness in capturing overarching trends and seasonal patterns, though its accuracy diminishes in predicting sudden anomalies and small variations. The diminishing predictive power of lag and rolling features over longer horizons, combined with the inherent limitations of tree-based models in extrapolation, likely contribute to this reduction in performance. Despite these challenges, the model benefits from its ability to prioritize relevant features during training, ensuring that the most critical information drives the predictions.

\clearpage

\begin{figure*}[htbp]
    \centering
    \begin{minipage}{0.48\textwidth}
        \centering
        \includegraphics[width=\textwidth]{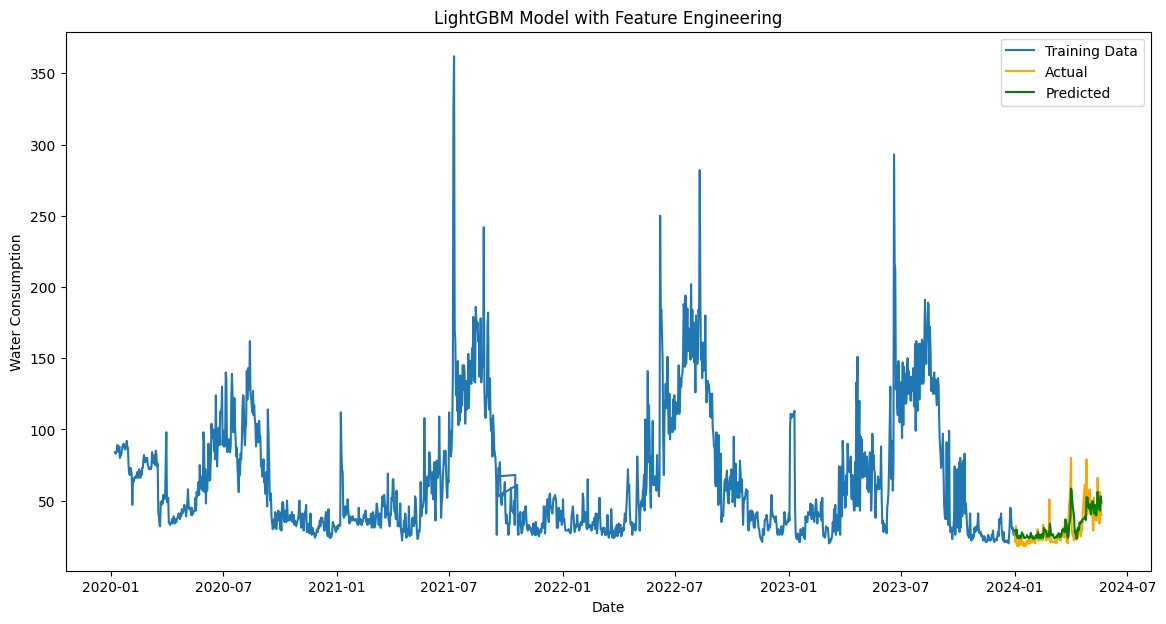}
        \caption*{(a) Six-month forecasting}
        \label{fig:LMFE6M}
    \end{minipage}\hfill
    \begin{minipage}{0.48\textwidth}
        \centering
        \includegraphics[width=\textwidth]{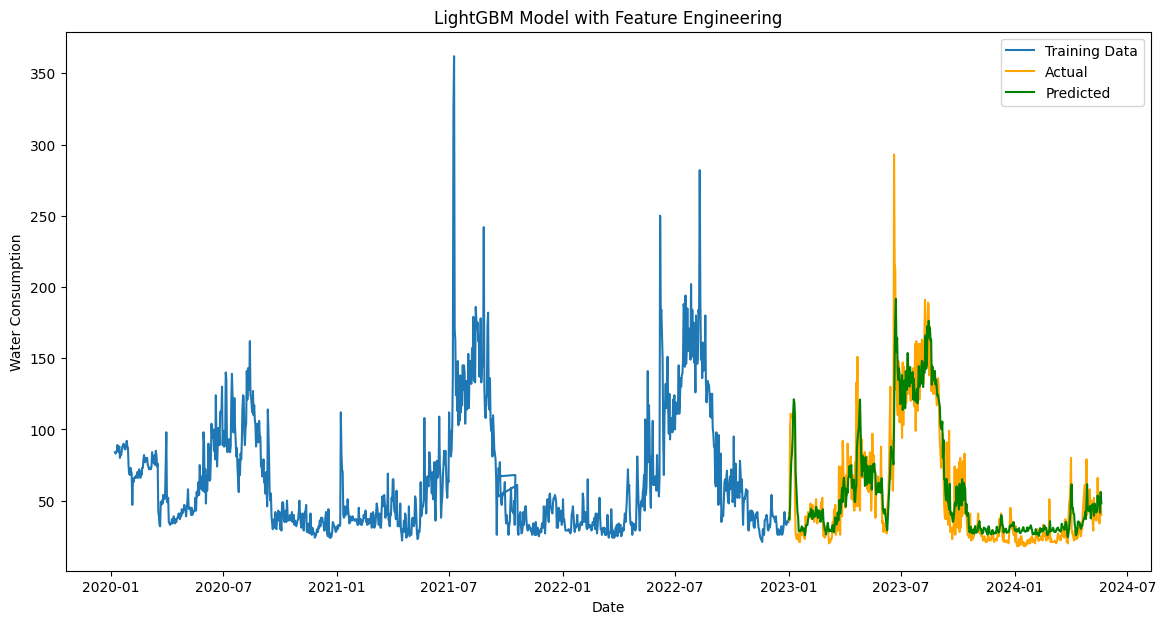}
        \caption*{(b) Eighteen-month forecasting}
        \label{fig:LMFE18M}
    \end{minipage}
    \caption{LightGBM Model with Feature Engineering (six and eighteen months forecasting). Subfigure (a) highlights short-term precision, while subfigure (b) illustrates long-term trend consistency.}
    \label{LMFE}
\end{figure*}

The LightGBM model with feature engineering is showed in \ref{LMFE} offers a nuanced understanding of water consumption dynamics, particularly over short and medium-term horizons. For the 6-month forecast, the model demonstrates remarkable precision in tracking the actual water consumption trends, effectively capturing both abrupt shifts and recurring seasonal patterns. This performance can be attributed to its reliance on well-designed features, including lag variables and rolling means, which encapsulate historical consumption behaviors and smooth out irregularities. Moreover, LightGBM's ability to assign feature importance dynamically ensures that critical variables, such as temperature and temporal markers like day-of-week or weekend indicators, are prioritized during the learning process. However, minor deviations during peak consumption periods suggest the potential for further optimization, perhaps through the inclusion of additional external drivers. Over the 18-month horizon, the model exhibits consistent performance in maintaining the overall trend alignment and seasonal pattern recognition, although some accuracy is sacrificed in capturing localized anomalies. This decline can be linked to the fading influence of lagged variables over extended periods and the inherent difficulty of modeling long-term variability with a gradient-boosting framework. Despite slight mismatches in predicting extreme events, the LightGBM model's efficiency in handling large datasets and its capacity to generalize well over different temporal scales make it a compelling choice for forecasting tasks where both accuracy and computational efficiency are crucial.

\begin{figure*}[htbp]
    \centering
    \begin{minipage}{0.48\textwidth}
        \centering
        \includegraphics[width=\textwidth]{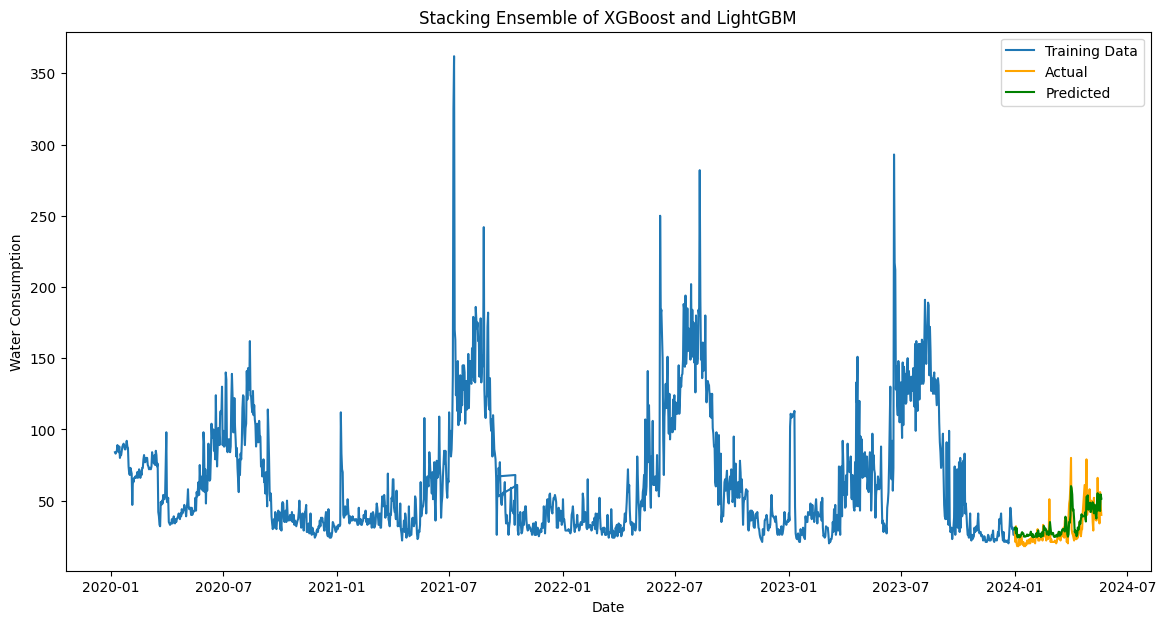}
        \caption*{(a) Six-month forecasting}
        \label{fig:SEXL6M}
    \end{minipage}\hfill
    \begin{minipage}{0.48\textwidth}
        \centering
        \includegraphics[width=\textwidth]{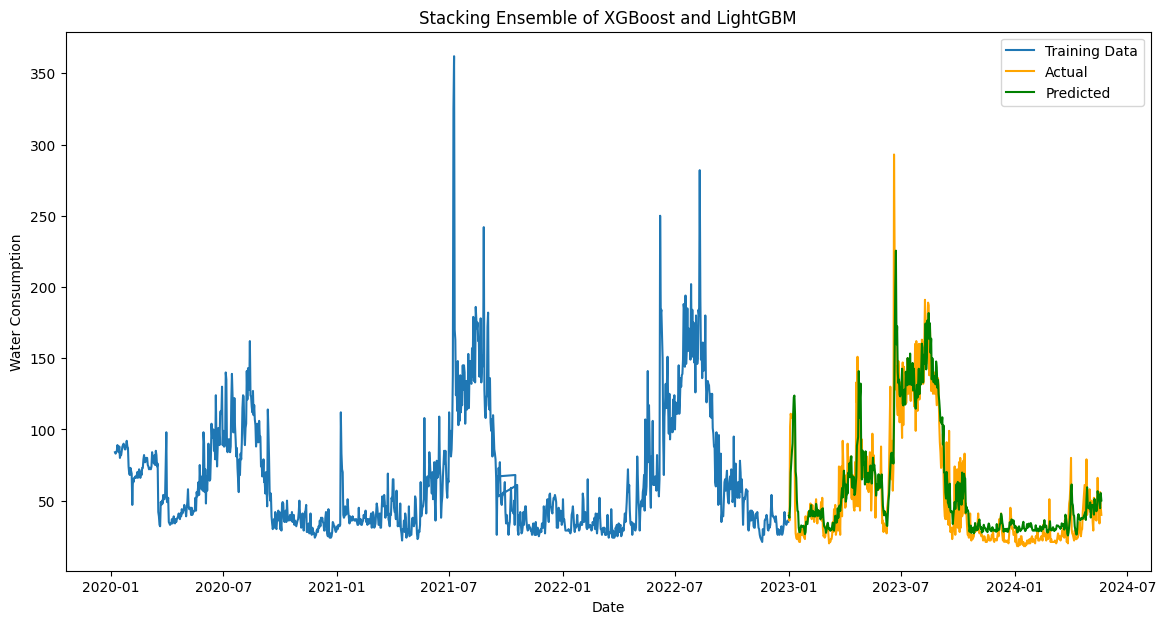}
        \caption*{(b) Eighteen-month forecasting}
        \label{fig:SEXL18M}
    \end{minipage}
    \caption{Stacking Ensemble of XGBoost and LightGBM (six and eighteen months forecasting). Subfigure (a) shows short-term accuracy, while subfigure (b) highlights trend stability in long-term forecasts.}
    \label{SEXL}
\end{figure*}

By the use of an Stacking Ensemble model, our research investigated water consumption forecasting using a hybrid approach that combines multiple prediction techniques. Results are displayed in Figure \ref{SEXL}. Analysis of 6-month forecasts revealed that merging different statistical approaches produced more accurate predictions than using any single method alone. The model successfully identified both day-to-day changes and seasonal trends in water usage patterns. While most predictions closely matched actual consumption data, some extreme usage peaks showed minor discrepancies, suggesting room for future refinement. When extended to 18-month predictions, the hybrid approach maintained its effectiveness in capturing both short-term fluctuations and long-term consumption patterns. The model proved particularly adept at identifying recurring seasonal changes in water usage. Empirical testing showed improved accuracy compared to traditional forecasting methods, especially when analyzing extended time periods. Though the model occasionally struggled to predict quick changes in consumption, it consistently provided reliable forecasts across various testing scenarios. These findings demonstrate the practical value of combining multiple forecasting techniques for water consumption prediction.

\begin{figure*}[htbp]
    \centering
    \begin{minipage}{0.48\textwidth}
        \centering
        \includegraphics[width=\textwidth]{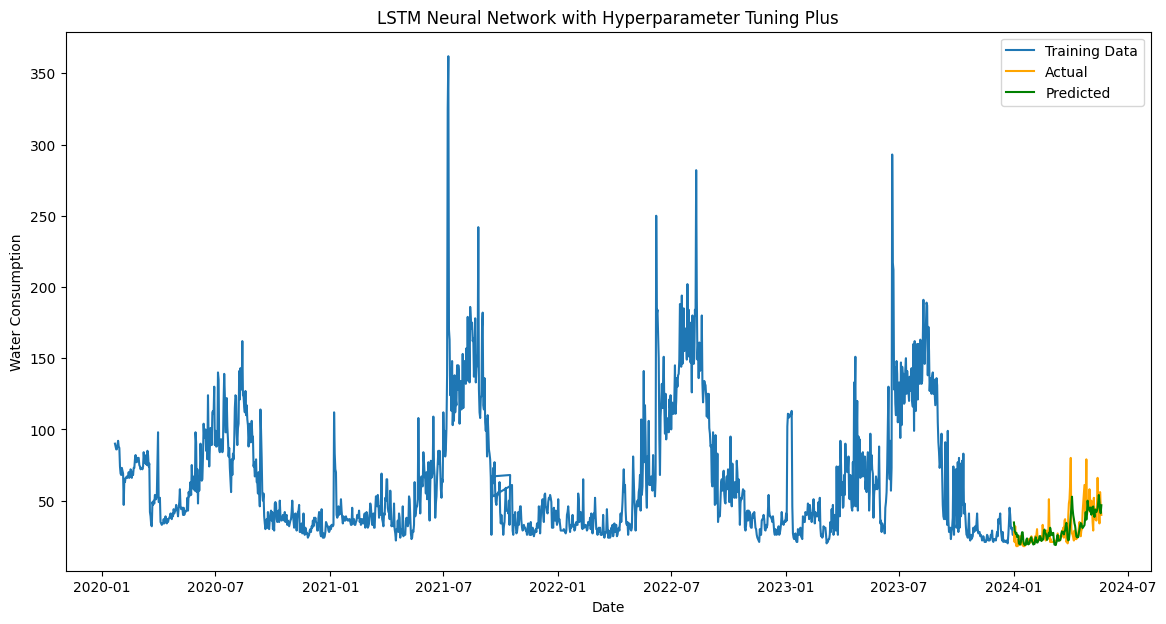}
        \caption*{(a) Six-month forecasting}
        \label{fig:LNNHTP6M}
    \end{minipage}\hfill
    \begin{minipage}{0.48\textwidth}
        \centering
        \includegraphics[width=\textwidth]{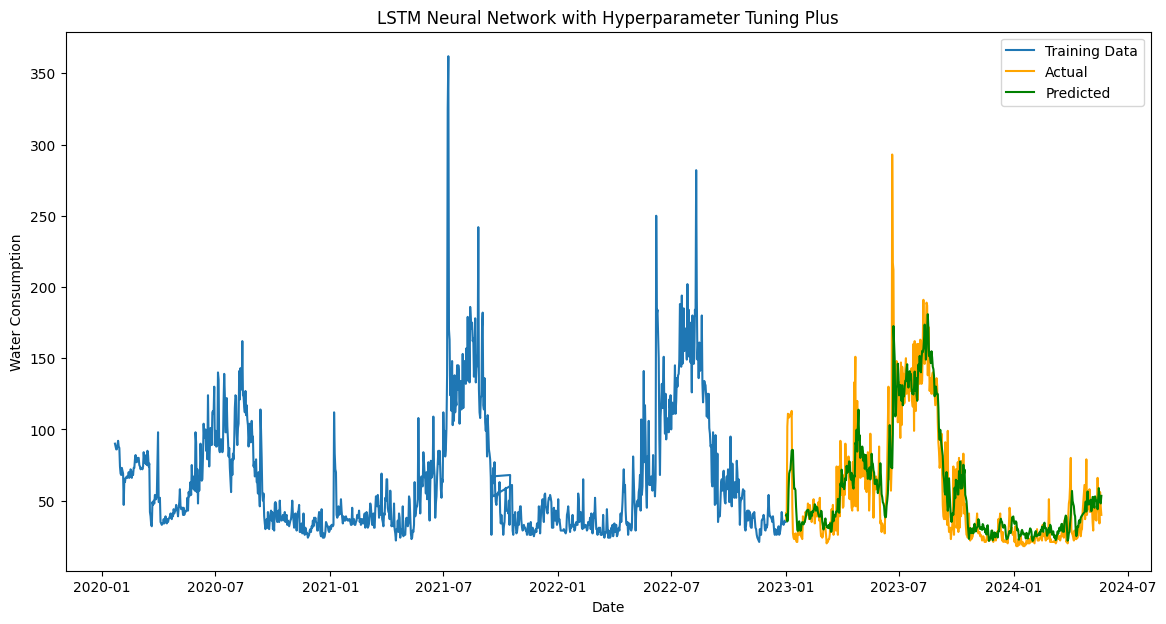}
        \caption*{(b) Eighteen-month forecasting}
        \label{fig:LNNHTP18M}
    \end{minipage}
    \caption{LSTM Neural Network (six and eighteen months forecasting). Subfigure (a) focuses on short-term accuracy, while subfigure (b) shows consistent long-term trends.}
    \label{LNNHTP}
\end{figure*}

This LSTM Neural Network gives good results as showed in Figure \ref{LNNHTP} in the forecasting of water consumption by using its strengths in portraying temporal dependencies within sequential data. The model stays close to the actual values during the 6-month forecast, showing good capture of short-term fluctuations and seasonality in the data. The LSTM architecture's recurrent structure allows storing and using previously computed information at later time steps, which gives the network the capability to represent dynamic and nonlinear relations peculiar to water consumption patterns. Small deviations in the case of important peaks suggest potential influences from external factors not represented in the input features; however, the predictions generally show a high level of reliability. In the 18-month forecast, the LSTM continues to perform well in maintaining long-term seasonal trends and general consumption patterns. It is good at capturing dependencies over long sequences, which helps in sustaining its predictive accuracy over time. However, the model displays some limitation in dealing with things that represent abrupt changes or localized anomalies, as most long-horizon predictions are bound to do, because of the attenuation of temporal dependencies over such periods.

\clearpage

\begin{figure*}[htbp]
    \centering
    \begin{minipage}{0.48\textwidth}
        \centering
        \includegraphics[width=\textwidth]{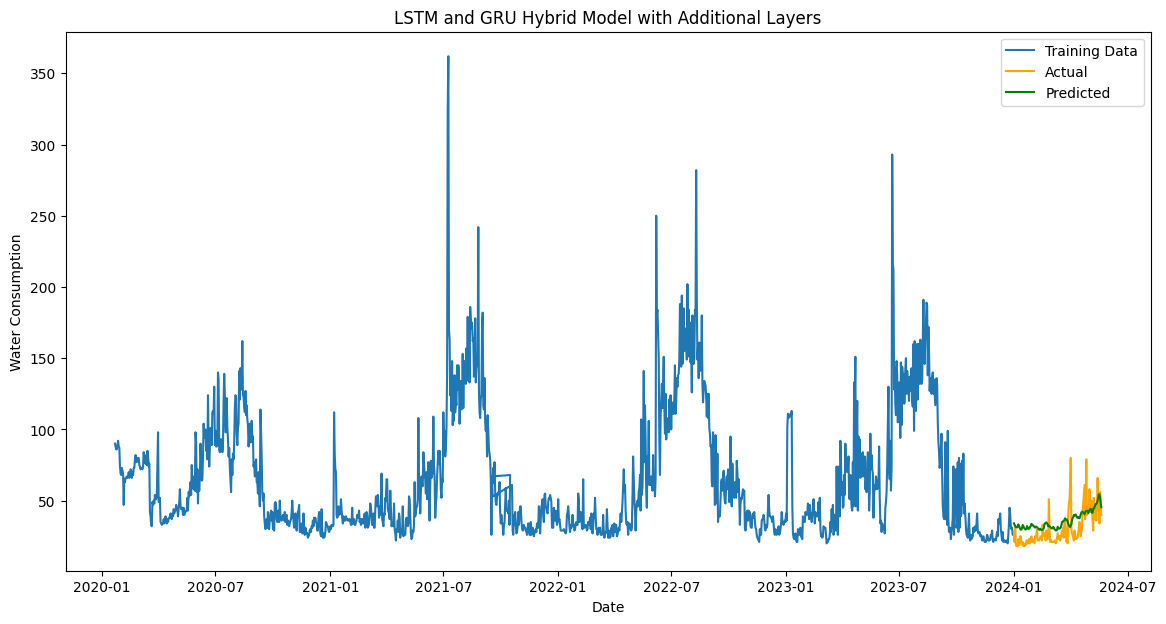}
        \caption*{(a) Six-month forecasting}
        \label{fig:LGHMAL6M}
    \end{minipage}\hfill
    \begin{minipage}{0.48\textwidth}
        \centering
        \includegraphics[width=\textwidth]{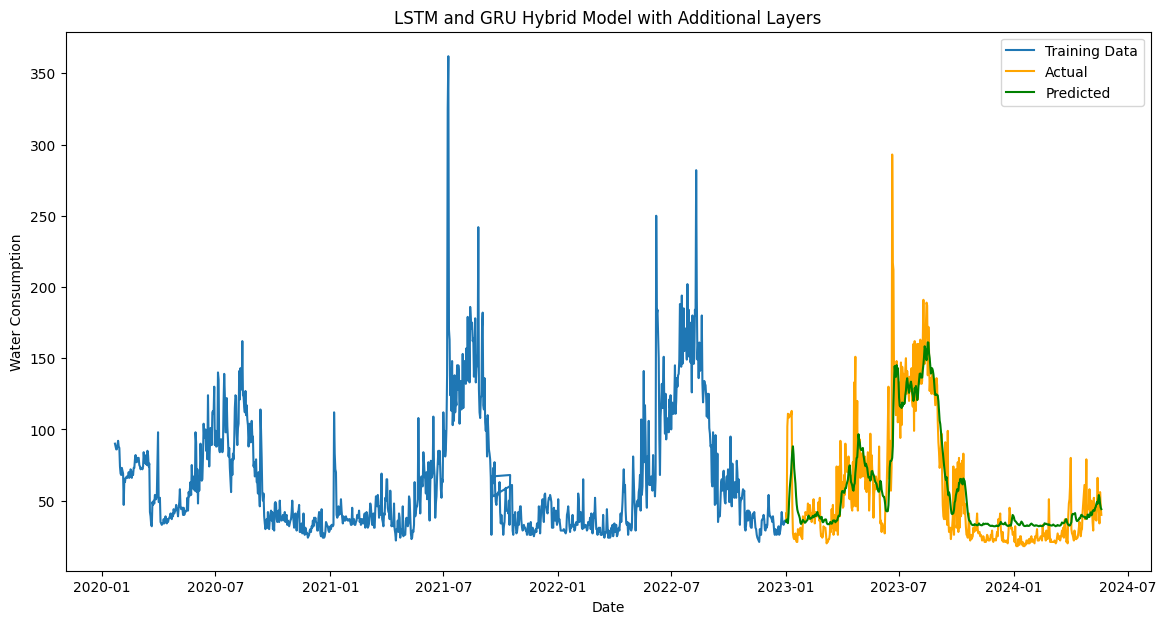}
        \caption*{(b) Eighteen-month forecasting}
        \label{fig:LGHMAL18M}
    \end{minipage}
    \caption{LSTM-GRU Hybrid Model with Additional Layers (six and eighteen months forecasting). Subfigure (a) improves short-term precision, while subfigure (b) focuses on long-term seasonal trends.}
    \label{LGHMAL}
\end{figure*}

The LSTM and GRU hybrid model's predictions are plotted in Figure \ref{LGHMAL}, which has been good at capturing the short- and long-term dependencies of the water consumption dataset. In 6-month forecasting, the combination of LSTM and GRU architectures really models the sequential dependencies well and gives accurate predictions very close to real consumption patterns. The LSTM layers learn long-term trends, while the GRU layers introduce flexibility to account for more immediate temporal changes. Further, the inclusion of dense layers helps fine-tune interactions between features, and the dropout mechanism decreases overfitting, especially during periods of high variance. The hybrid structure ascertains a balanced approach to dealing with the complexities related to short-term fluctuations and periodic trends. The hybrid model demonstrates a high ability to maintain and extrapolate learned patterns throughout a long forecasting period of 18 months. While long-term prediction naturally shows some variances due to the nature of sudden changes in water demand, the model can maintain consistency with both seasonal and trend components. The addition of more dense layers increases the generalization capability of the model over long sequences, while dropout regularization prevents overfitting common in neural networks with high-capacity architecture.

\begin{figure*}[htbp]
    \centering
    \begin{minipage}{0.48\textwidth}
        \centering
        \includegraphics[width=\textwidth]{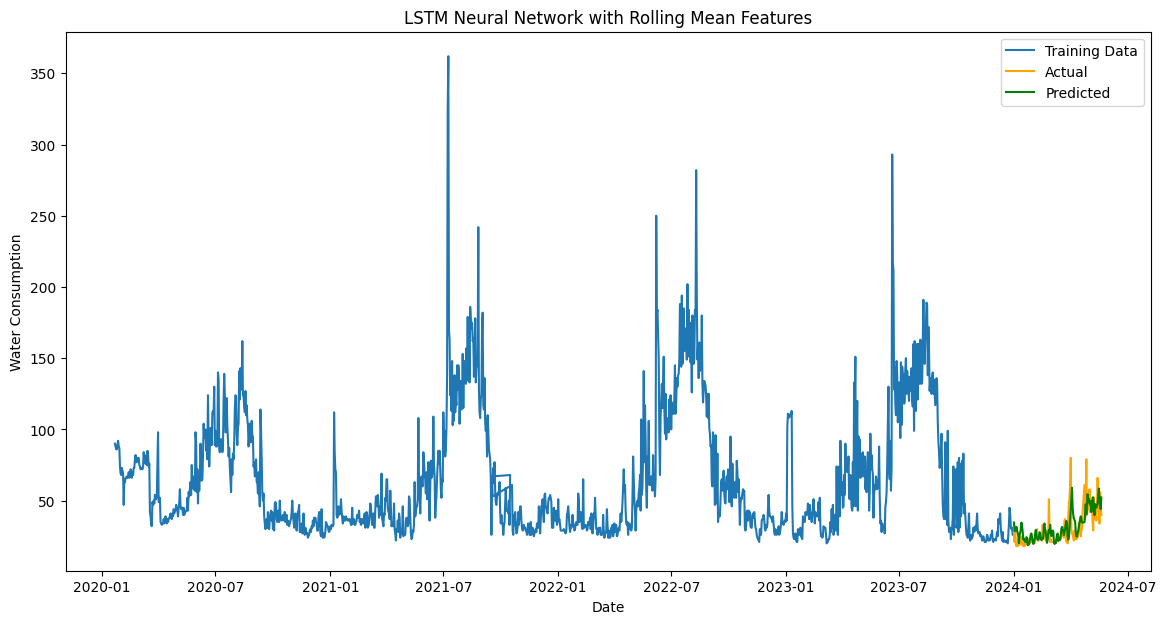}
        \caption*{(a) Six-month forecasting}
        \label{fig:LNNRMF6M}
    \end{minipage}\hfill
    \begin{minipage}{0.48\textwidth}
        \centering
        \includegraphics[width=\textwidth]{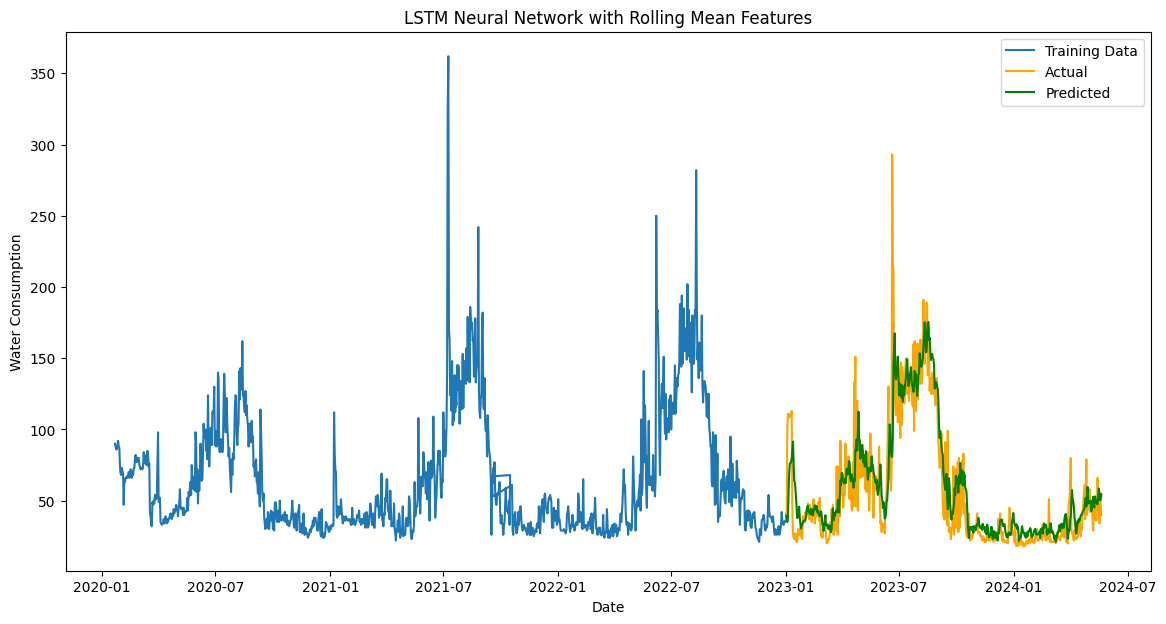}
        \caption*{(b) Eighteen-month forecasting}
        \label{fig:LNNRMF18M}
    \end{minipage}
    \caption{LSTM Neural Network with Rolling Mean Features (six and eighteen months forecasting). Subfigure (a) enhances short-term performance, while subfigure (b) focuses on seasonal patterns.}
    \label{LNNRMF}
\end{figure*}

The LSTM Neural Network with Rolling Mean Features takes a more simplified approach compared to the hybrid LSTM and GRU model with additional layers. Predictions obtained with this model are presented under Figure \ref{LNNRMF} While rolling mean features succeed in smoothing out short-term fluctuations and providing a clearer representation of underlying trends, this model's ability to adapt to sudden changes or to model complicated temporal patterns is limited by its lack of structural complexity—most specifically, the exclusion of GRU layers or additional dense layers. In the 6-month forecast, the model does well in generalizing trends and cyclical behavior, with the rolling mean features helping to stabilize the predictions. However, since this has no GRU layers, the model is less responsive to sharp, localized fluctuations; hence, slightly larger errors during abrupt changes in consumption compared to the hybrid model. To a certain extent, the rolling mean feature approach is able to expose more and more weaknesses in the 18-month ahead forecast. Although good at preserving long-term seasonal patterns, the model struggles more with smaller-scale oscillations and adaptive changes in water usage patterns. On the other hand, the hybrid model allows for much finer handling of the temporal dependencies over extended periods due to the combined strengths of LSTM and GRU layers.

\begin{figure*}[htbp]
    \centering
    \begin{minipage}{0.48\textwidth}
        \centering
        \includegraphics[width=\textwidth]{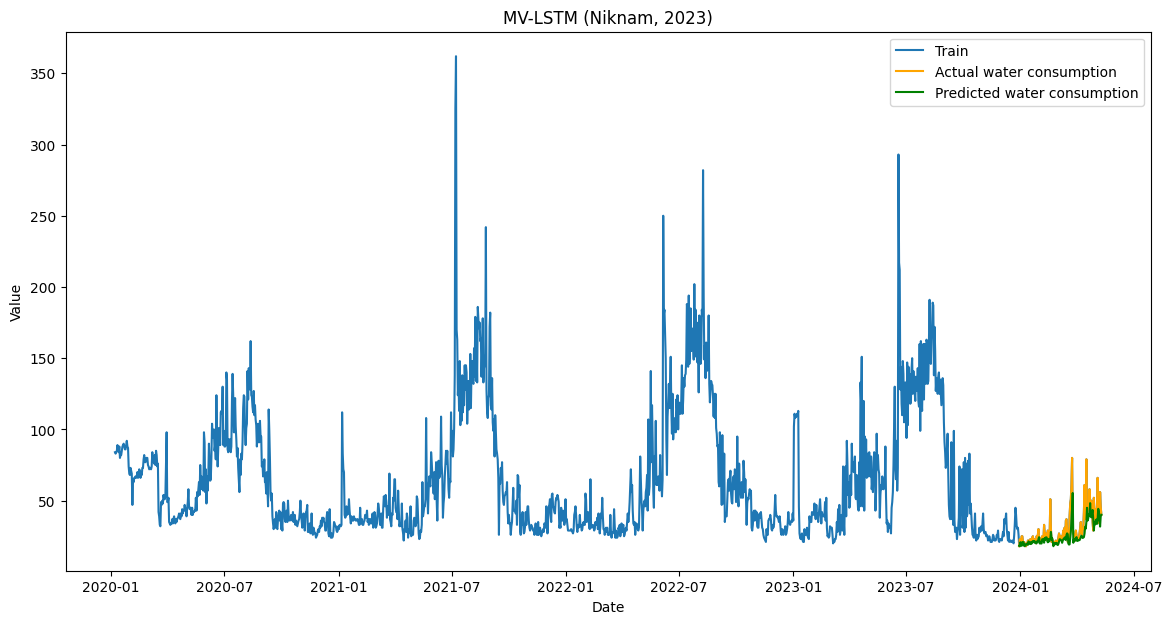}
        \caption*{(a) Six-month forecasting}
        \label{fig:NIKNAMLSTM6M}
    \end{minipage}\hfill
    \begin{minipage}{0.48\textwidth}
        \centering
        \includegraphics[width=\textwidth]{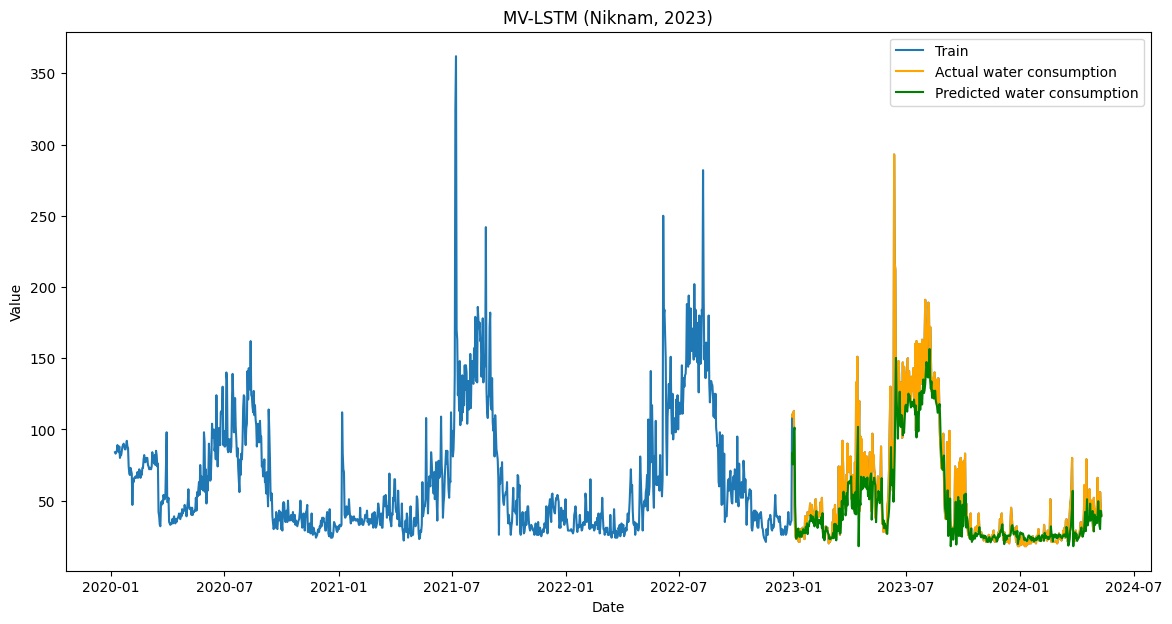}
        \caption*{(b) Eighteen-month forecasting }
        \label{fig:NIKNAMLSTM18M}
    \end{minipage}
    \caption{MV-LSTM Niknam (six and eighteen months forecasting). Subfigure (a) demonstrates short-term robustness, while subfigure (b) ensures long-term trend consistency.}
    \label{NIKNAMLSTM}
\end{figure*}

The MV-LSTM model, presented by Niknam \cite{Niknam2023}, exhibits its ability to fuse multivariate interdependencies by integrating external meteorological parameters with water consumption data. Predictions of water consumption are plotted in Figure \ref{NIKNAMLSTM}. This enables the model to learn complicated relationships between climatic variables and water demand, which consequently enhances its performance in forecasting short-term consumption variability. The combination of these different input features makes the model more capable of learning temporal dependencies and complicated interactions, which guarantees stable predictive performance in shorter time frames. In longer horizons, MV-LSTM preserves the possibility of accurate predictions by maintaining seasonal patterns and general consumption trends. Its sensitivity to new changes decreases, yet it still reacts to dynamic external conditions due to the inclusion of the variables with lags and those representing meteorological data. By integrating multivariate inputs with sophisticated recurrent architectures, the MV-LSTM proves suitable for applications in which internal and external variables that impact water consumption need to be evaluated.

This study conducts an in-depth evaluation of a variety of forecasting models for the task of water demand prediction, with a special focus on their performance over both the short and long terms. Models considered range from various variants of Prophet to gradient boosting models using LightGBM and XGBoost and to state-of-the-art neural network architectures, including LSTM-based models and the lately proposed multivariate LSTM (MV-LSTM) \cite{Niknam2023}. Results in Table \ref{tab:comprehensive_model_comparison} indicate that the performances of these models differ substantially. Each of the methods thus has its pros and cons: while Prophet models show stable performance across the forecasting horizon, they reveal large improvements when advanced feature engineering is applied. The Prophet with Advanced Feature Engineering variant performs the best among the Prophet models: MAE of 5.76 and RMSE of 8.31 for the 6-month horizon, and MAE of 10.07 and RMSE of 15.02 for the 18-month horizon. Such improvement highlights the need to include domain-specific features, such as lagged variables and custom seasonalities, as these increase the model's ability to recognize temporal patterns and seasonal trends. While Prophet models are computationally efficient and interpretable, they present limitations in adapting quickly to, for example, sudden changes in consumption for the longer forecast horizons.

Among all the gradient boosting models, LightGBM outperforms XGBoost for both short- and long-term horizons, with an MAE of 5.90 and RMSE of 8.25 for the 6-month horizon, and an MAE of 11.77 and RMSE of 18.31 for the 18-month horizon. This improved performance of LightGBM may be attributed to its capability of handling large datasets efficiently while capturing complex interactions among features by its gradient boosting framework. XGBoost is competitive but has slightly higher errors and more so with the longer-term horizon. Ensemble methods, such as stacking LightGBM and XGboost, are more robust but do not outperform LightGBM alone in terms of general accuracy. This underlines the stand-alone strength that LightGBM possesses for time series forecasting.

The neural network approaches, especially those based on LSTM, are capable of capturing complex temporal dependencies. The LSTM model performs well, with an MAE of 5.96 and RMSE of 9.38 for 6 months, showing its strength in short-term forecasting. However, the model's performance slightly deteriorates for the 18-month horizon, with an MAE of 12.63 and RMSE of 20.66, pointing out some challenges in keeping high accuracy over longer time frames. The hybrid model LSTM + GRU, despite its architectural complexity, does not outperform the simpler configurations of LSTM, indicating that deeper layers and more components may introduce overfitting or a higher sensitivity to noise. The MV-LSTM proposed by Niknam outperforms all the other models with respect to accuracy in capturing multivariate dependencies. It realizes the lowest MAPE 15.48\% for the 6-month horizon and performs well in the long-term horizon with an MAPE of 19.30\%, which is the most accurate model concerning the relative percentage error. These results have highlighted the potential of multivariate approaches in leveraging external regressors to enhance predictive precision.

\begin{table}[H]
\centering
\caption{Comprehensive Comparison of Forecasting Models Across Different Time Frames}
\label{tab:comprehensive_model_comparison}
\scriptsize 
\begin{tabular}{@{}lcccccc@{}}
\toprule
 & \multicolumn{3}{c}{6 Months} & \multicolumn{3}{c}{18 Months} \\
\cmidrule(lr){2-4} \cmidrule(lr){5-7}
\textbf{Model} & \textbf{MAE} & \textbf{RMSE} & \textbf{MAPE} & \textbf{MAE} & \textbf{RMSE} & \textbf{MAPE} \\
\midrule
Prophet Basic & 10.37 & 13.66 & 22.45\% & 19.70 & 28.68 & 22.45\% \\
Prophet + Seasonality & 12.39 & 14.91 & 22.45\% & 24.25 & 35.02 & 22.45\% \\
Advanced Prophet & 6.24 & 8.78 & 20.77\% & 11.14 & 18.02 & 22.34\% \\
\textbf{Prophet Adv. Engineering} & \textbf{5.76} & \textbf{8.31} & \textbf{18.61}\% & \textbf{10.07} & \textbf{15.02 }& \textbf{20.12}\% \\
\hline
XGBoost & 7.02 & 8.74 & 24.93\% & 12.34 & 18.50 & 27.49\% \\
\textbf{LightGBM} & \textbf{5.90} & \textbf{8.25} & \textbf{19.64}\% & \textbf{11.77} & \textbf{18.31} & \textbf{24.98}\% \\
Stacking XGBoost + LightGBM & 6.57 & 8.70 & 22.45\% & 12.48 & 18.94 & 27.62\% \\
\hline
\textbf{LSTM} & 5.96 & \textbf{9.38} & 18.64\% & 12.63 & 20.66 & 25.61\% \\
LSTM + GRU Hybrid & 8.18 & 10.10 & 30.10\% & 14.64 & 22.32 & 34.06\% \\
LSTM Rolling Mean Features & 7.94 & 10.82 & 27.59\% & 12.33 & \textbf{20.57} & 24.67\% \\
MV-LSTM \cite{Niknam2023} & 5.91 & 10.03 & \textbf{15.48\%} & \textbf{12.30} & 22.53 & \textbf{19.30}\%
 \\
\bottomrule
\end{tabular}
\end{table}

\newpage

\section{Optimizing WDS Maintenance}\label{sec:OptimisationWDS}

In many rural water distribution networks, operators face challenges in determining the most effective routing and scheduling of maintenance activities. Such decisions are usually based on subjective judgment, which may lead to inefficiencies, especially where many activities with varying priorities and dependencies must be coordinated simultaneously. Without a systematic approach, operators are left to grapple with how to prioritize tasks in a manner that minimizes operational costs—travel time, CO$_2$ emissions, distance traveled—to the end user while resulting in longer times to complete tasks and suboptimal resource utilization. The problem becomes even more complicated when high-priority emergency tasks arrive at times that don't agree with the existing schedule presented to the operator and when tasks depend on the execution of each other. This calls for more structured and dynamic scheduling with a view to incorporating uncertainties and task dependencies.

We are dealing with an NP-hard problem known as \textbf{\textit{Single Machine Scheduling with Preemptive Jobs, Variable Release Times, and Task Dependencies}}, in which there is a single resource that has to deal with multiple tasks, which may be interrupted (preemptions), new tasks may arrive at any moment, and some tasks may depend on the completion of others. This model incorporates, in each day, random assignments of dependencies between tasks to reflect real-world operational complexities where not all tasks are independent. The objective is to minimize general operational costs, including the total completion time, fuel consumption in liters, emissions of CO$_2$, and delays of tasks, and optimally manage tasks with different priority levels, including high-priority emergency tasks, also known as emergency response activities while satisfying task dependencies.

Apart from saving time and delays in the completion of the tasks, the model also accounts for fuel consumption and CO$_2$ emissions by analyzing various task characteristics such as location, processing time, and gradients that affect fuel efficiency. With this scheduling model in place, the system is able to cope with the challenges in dynamic task arrivals, task dependencies, and preemptions while ensuring the proper execution of prioritized tasks. Given that the problem is NP-hard, we consider using CP as an approach that provides an optimized solution. In the following sections, we describe the mathematical formulation of the problem.

\begin{itemize}

\item \textbf{Mathematical Model}

We have adapted the model to be compatible with CP, focusing on deterministic parameters. Uncertainties in processing and travel times are handled using expected or conservative estimates. The model aims to optimize the objectives while satisfying all constraints.

The objectives are to:

\begin{itemize} 
    \item Minimize the total completion time ($C_{\text{max}}$). 
    \item Minimize the total fuel consumption ($F_{\text{total}}$). 
    \item Minimize the total CO$_2$ emissions ($C_{\text{total}}$). 
    \item Minimize the total delays and penalties ($D_{\text{total}}$). 
\end{itemize}

\item \textbf{Sets and Indices}

\begin{itemize} 
    \item Let $T$ be the set of all tasks indexed by $i$. 
    \item Let $K_i$ be the set of processing segments for task $i$ due to preemption, indexed by $k$. 
    \item Let $D$ be the set of task dependencies, where $(i, j) \in D$ means task $j$ depends on task $i$.
\end{itemize}

An example of task dependencies is provided in Table~\ref{tab:task_dependencies}.

\begin{table}[H]
\scriptsize
\centering
\caption{Task Dependencies}
\label{tab:task_dependencies}
\begin{tabular}{cc}
\toprule
Dependency $(i, j)$ & Description \\
\midrule
(1, 3) & Task 3 depends on Task 1 \\
(2, 4) & Task 4 depends on Task 2 \\
(5, 3) & Task 3 depends on Task 5 \\
\bottomrule
\end{tabular}
\end{table}

\item \textbf{Parameters}

For each task $i \in T$:

\begin{itemize} 
    \item $p_i$: Deterministic processing time required for task $i$ (in hours), based on expected or conservative estimates. 
    \item $d_{ij}$: Deterministic travel time from task $i$ to task $j$ (in hours), based on expected or conservative estimates. 
    \item $f_i$: Fuel consumption for processing task $i$ (in liters). 
    \item $c_i$: CO$_2$ emissions for processing task $i$ (in kg). 
    \item $\text{loc}_i$: Location coordinates $(\text{lat}_i, \text{lon}_i)$ of task $i$. 
    \item $\text{priority}_i$: Priority level of task $i$ (higher value indicates higher priority). 
    \item $r_i$: Task's release time (arrival time). For regular tasks, $r_i = 0$; for emergency tasks, $r_i \geq 0$. 
    \item $l_{ij}$: Distance from task $i$ to task $j$ (in kilometers). 
    \item $\text{VE}_v$: Fuel efficiency of vehicle type $v$ (in km per liter). 
    \item $\text{EF}_v$: Emission factor of vehicle type $v$ (in kg CO$_2$ per liter of fuel). 
    \item $S$: Start of the working day (e.g., 8:00 AM). 
    \item $E$: End of the working day (e.g., 3:00 PM). 
    \item $\text{Vehicle}_i$: Vehicle type required for task $i$ (e.g., Van, Small Truck). 
    \item $\text{MaxPreemptions}$: Maximum allowed number of preemptions per task. 
\end{itemize}

Table~\ref{tab:tasks_parameters} provides an example of tasks and associated parameters.

\begin{table}[H]
\scriptsize
\centering
\caption{Example Tasks and Parameters}
\label{tab:tasks_parameters}
\begin{tabular}{ccccccccc}
\toprule
Task $i$ & $p_i$ (hrs) & $f_i$ (L) & $c_i$ (kg) & $\text{loc}_i$ (lat, lon) & $\text{priority}_i$ & $r_i$ (hrs) & $\text{Vehicle}_i$ & $\text{MaxPreemptions}$ \\
\midrule
1 & 2.0 & 5.0 & 13.2 & (51.5074, -0.1278) & 2 & 0 & Van & 1 \\
2 & 1.5 & 3.5 & 9.24 & (51.5155, -0.1410) & 3 & 0 & Van & 2 \\
3 & 2.5 & 6.0 & 15.84 & (51.5237, -0.1585) & 1 & 2 & Small Truck & 1 \\
4 & 1.0 & 2.5 & 6.6 & (51.5308, -0.1208) & 4 & 0 & Van & 1 \\
5 & 3.0 & 7.5 & 19.8 & (51.4975, -0.1357) & 5 & 1 & Small Truck & 2 \\
\bottomrule
\end{tabular}
\end{table}

The travel times and distances between tasks are provided in Table~\ref{tab:travel_times_distances}.

\begin{table}[H]
\scriptsize
\centering
\caption{Travel Times and Distances Between Tasks}
\label{tab:travel_times_distances}
\begin{tabular}{cccc}
\toprule
From Task $i$ & To Task $j$ & $d_{ij}$ (hrs) & $l_{ij}$ (km) \\
\midrule
1 & 2 & 0.5 & 10 \\
1 & 3 & 0.7 & 14 \\
1 & 4 & 0.4 & 8 \\
1 & 5 & 0.6 & 12 \\
2 & 3 & 0.6 & 12 \\
2 & 4 & 0.3 & 6 \\
2 & 5 & 0.7 & 14 \\
3 & 4 & 0.8 & 16 \\
3 & 5 & 0.5 & 10 \\
4 & 5 & 0.6 & 12 \\
\bottomrule
\end{tabular}
\end{table}

The fuel efficiency and emission factors for the vehicle types used are given in Table~\ref{tab:vehicle_specifications}.

\begin{table}[H]
\scriptsize
\centering
\caption{Vehicle Types and Specifications}
\label{tab:vehicle_specifications}
\begin{tabular}{ccc}
\toprule
Vehicle Type $v$ & $\text{VE}_v$ (km/L) & $\text{EF}_v$ (kg CO$_2$/L) \\
\midrule
Van & 12 & 2.64 \\
Small Truck & 8 & 2.68 \\
\bottomrule
\end{tabular}
\end{table}

\item \textbf{Decision Variables}

\begin{itemize}

\item \textbf{Task Scheduling Variables}: Variables that define when each task or task segment starts and ends.

\begin{itemize}
    \item $s_{ik}$: Scheduled start time of segment $k$ of task $i$.
    \item $C_{ik}$: Scheduled completion time of segment $k$ of task $i$.
    \item $p_{ik}$: Scheduled processing time of segment $k$ of task $i$.
    \item $K_i$: Number of segments into which task $i$ is divided due to preemption, where $K_i \leq \text{MaxPreemptions}$.
\end{itemize}

\item \textbf{Sequencing Variables}: Binary variables that determine the order in which tasks are performed relative to each other.

\begin{itemize}
    \item $y_{ij}$: Binary variable; $y_{ij} = 1$ if task $i$ is scheduled immediately before task $j$, 0 otherwise.
\end{itemize}

\item \textbf{Auxiliary Variables}:

\begin{itemize}
    \item $\delta_i$: Binary variable; $\delta_i = 1$ if task $i$ is preempted, 0 otherwise.
    \item $\theta_{ij}$: Binary variable; $\theta_{ij} = 1$ if task $j$ depends on task $i$, 0 otherwise.
    \item $x_{ijkl}$: Binary variable indicating if segment $k$ of task $i$ is scheduled before segment $l$ of task $j$.
\end{itemize}

\end{itemize}

\item \textbf{Objective Function}

\begin{equation} \label{eq:objective_function}
\min Z = w_t \times (C_{\text{max}} - S) + w_f \times F_{\text{total}} + w_c \times C_{\text{total}} + w_d \times D_{\text{total}}
\end{equation}

Where:

\begin{itemize} 
    \item $w_t, w_f, w_c, w_d$ are the weights for time, fuel consumption, CO$_2$ emissions, and delays and penalties, respectively. 
    \item $C_{\text{max}} = \max_{i,k} C_{ik}$: Completion time of the last task segment. 
    \item $F_{\text{total}} = \sum_{i \in T} f_i + \sum_{i \in T} \sum_{j \in T} y_{ij} \cdot f_{ij}$: Total fuel consumption for processing and traveling. 
    \item $C_{\text{total}} = \sum_{i \in T} c_i + \sum_{i \in T} \sum_{j \in T} y_{ij} \cdot c_{ij}$: Total CO$_2$ emissions for processing and traveling. 
    \item $D_{\text{total}} = \sum_{\text{emergency } i} (C_{iK_i} - r_i)$: Total delays for emergency tasks beyond their release times. 
\end{itemize}

\item \textbf{Fuel Consumption and CO$_2$ Emissions for Traveling}

The fuel consumption and CO$_2$ emissions for traveling between tasks are calculated using the following equations:

\begin{equation} \label{eq:fuel_consumption_travel}
f_{ij} = \frac{l_{ij}}{\text{VE}_{\text{Vehicle}_i}}
\end{equation}

\begin{equation} \label{eq:co2_emissions_travel}
c_{ij} = f_{ij} \times \text{EF}_{\text{Vehicle}_i}
\end{equation}

Using the data from Tables~\ref{tab:tasks_parameters}, \ref{tab:vehicle_specifications}, and \ref{tab:travel_times_distances}, the calculated fuel consumption and CO$_2$ emissions for traveling are presented in Table~\ref{tab:fuel_emissions_travel}.

\begin{table}[H]
\scriptsize
\centering
\caption{Calculated Fuel Consumption and CO$_2$ Emissions for Traveling}
\label{tab:fuel_emissions_travel}
\begin{tabular}{cccccc}
\toprule
From Task $i$ & To Task $j$ & Vehicle & $l_{ij}$ (km) & $f_{ij}$ (L) & $c_{ij}$ (kg) \\
\midrule
1 & 2 & Van & 10 & $\frac{10}{12} \approx 0.83$ & $0.83 \times 2.64 \approx 2.19$ \\
1 & 3 & Small Truck & 14 & $\frac{14}{8} = 1.75$ & $1.75 \times 2.68 \approx 4.69$ \\
1 & 4 & Van & 8 & $\frac{8}{12} \approx 0.67$ & $0.67 \times 2.64 \approx 1.77$ \\
1 & 5 & Small Truck & 12 & $\frac{12}{8} = 1.5$ & $1.5 \times 2.68 = 4.02$ \\
\bottomrule
\end{tabular}
\end{table}

\item \textbf{Constraints}

\begin{itemize} 
\item \textbf{Processing Time Constraints}: Ensure that each task's total scheduled processing time, including preempted segments, matches the required time.

\begin{equation} \label{eq:processing_time_constraint}
\sum_{k=1}^{K_i} p_{ik} = p_i
\end{equation}

\item \textbf{Segment Completion Constraints}:

For all segments $k$ of task $i$:

\begin{equation} \label{eq:segment_completion_constraint}
C_{ik} = s_{ik} + p_{ik}
\end{equation}

\item \textbf{Precedence Constraints for Task Dependencies}:

If task $j$ depends on task $i$ (i.e., $(i, j) \in D$):

\begin{equation} \label{eq:precedence_constraint}
s_{j1} \geq C_{iK_i}
\end{equation}

This ensures that task $j$ cannot start before task $i$ is completed.

\item \textbf{Travel Time Constraints}:

When task $i$ is scheduled immediately before task $j$:

\begin{equation} \label{eq:travel_time_constraint}
s_{j1} \geq C_{iK_i} + d_{ij}
\end{equation}

\item \textbf{Non-Overlap Constraints (Single-Machine Constraint)}:

To prevent overlapping of processing times on the single machine, we include the following constraints for all tasks \( i \neq j \) and their segments \( k \) and \( l \):

    \begin{equation}
    C_{ik} \leq s_{jl} + M (1 - x_{ijkl})
    \label{eq:non_overlap_corrected_1}
    \end{equation}

    \begin{equation}
    C_{jl} \leq s_{ik} + M x_{ijkl}
    \label{eq:non_overlap_corrected_2}
    \end{equation}

Where:

\begin{itemize}
    \item \( C_{ik} \) is the completion time of segment \( k \) of task \( i \).
    \item \( s_{ik} \) is the start time of segment \( k \) of task \( i \).
    \item \( x_{ijkl} \) is a binary variable defined as:
    \[
    x_{ijkl} =
    \begin{cases}
    1, & \text{if segment } k \text{ of task } i \text{ is scheduled before segment } l \text{ of task } j; \\
    0, & \text{otherwise.}
    \end{cases}
    \]
    \item \( M \) is a sufficiently large positive constant.
\end{itemize}

\item \textbf{Work Hours Constraints}:

Ensure tasks are scheduled within working hours:

\begin{equation} \label{eq:work_hours_constraint}
S \leq s_{ik} \leq C_{ik} \leq E
\end{equation}

\item \textbf{Emergency Task Constraints}:

\begin{itemize}
    \item \textbf{Release Time Constraint}:

    \begin{equation} \label{eq:release_time}
    s_{i1} \geq r_i
    \end{equation}

    \item \textbf{Delay Penalties}: Delays for emergency tasks are included in the objective function through $D_{\text{total}}$ as shown in Equation \eqref{eq:objective_function}.
\end{itemize}

\item \textbf{Limit on Preemptions}:

\begin{equation} \label{eq:preemption_limit}
K_i \leq \text{MaxPreemptions}
\end{equation}

\item \textbf{Sequencing Constraints}:

Ensure that each task is preceded and succeeded by at most one other task:

\begin{equation} \label{eq:sequencing_constraint_1}
\sum_{j \in T} y_{ij} = 1 \quad \forall i \in T
\end{equation}

\begin{equation} \label{eq:sequencing_constraint_2}
\sum_{i \in T} y_{ij} = 1 \quad \forall j \in T
\end{equation}

\item \textbf{Subtour Elimination Constraints} (Miller-Tucker-Zemlin constraints):

Introduce variables $u_i$ for each task $i$:

\begin{equation} \label{eq:subtour_elimination_1}
u_i \geq 1 \quad \forall i \in T
\end{equation}

\begin{equation} \label{eq:subtour_elimination_2}
u_i - u_j + |T| y_{ij} \leq |T| - 1 \quad \forall i \neq j
\end{equation}

\end{itemize}

\item \textbf{Performance Metrics}

In addition to the objective function, we define the following performance metrics to evaluate the scheduling model:

\begin{itemize} 
    \item \textbf{Total Delays and Penalties ($D_{\text{total}}$)}:

    \begin{equation} \label{eq:total_delays}
    D_{\text{total}} = \sum_{i \in T} \max(0, C_{iK_i} - r_i)
    \end{equation}

    This represents the total delay beyond the release times $r_i$ for all tasks.

    \item \textbf{Efficiency and Utilization ($E_{\text{eff}}$)}:

    \begin{equation} \label{eq:efficiency}
    E_{\text{eff}} = \left( \frac{\sum_{i \in T} p_i}{C_{\text{max}} - S} \right) \times 100\%
    \end{equation}

    This represents the percentage of time spent on processing tasks relative to the total time from the start to the completion of all tasks.
\end{itemize}

\end{itemize}

\begin{algorithm}[H]
\scriptsize
\caption{CP for Scheduling Problem}
\label{alg:cp_scheduling}
\begin{algorithmic}[1]
\Require Set of tasks $T$, release times $r_i$, processing times $p_i$, dependencies $D$, preemption limit $\text{MaxPreemptions}$, objective weights $w_t, w_f, w_c, w_d$
\Ensure Optimized schedule with minimized completion time, fuel consumption, CO$_2$ emissions, and task delays
\State Initialize constraint model \texttt{CPModel}
\For{each task $i \in T$}
    \State Define start time variables $s_{ik}$ and completion time variables $C_{ik}$ for each segment $k$ of task $i$
    \State Define preemption segments $K_i \leq \text{MaxPreemptions}$, with $\sum_{k=1}^{K_i} p_{ik} = p_i$
    \State Set release time constraint: $s_{i1} \geq r_i$
    \If{task $i$ has dependencies}
        \For{each $(i, j) \in D$}
            \State Add precedence constraint: $s_{j1} \geq C_{iK_i}$
        \EndFor
    \EndIf
\EndFor
\For{each task pair $(i, j) \in T$ where $i \neq j$}
    \State Add sequencing constraints to prevent overlapping: 
    \State \quad $C_{ik} \leq s_{jl} + M(1 - x_{ijkl})$
    \State \quad $C_{jl} \leq s_{ik} + M x_{ijkl}$
\EndFor
\For{each task $i \in T$}
    \State Set within working hours constraints: $S \leq s_{ik} \leq C_{ik} \leq E$
\EndFor
\State Define objective function to minimize:
\State \quad $Z = w_t \times (C_{\text{max}} - S) + w_f \times F_{\text{total}} + w_c \times C_{\text{total}} + w_d \times D_{\text{total}}$
\State Solve \texttt{CPModel} using a CP solver (e.g., Google OR-Tools)
\If{solution found}
    \State Extract optimized schedule with task start and completion times
\Else
    \State Report that no feasible solution was found
\EndIf
\end{algorithmic}
\end{algorithm}

The above model addresses uncertainties in processing and travel times by using deterministic estimates, such as expected or conservative values. By adopting CP, we optimize the schedule while satisfying all constraints and providing robust solutions that effectively handle task dependencies, variable release times, preemptions, and multiple objectives (see Algorithm \ref{alg:cp_scheduling} for the pseudocode of our CP-based approach).

Note: The deterministic approach simplifies the model for CP solvers while still practically accounting for variability. This allows for efficient computation and implementation using tools like Google OR-Tools, which are well-suited for handling complex scheduling problems with the defined constraints.

\begin{figure}[h!] 
    \centering
    \begin{minipage}{0.50\textwidth}
        \centering
        \includegraphics[width=\textwidth]{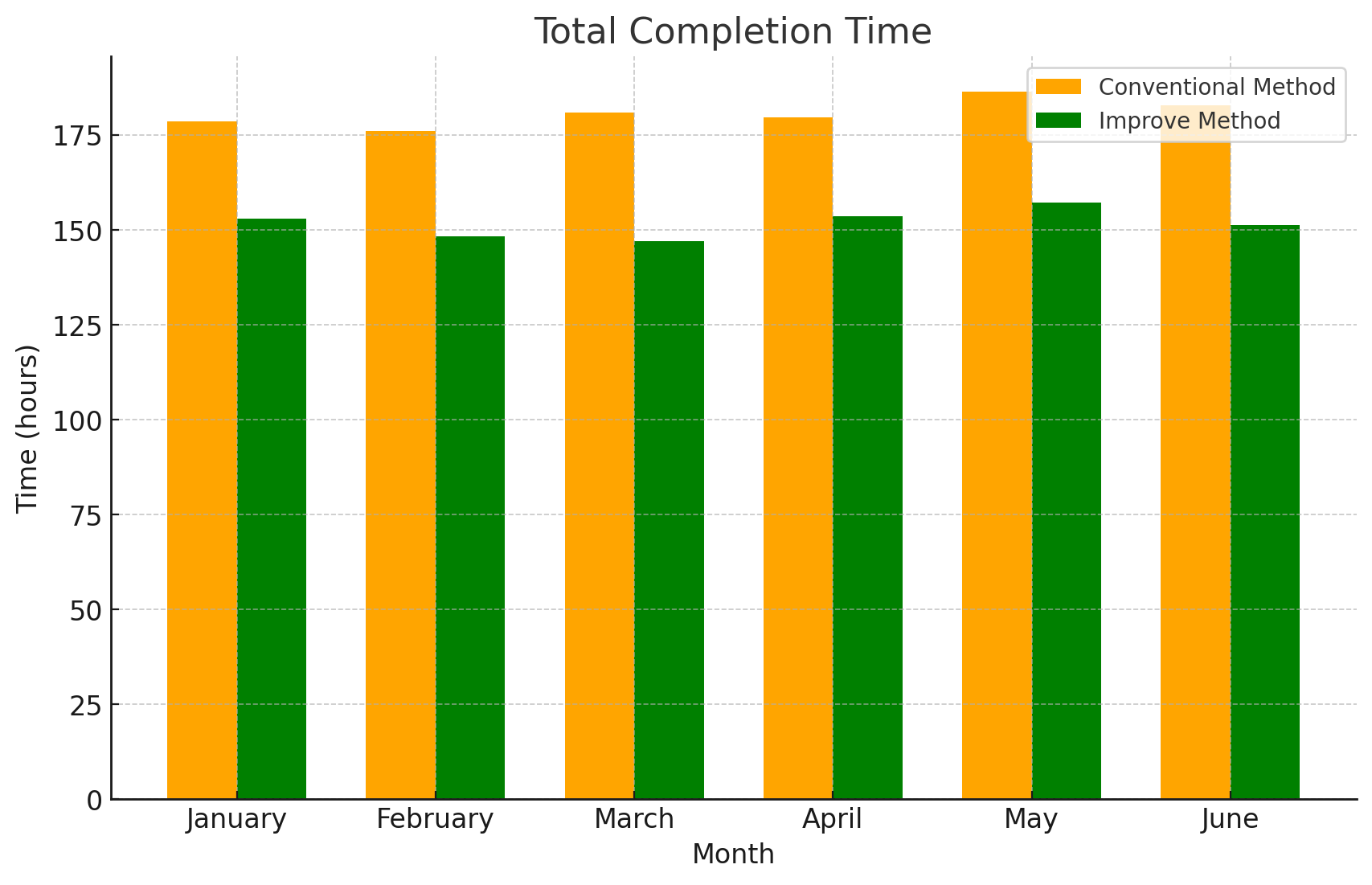}
        \raisebox{1ex}{(A)}
    \end{minipage}\hfill
    \begin{minipage}{0.50\textwidth}
        \centering
        \includegraphics[width=\textwidth]{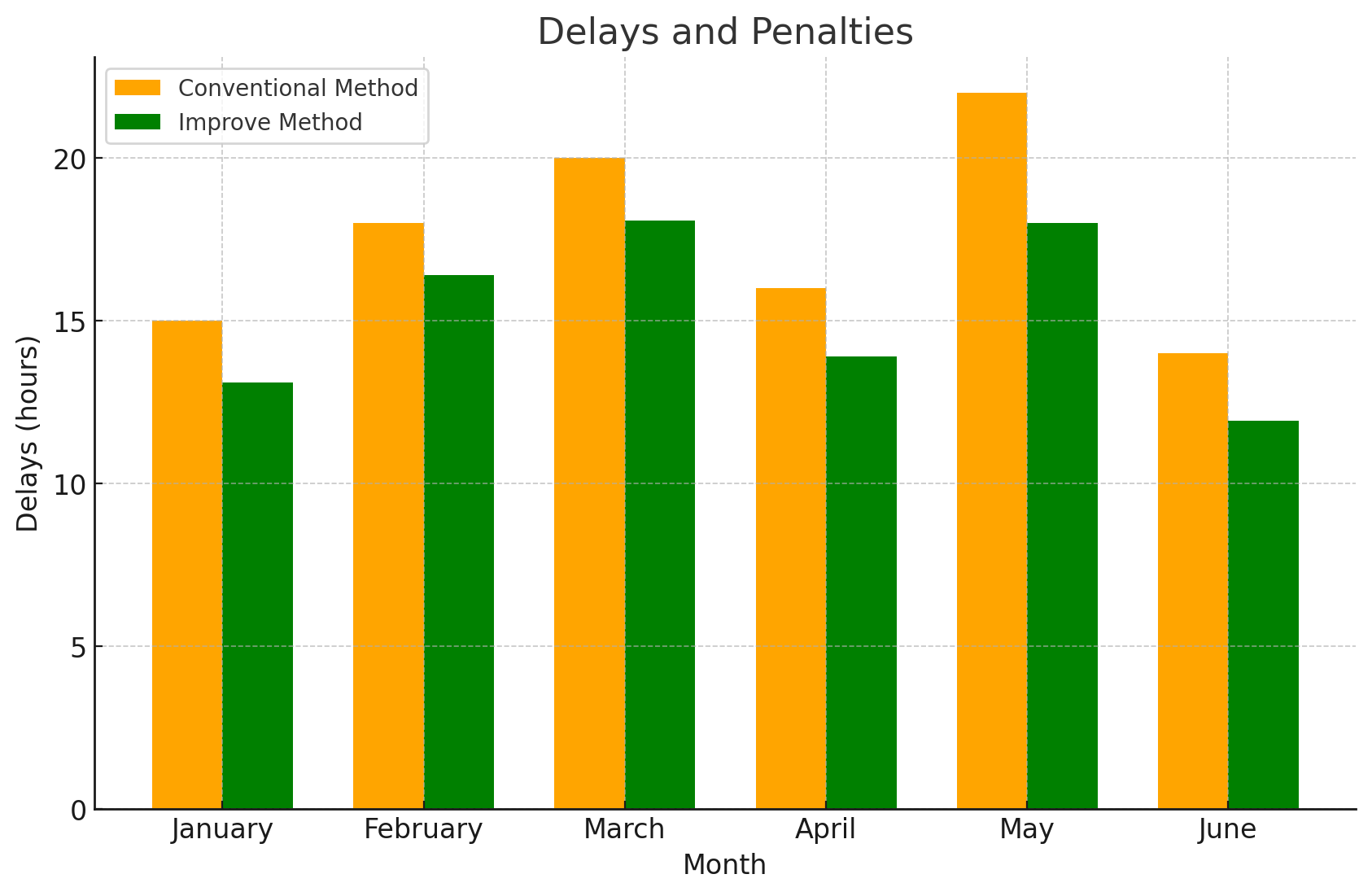}
        \raisebox{1ex}{(B)}
    \end{minipage}
    
    \vspace{0.5cm} 

    \begin{minipage}{0.50\textwidth}
        \centering
        \includegraphics[width=\textwidth]{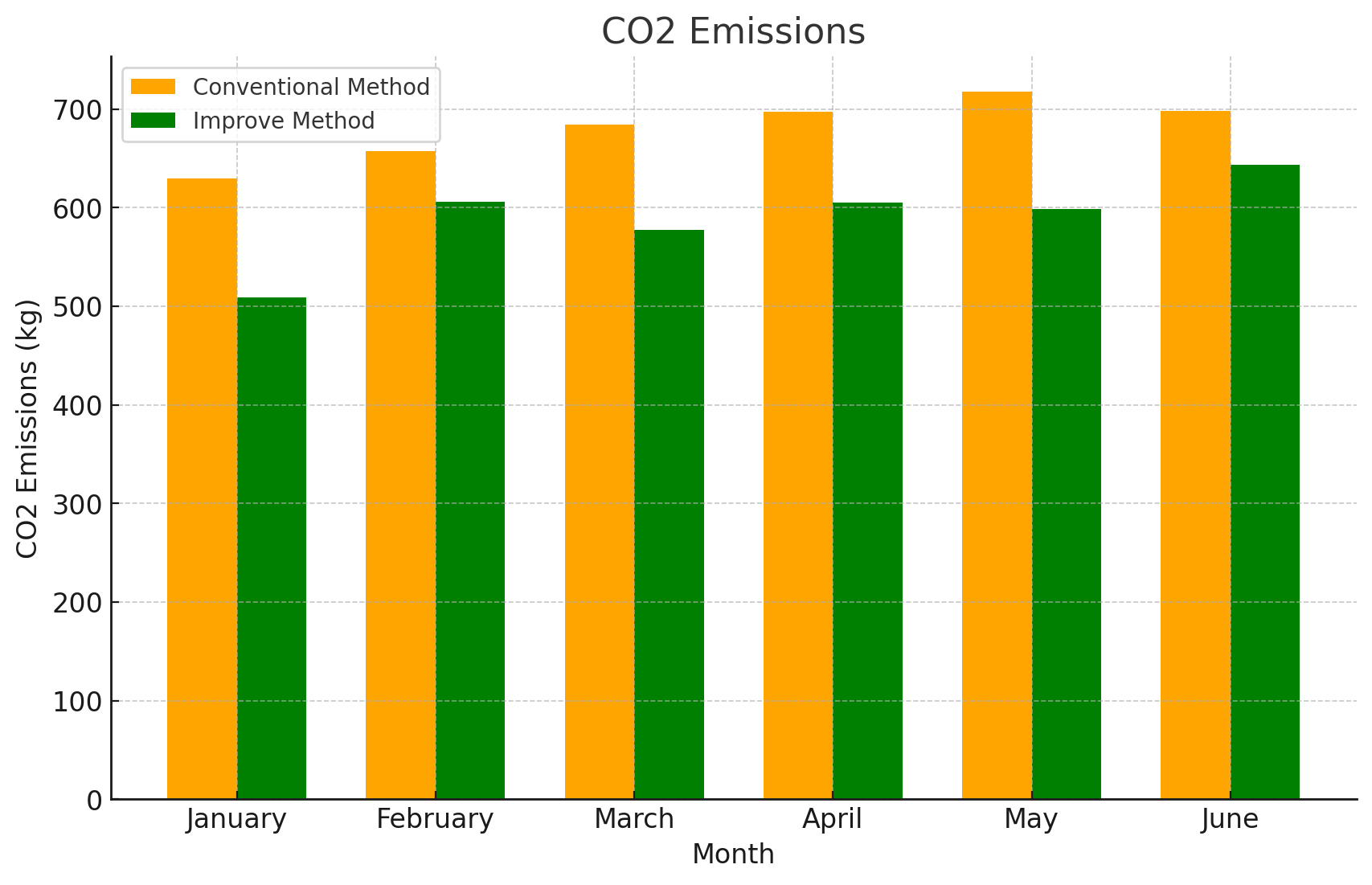}
        \raisebox{1ex}{(C)}
    \end{minipage}\hfill
    \begin{minipage}{0.50\textwidth}
        \centering
        \includegraphics[width=\textwidth]{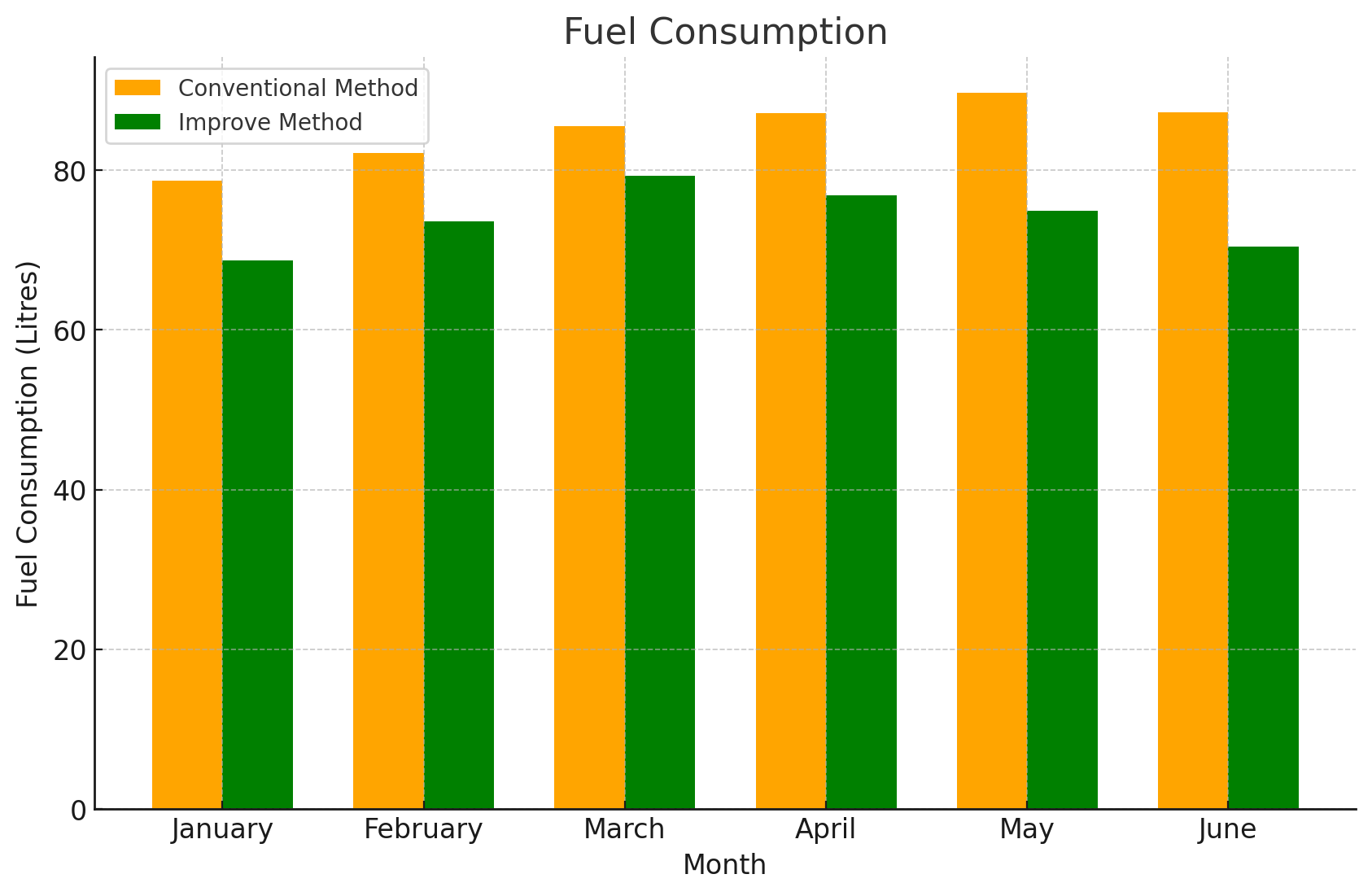}
        \raisebox{1ex}{(D)}
    \end{minipage}
    
    \vspace{0.5cm} 

    \begin{minipage}{0.50\textwidth}
        \centering
        \includegraphics[width=\textwidth]{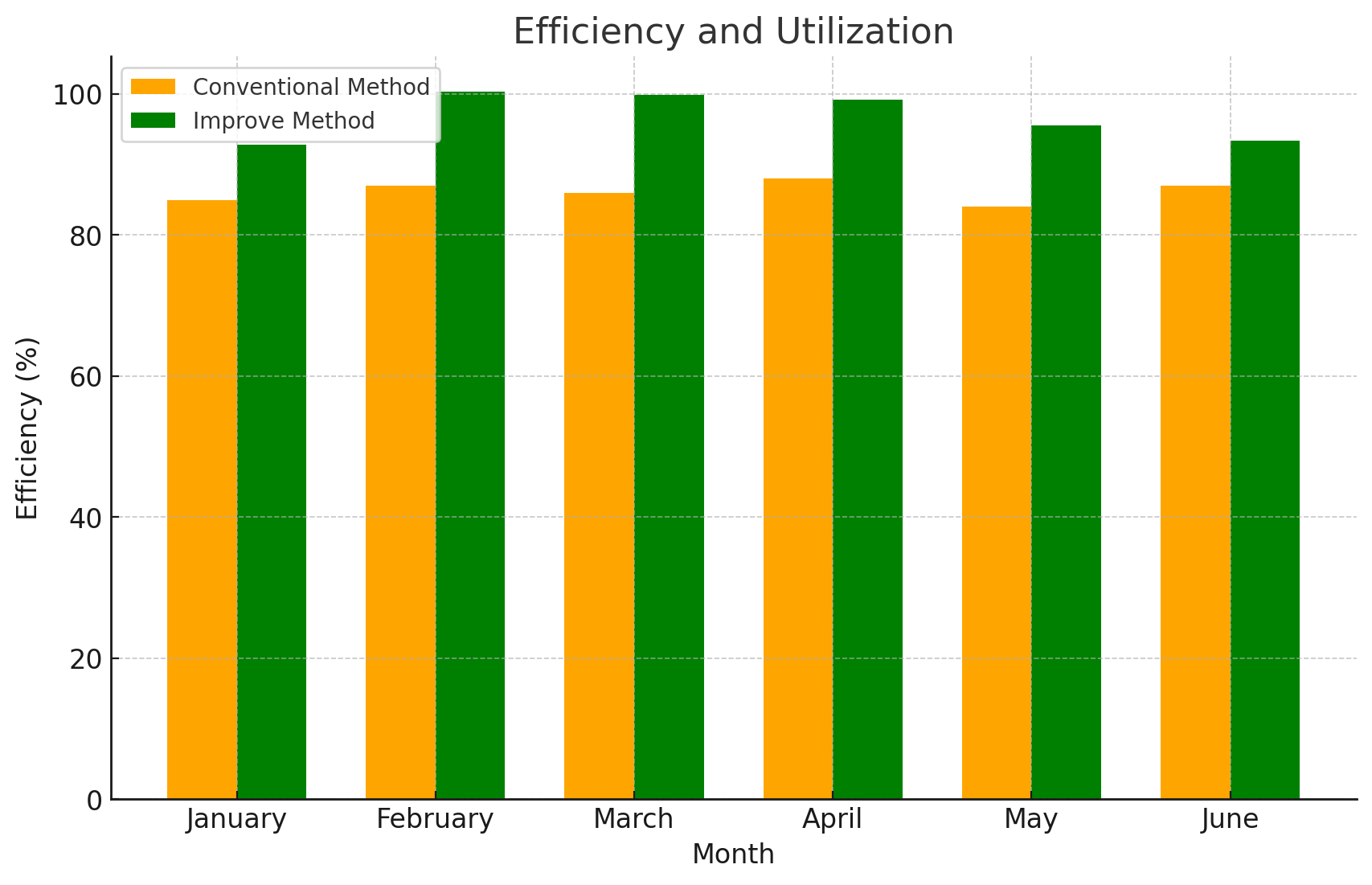}
        \raisebox{1ex}{(E)}
    \end{minipage}

    \caption{Overview of key operational metrics in system performance analysis. Figure A shows the \textit{Completion Time}, reflecting the efficiency of task completion across varying conditions. Figure B illustrates \textit{Delays and Penalties}, representing the penalties incurred due to task delays and their impact on overall costs. Figure C presents data on \textit{CO$_2$ Emissions}, highlighting environmental impacts related to system operations. Figure D provides insights into \textit{Fuel Consumption}, tracking resource usage efficiency. Lastly, Figure E displays \textit{Efficiency and Utilization}, summarizing the effectiveness of resource utilization in the system.}

    \label{fig:optimization}
\end{figure}

The comparison and improvement of the proposed scheduling model against conventional operator methods are illustrated in Table \ref{tab:performance_comparison} and Figure \ref{fig:optimization}. These results represent the average performance obtained from 20 independent executions of the algorithm to ensure statistical robustness and reliability.

\begin{table}[h]
 \caption{Comparison of Proposed Scheduling Model Against Conventional Operator Methods}
 \scriptsize
 \centering
  \begin{tabular}[htbp]{@{} c c c c @{}}
    \hline
    Metric & Conventional Method & Proposed Model & Improvement (\%) \\ 
    \hline
    Total Completion Time ($\mathbb{E}[C_{\text{max}}]$) & 180.58 hours & 155.24 hours & 14\% \\ 
    \hline
    Delays and Penalties ($\mathbb{E}[D_{\text{total}}]$) & 17.5 hours & 13.15 hours & 25\% \\ 
    \hline
    CO$_2$ Emissions ($\mathbb{E}[C_{\text{total}}]$) & 660.8 kg & 545.7 kg & 17\% \\ 
    \hline
    Fuel Consumption ($\mathbb{E}[F_{\text{total}}]$) & 85.58 Litres & 71.98 Litres & 16\% \\ 
    \hline
    Efficiency and Utilization ($\mathbb{E}[E_{\text{eff}}]$) & 86.17\% & 92.23\% & 7\% \\ 
    \hline
  \end{tabular}
\label{tab:performance_comparison}
\end{table}

\clearpage

\section{Security Layer in DTs Platform}\label{sec:Security measures}

In the digital transformation era, DTs have become a cornerstone for enhancing operational efficiency and decision-making across industries \cite{Homaei2024, Guikema_2024}. These virtual replicas, integral to systems from manufacturing to smart cities, leverage real-time data to mirror and predict the physical world's behavior. However, the complexity of DTs introduces multiple layers of cybersecurity risks that must be meticulously managed. Cybersecurity for DTs is not just an add-on but a foundational component that ensures the safe functioning of these systems. Each aspect, from hardware and software to models and algorithms, requires comprehensive protection to defend against cyber threats. Without robust cybersecurity measures, DTs could become liabilities, offering cyber criminals potential backdoors to critical infrastructure \cite{Homaei2022, Wei2022b}.

To mitigate these vulnerabilities, implementing predefined security layers is essential. As detailed in Table \ref{tab:Implemented Cybersecurity Strategies for CAUCCES Project}, CAUCCES integrates a range of tailored cybersecurity strategies specifically for the context of smart water management. These strategies include advanced encryption methods for securing data at rest and in transit, rigorous access controls to limit interactions with DT systems to authorized personnel, and continuous monitoring to detect and respond to threats in real-time. The table also outlines the specific cybersecurity approaches applied in the CAUCCES project to protect DTs used in managing WDSs. This includes securing communication channels that transmit sensitive data, employing anomaly detection techniques to quickly identify potential threats, and implementing robust data flow integrity measures.

The integration of these cybersecurity measures is critical to maintaining the integrity and reliability of DTs. As these systems increasingly support essential infrastructure—from healthcare to public utilities—the stakes for cybersecurity can hardly be overstated. Ensuring the resilience of DTs against cyber threats is not merely about protecting information but about safeguarding public welfare and the environments these systems serve.

By proactively addressing cybersecurity and embedding it within DT technologies, organizations can prevent potential vulnerabilities, ensuring that DTs enhance, rather than compromise, the digital future. The CAUCCES project underscores the criticality of incorporating robust cybersecurity strategies within the DT framework for water management. By addressing key areas, including smart meter security, communication network protection, employee training, database security, anomaly detection, data privacy, real-time monitoring, and collaborative approaches, the project sets a benchmark in fortifying water infrastructure resilience against evolving cyber threats. This comprehensive approach strengthens infrastructure while fostering a culture of security awareness and preparedness, serving as a model for future DT and cybersecurity integration initiatives.

\begin{table}[h!]
\caption{Implemented Cybersecurity Strategies for CAUCCES Project based on ISO 27001 Compliance}
\centering
\scriptsize
\begin{tabular}{| p{2.6cm} | p{9cm} | p{2.6cm} |}
\hline
\textbf{Component} & \textbf{Cybersecurity Measure} & \textbf{Security Standard/Protocol} \\
\hline
\multicolumn{3}{|c|}{\textbf{Cybersecurity for Smart Water Meters}} \\
\hline
Private Key Encryption & AES-128 encryption for data transmission, with private keys ensuring authorized decryption. & AES-128, LoRaWAN, NB-IoT \\
\hline
LoRaWAN and NB-IoT Compatibility & Supports secure, long-range data transmission suitable for rural areas. & LoRaWAN, NB-IoT \\
\hline
Data Logging and IP-68 Protection & IP-68 protection and internal datalogger for secure, resilient data storage. & IP-68, Data Integrity \\
\hline
\multicolumn{3}{|c|}{\textbf{Secure Data Transmission via Gateway and ChirpStack Platform}} \\
\hline
LoRaWAN Gateway & IP-67 gateway securely transmits encrypted data via MQTT to the cloud. & IP-67, MQTT, Data Integrity \\
\hline
ChirpStack Platform Security & ChirpStack manages devices with unique keys, MAC-based authentication, and data validation. & Device Authentication, MAC Address Validation \\
\hline
SSL/TLS Protocols & SSL/TLS secures all data transmission between services, ensuring encryption in transit. & SSL/TLS \\
\hline
\multicolumn{3}{|c|}{\textbf{Data Flow, Decryption, and Secure Transmission}} \\
\hline
Encrypted Data Transmission & Transmits data in AES-128 encrypted format via MQTT to cloud storage. & AES-128, MQTT, Data Encryption \\
\hline
Data Decryption & Decrypted and stored in main and backup databases for integrity and redundancy. & Data Integrity, Redundancy \\
\hline
\multicolumn{3}{|c|}{\textbf{Secure Access, API Integration, and Web Application Security}} \\
\hline
API Data Loading & Secure APIs (FastAPI, Django) with HTTPS prevent data interception. & HTTPS, API Security \\
\hline
Secure Login Platform & Two-step authentication secures access to platform data. & Two-Step Authentication \\
\hline
Password Encryption with Salting & Salting and hashing for secure password and sensitive data storage. & Salting, Hashing, Data Encryption \\
\hline
Database Security (PostgreSQL) & Access controls, encryption at rest, and regular audits protect database data. & Data Encryption, Access Control \\
\hline
\multicolumn{3}{|c|}{\textbf{Backup, Data Integrity, and ISO 27001 Compliance}} \\
\hline
Backup Protocols & Regular backups ensure data availability and integrity. & Data Backup, ISO 27001 \\
\hline
ISO 27001 Compliance & Aligns cybersecurity measures with ISO 27001 standards; regular audits verify compliance. & ISO 27001 \\
\hline
Data Privacy Compliance & GDPR-compliant protocols and audits for data privacy. & GDPR, Data Privacy \\
\hline
\multicolumn{3}{|c|}{\textbf{Real-Time Monitoring and Threat Detection}} \\
\hline
Zabbix Monitoring & Monitors network and infrastructure performance, detecting anomalies. & Real-Time Monitoring \\
\hline
Wazuh for Threat Detection & Detects intrusions and abnormal activity through log analysis. & Threat Detection, Log Analysis \\
\hline
\end{tabular}
\label{tab:Implemented Cybersecurity Strategies for CAUCCES Project}
\end{table}

\clearpage
\section{Conclusion}\label{sec:Conclusion}
This paper has explored the transformative potential of DTs in the water distribution sector, underscoring their pivotal role in enhancing system efficiency, reliability, and sustainability. By integrating IoT devices, advanced AI, and ML algorithms, DTs offer a dynamic and robust platform for simulating real-world water systems, enabling predictive maintenance, real-time monitoring, and strategic decision-making.

We presented a novel DT platform within the CAUCCES project, which integrates sophisticated AI/ML models—including LSTM networks, Prophet, LightGBM, and XGBoost—for accurate forecasting of water consumption based on historical and meteorological data. The model evaluations demonstrated that incorporating advanced feature engineering and hyperparameter tuning significantly improves forecasting accuracy, which is essential for effective water resource management. For instance, the Prophet model with advanced feature engineering achieved a MAE of 5.76 and a MAPE of 18.61\% over a 6-month forecasting horizon, outperforming basic models.

Additionally, we addressed the optimization of WDS maintenance by formulating a scheduling problem using CP. The proposed model effectively minimizes total completion time, fuel consumption, CO\textsubscript{2} emissions, and delays, enhancing operational efficiency and reducing environmental impact. The results indicated a 14\% reduction in total completion time and a 17\% decrease in CO\textsubscript{2} emissions compared to conventional methods, highlighting the efficacy of the CP-based approach in handling complex scheduling problems in WDS maintenance.

Moreover, the platform's emphasis on cybersecurity ensures the integrity and confidentiality of data, a critical aspect given the increasing threats in the digital landscape. Implementing robust cybersecurity measures aligned with standards like ISO 27001 protects the system against potential cyber-attacks, ensuring continuous and reliable service delivery.

Future research lines could prioritize the integration of more advanced artificial intelligence and machine learning methodologies, including deep learning frameworks and ensemble techniques, with the focus of improving the accuracy of the prediction of water consumption and anomaly detection. Combination of real-time data analytics with optimization techniques may, therefore, yield a possible way to shift toward more dynamic and adaptive management of water distribution systems, one that reacts rapidly to change and unexpected events. Likewise, the expansion of this platform to greater and complex water networks and integration with other utility systems can provide urban resource management in an altogether more holistic approach. In addition, a better user interface and extensive training programs for utility operators and stakeholders could help achieve greater adoption of the platform. There is a need for continuing compliance with evolving cybersecurity measures and regulatory requirements in order to maintain data security and system integrity. Lastly, future releases of the platform could contain elements focused on assessing and reducing environmental impacts, such as water loss and energy consumption, and thereby contribute significantly to sustainable development goals.

\section*{Declarations}

\begin{itemize}
\item \textbf{Funding:} This project is carried out within the framework of the funds of the Recovery, Transformation and Resilience Plan, financed by the European Union (Next Generation). The publication is part of the Spanish Strategic Cybersecurity Project “Artificial Intelligence applied to Cybersecurity in Critical Water and Sanitation Infrastructures (***/**)” funded by Instituto Nacional de Ciberseguridad de España (INCIBE).
\item \textbf{Conflict of interest:} The authors declare no conflicts of interest relevant to the content of this article.
\item \textbf{Ethics approval and consent to participate:} Not applicable.
\item \textbf{Consent for publication:} Not applicable.
\item \textbf{Data and code availability:} All code and datasets used in this study are publicly available at \url{https://github.com/Homaei/DigitalTwin-Water-ML}.
\item \textbf{Materials availability:} Not applicable.
\item \textbf{Author contributions:} All authors contributed equally to the conception, design, drafting, and revision of the manuscript.
\end{itemize}

 \bibliographystyle{elsarticle-num} 

\begin{thebibliography}{00}

\bibitem{Hu2021}
    Zukang Hu, Beiqing Chen, Wenlong Chen, Debao Tan, Dingtao Shen.
    \textit{Review of model-based and data-driven approaches for leak detection and location in water distribution systems}.
    Water Supply, IWA Publishing, Volume: 21, Number: 7, Pages: 3282--3306, April 2021.
    DOI: \href{https://doi.org/10.2166/ws.2021.101}{10.2166/ws.2021.101}.

\bibitem{DAYIOLU2021}
    Mehmet Ali DAYIOĞLU, Ufuk TURKER.
    \textit{Digital Transformation for Sustainable Future - Agriculture 4.0: A review}.
    Tarım Bilimleri Dergisi, Ankara University Faculty of Agriculture, November 2021.
    DOI: \href{https://doi.org/10.15832/ankutbd.986431}{10.15832/ankutbd.986431}.

\bibitem{Bauer2021}
    Peter Bauer, Bjorn Stevens, Wilco Hazeleger.
    \textit{A digital twin of Earth for the green transition}.
    Nature Climate Change, Springer Science and Business Media LLC, Volume: 11, Number: 2, Pages: 80--83, February 2021.
    DOI: \href{https://doi.org/10.1038/s41558-021-00986-y}{10.1038/s41558-021-00986-y}.

\bibitem{Beji2022}
    Haifa Beji, Markus Lade.
    \textit{Impact of Digital Transformation on Carbon Emissions Reductions in the Water Industry}.
    Lecture Notes in Energy, Springer International Publishing, Pages: 117--127, 2022.
    DOI: \href{https://doi.org/10.1007/978-3-030-86215-2\_13}{10.1007/978-3-030-86215-2\_13}.

\bibitem{Khanna2020}
    Madhu Khanna.
    \textit{Digital Transformation of the Agricultural Sector: Pathways, Drivers and Policy Implications}.
    Applied Economic Perspectives and Policy, Wiley, Volume: 43, Number: 4, Pages: 1221--1242, October 2020.
    DOI: \href{https://doi.org/10.1002/aepp.13103}{10.1002/aepp.13103}.

\bibitem{Agostinelli2021}
    Sofia Agostinelli, Fabrizio Cumo, Giambattista Guidi, Claudio Tomazzoli.
    \textit{Cyber-Physical Systems Improving Building Energy Management: Digital Twin and Artificial Intelligence}.
    Energies, MDPI AG, Volume: 14, Number: 8, Pages: 2338, April 2021.
    DOI: \href{https://doi.org/10.3390/en14082338}{10.3390/en14082338}.

\bibitem{Ciliberti2021}
    Francesco Gino Ciliberti, Luigi Berardi, Daniele Biagio Laucelli, Orazio Giustolisi.
    \textit{Digital Transformation Paradigm for Asset Management in Water Distribution Networks}.
    2021 10th International Conference on ENERGY and ENVIRONMENT (CIEM), IEEE, Pages: 760--765, October 2021.
    DOI: \href{https://doi.org/10.1109/ciem52821.2021.9614864}{10.1109/ciem52821.2021.9614864}.

\bibitem{Jiang2021}
    Yuchen Jiang, Shen Yin, Kuan Li, Hao Luo, Okyay Kaynak.
    \textit{Industrial applications of digital twins}.
    Philosophical Transactions of the Royal Society A: Mathematical, Physical and Engineering Sciences, The Royal Society, Volume: 379, Number: 2207, Pages: 20200360, August 2021.
    DOI: \href{https://doi.org/10.1098/rsta.2020.0360}{10.1098/rsta.2020.0360}.

\bibitem{Zekri2022}
    Slim Zekri, Nafaa Jabeur, Hana Gharrad.
    \textit{Smart Water Management Using Intelligent Digital Twins}.
    Computing and Informatics, Central Library of the Slovak Academy of Sciences, Volume: 41, Number: 1, Pages: 135--153, 2022.
    DOI: \href{https://doi.org/10.31577/cai\_2022\_1\_135}{10.31577/cai\_2022\_1\_135}.

\bibitem{Pylianidis2021}
    Christos Pylianidis, Sjoukje Osinga, Ioannis N. Athanasiadis.
    \textit{Introducing digital twins to agriculture}.
    Computers and Electronics in Agriculture, Elsevier BV, Volume: 184, Pages: 105942, May 2021.
    DOI: \href{https://doi.org/10.1016/j.compag.2020.105942}{10.1016/j.compag.2020.105942}.

\bibitem{Boyle2022}
    Carol Boyle, Greg Ryan, Pratik Bhandari, Kris M. Y. Law, Jinzhe Gong, Douglas Creighton.
    \textit{Digital Transformation in Water Organizations}.
    Journal of Water Resources Planning and Management, American Society of Civil Engineers (ASCE), Volume: 148, Number: 7, July 2022.
    DOI: \href{https://doi.org/10.1061/(asce)wr.1943-5452.0001555}{10.1061/(asce)wr.1943-5452.0001555}.

\bibitem{Callaway2024}
    Stephen Callaway, Narges Mashhadi Nejad.
    \textit{Innovating Toward CSR: Creating Value by Empowering Employees, Customers, and Stockholders}.
    SSRN Electronic Journal, Elsevier BV, 2024.
    DOI: \href{https://doi.org/10.2139/ssrn.5015070}{10.2139/ssrn.5015070}.

\bibitem{Haaker2021}
    Timber Haaker, Pham Thi Minh Ly, Nhan Nguyen-Thanh, Hanh Thi Hong Nguyen.
    \textit{Business model innovation through the application of the Internet-of-Things: A comparative analysis}.
    Journal of Business Research, Elsevier BV, Volume: 126, Pages: 126--136, March 2021.
    DOI: \href{https://doi.org/10.1016/j.jbusres.2020.12.034}{10.1016/j.jbusres.2020.12.034}.

\bibitem{Feroz2021}
    Abdul Karim Feroz, Hangjung Zo, Ananth Chiravuri.
    \textit{Digital Transformation and Environmental Sustainability: A Review and Research Agenda}.
    Sustainability, MDPI AG, Volume: 13, Number: 3, Pages: 1530, February 2021.
    DOI: \href{https://doi.org/10.3390/su13031530}{10.3390/su13031530}.

\bibitem{Pivoto2021}
    Diego G.S. Pivoto, Luiz F.F. de Almeida, Rodrigo da Rosa Righi, Joel J.P.C. Rodrigues, Alexandre Baratella Lugli, Antonio M. Alberti.
    \textit{Cyber-physical systems architectures for industrial internet of things applications in Industry 4.0: A literature review}.
    Journal of Manufacturing Systems, Elsevier BV, Volume: 58, Pages: 176--192, January 2021.
    DOI: \href{https://doi.org/10.1016/j.jmsy.2020.11.017}{10.1016/j.jmsy.2020.11.017}.

\bibitem{Wjcicki2022}
    Krzysztof Wójcicki, Marta Biegańska, Beata Paliwoda, Justyna Górna.
    \textit{Internet of Things in Industry: Research Profiling, Application, Challenges and Opportunities—A Review}.
    Energies, MDPI AG, Volume: 15, Number: 5, Pages: 1806, February 2022.
    DOI: \href{https://doi.org/10.3390/en15051806}{10.3390/en15051806}.

\bibitem{Jagani2023}
    Sandeep Jagani, Xiyue Deng, Paul C. Hong, Narges Mashhadi Nejad.
    \textit{Adopting sustainability business models for value creation and delivery: an empirical investigation of manufacturing firms}.
    Journal of Manufacturing Technology Management, Emerald, Volume: 35, Number: 2, Pages: 360–382, December 2023.
    DOI: \href{https://doi.org/10.1108/jmtm-03-2023-0099}{10.1108/jmtm-03-2023-0099}.

\bibitem{BotnSanabria2022}
    Diego M. Botín-Sanabria, Adriana-Simona Mihaita, Rodrigo E. Peimbert-García, Mauricio A. Ramírez-Moreno, Ricardo A. Ramírez-Mendoza, Jorge de J. Lozoya-Santos.
    \textit{Digital Twin Technology Challenges and Applications: A Comprehensive Review}.
    Remote Sensing, MDPI AG, Volume: 14, Number: 6, Pages: 1335, March 2022.
    DOI: \href{https://doi.org/10.3390/rs14061335}{10.3390/rs14061335}.

\bibitem{Futai2022}
    Marcos Massao Futai, Túlio N. Bittencourt, Hermes Carvalho, Duperron M. Ribeiro.
    \textit{Challenges in the application of digital transformation to inspection and maintenance of bridges}.
    Structure and Infrastructure Engineering, Informa UK Limited, Volume: 18, Number: 10-11, Pages: 1581--1600, April 2022.
    DOI: \href{https://doi.org/10.1080/15732479.2022.2063908}{10.1080/15732479.2022.2063908}.

\bibitem{khazrak2023}
    Iman Khazrak.
    \textit{A Study on Corporate Carbon Footprint Using Panel Data Analysis}.
    Master’s Thesis, Bowling Green State University, Pages: 59, 2023.
    Note: Committee: Yuhang Xu, Ph.D. (Committee Chair); Shuchismita Sarkar, Ph.D. (Committee Member); Sophie Song, Ph.D. (Committee Member).

\bibitem{Singh2021}
    Maulshree Singh, Evert Fuenmayor, Eoin Hinchy, Yuansong Qiao, Niall Murray, Declan Devine.
    \textit{Digital Twin: Origin to Future}.
    Applied System Innovation, MDPI AG, Volume: 4, Number: 2, Pages: 36, May 2021.
    DOI: \href{https://doi.org/10.3390/asi4020036}{10.3390/asi4020036}.

\bibitem{arena2022novel}
    Simone Arena, Eleonora Florian, Ilenia Zennaro, P.F. Orrù, Fabio Sgarbossa.
    \textit{A novel decision support system for managing predictive maintenance strategies based on machine learning approaches}.
    Safety Science, Elsevier, Volume: 146, Pages: 105529, 2022.
    DOI: \href{https://doi.org/10.1016/j.ssci.2021.105529}{10.1016/j.ssci.2021.105529}.

\bibitem{Nadkarni2020}
    Swen Nadkarni, Reinhard Prügl.
    \textit{Digital transformation: a review, synthesis and opportunities for future research}.
    Management Review Quarterly, Springer Science and Business Media LLC, Volume: 71, Number: 2, Pages: 233--341, April 2020.
    DOI: \href{https://doi.org/10.1007/s11301-020-00185-7}{10.1007/s11301-020-00185-7}.

\bibitem{Mendhurwar2019}
    Subodh Mendhurwar, Rajhans Mishra.
    \textit{Integration of social and IoT technologies: architectural framework for digital transformation and cyber security challenges}.
    Enterprise Information Systems, Informa UK Limited, Volume: 15, Number: 4, Pages: 565--584, April 2019.
    DOI: \href{https://doi.org/10.1080/17517575.2019.1600041}{10.1080/17517575.2019.1600041}.

\bibitem{Broo2021}
    Didem Gürdür Broo, Jennifer Schooling.
    \textit{Digital twins in infrastructure: definitions, current practices, challenges and strategies}.
    International Journal of Construction Management, Informa UK Limited, Volume: 23, Number: 7, Pages: 1254--1263, August 2021.
    DOI: \href{https://doi.org/10.1080/15623599.2021.1966980}{10.1080/15623599.2021.1966980}.

\bibitem{Shahi2020}
    Chinmay Shahi, Manish Sinha.
    \textit{Digital transformation: challenges faced by organizations and their potential solutions}.
    International Journal of Innovation Science, Emerald, Volume: 13, Number: 1, Pages: 17--33, December 2020.
    DOI: \href{https://doi.org/10.1108/ijis-09-2020-0157}{10.1108/ijis-09-2020-0157}.

\bibitem{Ko2021}
    Andrea Ko, Péter Fehér, Tibor Kovacs, Ariel Mitev, Zoltán Szabó.
    \textit{Influencing factors of digital transformation: management or IT is the driving force?}.
    International Journal of Innovation Science, Emerald, Volume: 14, Number: 1, Pages: 1--20, June 2021.
    DOI: \href{https://doi.org/10.1108/ijis-01-2021-0007}{10.1108/ijis-01-2021-0007}.

\bibitem{Li2020}
    Feng Li.
    \textit{Leading digital transformation: three emerging approaches for managing the transition}.
    International Journal of Operations and Production Management, Emerald, Volume: 40, Number: 6, Pages: 809--817, July 2020.
    DOI: \href{https://doi.org/10.1108/ijopm-04-2020-0202}{10.1108/ijopm-04-2020-0202}.

\bibitem{AguileraCastillo2021}
    Andres Aguilera Castillo.
    \textit{Digital Transformation and the Public Sector Workforce: An exploration and research agenda}.
    14th International Conference on Theory and Practice of Electronic Governance (ICEGOV 2021), ACM, Pages: 471--475, October 2021.
    DOI: \href{https://doi.org/10.1145/3494193.3494257}{10.1145/3494193.3494257}.

\bibitem{ConejosFuertes2020}
    P. Conejos Fuertes, F. Martínez Alzamora, M. Hervás Carot, J.C. Alonso Campos.
    \textit{Building and exploiting a Digital Twin for the management of drinking water distribution networks}.
    Urban Water Journal, Informa UK Limited, Volume: 17, Number: 8, Pages: 704--713, June 2020.
    DOI: \href{https://doi.org/10.1080/1573062x.2020.1771382}{10.1080/1573062x.2020.1771382}.


\bibitem{Haag2018}
    Sebastian Haag, Reiner Anderl.
    \textit{Digital twin — Proof of concept}.
    Manufacturing Letters, Elsevier BV, Volume: 15, Pages: 64--66, January 2018.
    DOI: \href{https://doi.org/10.1016/j.mfglet.2018.02.006}{10.1016/j.mfglet.2018.02.006}.

\bibitem{Ketzler2020}
    Bernd Ketzler, Vasilis Naserentin, Fabio Latino, Christopher Zangelidis, Liane Thuvander, Anders Logg.
    \textit{Digital Twins for Cities: A State of the Art Review}.
    Built Environment, Alexandrine Press, Volume: 46, Number: 4, Pages: 547--573, December 2020.
    DOI: \href{https://doi.org/10.2148/benv.46.4.547}{10.2148/benv.46.4.547}.

\bibitem{Linda2022}
    Linda Rice.
    \textit{Digital Twins of Smart Cities: Spatial Data Visualization Tools, Monitoring and Sensing Technologies, and Virtual Simulation Modeling}.
    Geopolitics, History, and International Relations, Addleton Academic Publishers, Volume: 14, Number: 1, Pages: 26, 2022.
    DOI: \href{https://doi.org/10.22381/ghir14120222}{10.22381/ghir14120222}.

\bibitem{YossefRavid2022}
    Batel Yossef Ravid, Meirav Aharon-Gutman.
    \textit{The Social Digital Twin: The Social Turn in the Field of Smart Cities}.
    Environment and Planning B: Urban Analytics and City Science, SAGE Publications, November 2022.
    DOI: \href{https://doi.org/10.1177/23998083221137079}{10.1177/23998083221137079}.

\bibitem{JalaliSepehr2024}
    Mehrdad Jalali Sepehr, Narges Mashhadi Nejad.
    \textit{Exploring Strategic National Research and Development Factors for Sustainable Adoption of Cellular Agriculture Technology}.
    Proceedings of the 2024 Midwest Decision Sciences Institute Conference, Decision Sciences Institute, 2024.
    DOI: \href{https://doi.org/10.2139/ssrn.5024700}{10.2139/ssrn.5024700}.

\bibitem{Bariah2022}
    Lina Bariah, Hikmet Sari, Merouane Debbah.
    \textit{Digital Twin-Empowered Smart Cities: A New Frontier of Wireless Networks}.
    TechRxiv, Institute of Electrical and Electronics Engineers (IEEE), July 2022.
    DOI: \href{https://doi.org/10.36227/techrxiv.20375325.v1}{10.36227/techrxiv.20375325.v1}.

    Manisha Vohra.
    \textit{Digital Twin in Smart Cities}.
    Wiley, Pages: 159--172, November 2022.
    DOI: \href{https://doi.org/10.1002/9781119842316.ch10}{10.1002/9781119842316.ch10}.

    Maulshree Singh, Rupal Srivastava, Evert Fuenmayor, Vladimir Kuts, Yuansong Qiao, Niall Murray, Declan Devine.
    \textit{Applications of Digital Twin across Industries: A Review}.
    Applied Sciences, MDPI AG, Volume: 12, Number: 11, Pages: 5727, June 2022.
    DOI: \href{https://doi.org/10.3390/app12115727}{10.3390/app12115727}.

    Rafael da Silva Mendonça, Sidney de Oliveira Lins, Iury Valente de Bessa, Florindo Antônio de Carvalho Ayres, Renan Landau Paiva de Medeiros, Vicente Ferreira de Lucena.
    \textit{Digital Twin Applications: A Survey of Recent Advances and Challenges}.
    Processes, MDPI AG, Volume: 10, Number: 4, Pages: 744, April 2022.
    DOI: \href{https://doi.org/10.3390/pr10040744}{10.3390/pr10040744}.

    Fei Tao, Bin Xiao, Qinglin Qi, Jiangfeng Cheng, Ping Ji.
    \textit{Digital twin modeling}.
    Journal of Manufacturing Systems, Elsevier BV, Volume: 64, Pages: 372--389, July 2022.
    DOI: \href{https://doi.org/10.1016/j.jmsy.2022.06.015}{10.1016/j.jmsy.2022.06.015}.

\bibitem{MashhadiNejad2024}
    Narges Mashhadi Nejad, Marcelo J. Alvarado-Vargas, Mehrdad Jalali Sepehr.
    \textit{Refining Literature Review Strategies: Analyzing Big Data Trends Across Journal Tiers}.
    Academy of Management Proceedings, Academy of Management, Volume: 2024, Number: 1, August 2024.
    DOI: \href{https://doi.org/10.5465/amproc.2024.14852abstract}{10.5465/amproc.2024.14852abstract}.

\bibitem{Savic2022}
    Dragan Savić.
    \textit{Digital water developments and lessons learned from automation in the car and aircraft industries}.
    Engineering, Elsevier, Volume: 9, Pages: 35--41, 2022.

\bibitem{Henriksen2022}
    Hans Jørgen Henriksen, Raphael Schneider, Julian Koch, Maria Ondracek, Lars Troldborg, Ida K. Seidenfaden, Søren Julsgaard Kragh, Eva Bøgh, Simon Stisen.
    \textit{A New Digital Twin for Climate Change Adaptation, Water Management, and Disaster Risk Reduction (HIP Digital Twin)}.
    Water, MDPI AG, Volume: 15, Number: 1, Pages: 25, December 2022.
    DOI: \href{https://doi.org/10.3390/w15010025}{10.3390/w15010025}.

\bibitem{Yousif2021}
    Ibrahim Yousif.
    \textit{Application of Digital Transformation in the Water Desalination Industry to Develop Smart Desalination Plants}.
    Master’s Thesis, College of Engineering and Computing, University of South Carolina, 2021.

\bibitem{Ramos2022}
    Helena M. Ramos, Maria Cristina Morani, Armando Carravetta, Oreste Fecarrotta, Kemi Adeyeye, P. Amparo López-Jiménez, Modesto Pérez-Sánchez.
    \textit{New Challenges towards Smart Systems' Efficiency by Digital Twin in Water Distribution Networks}.
    Water, MDPI AG, Volume: 14, Number: 8, Pages: 1304, April 2022.
    DOI: \href{https://doi.org/10.3390/w14081304}{10.3390/w14081304}.

\bibitem{Bonilla2022}
    Carlos A. Bonilla, Ariele Zanfei, Bruno Brentan, Idel Montalvo, Joaquín Izquierdo.
    \textit{A Digital Twin of a Water Distribution System by Using Graph Convolutional Networks for Pump Speed-Based State Estimation}.
    Water, MDPI AG, Volume: 14, Number: 4, Pages: 514, February 2022.
    DOI: \href{https://doi.org/10.3390/w14040514}{10.3390/w14040514}.

\bibitem{Wei2022a}
    Yuying Wei, Adrian Wing-Keung Law, Chun Yang, Di Tang.
    \textit{Combined Anomaly Detection Framework for Digital Twins of Water Treatment Facilities}.
    Water, MDPI AG, Volume: 14, Number: 7, Pages: 1001, March 2022.
    DOI: \href{https://doi.org/10.3390/w14071001}{10.3390/w14071001}.

\bibitem{CardilloAlbarrn2021}
    Juan Cardillo Albarrán, Edgar Chacón Ramírez, Luis Alberto Cruz Salazar, Yenny Alexandra Paredes Astudillo.
    \textit{Digital Twin in Water Supply Systems to Industry 4.0: The Holonic Production Unit}.
    Service Oriented, Holonic and Multi-Agent Manufacturing Systems for Industry of the Future, Springer International Publishing, Pages: 42--54, 2021.
    DOI: \href{https://doi.org/10.1007/978-3-030-80906-5\_4}{10.1007/978-3-030-80906-5\_4}.

\bibitem{GinoCiliberti2023}
    Francesco Gino Ciliberti, Luigi Berardi, Daniele Biagio Laucelli, Andres David Ariza, Laura Vanessa Enriquez, Orazio Giustolisi.
    \textit{From digital twin paradigm to digital water services}.
    Journal of Hydroinformatics, IWA Publishing, 2023.
    DOI: \href{https://doi.org/10.2166/hydro.2023.237}{10.2166/hydro.2023.237}.

\bibitem{valverde2021digital}
    Borja Valverde-Pérez, Bruce Johnson, Christoffer Wårff, Douglas Lumley, Elena Torfs, Ingmar Nopens, Lloyd Townley.
    \textit{Digital Water-Operational digital twins in the urban water sector: case studies}.
    International Water Association, London, UK, White paper, 2021.

\bibitem{Giudicianni2020}
    Carlo Giudicianni, Manuel Herrera, Armando di Nardo, Kemi Adeyeye, Helena M. Ramos.
    \textit{Overview of Energy Management and Leakage Control Systems for Smart Water Grids and Digital Water}.
    Modelling, MDPI AG, Volume: 1, Number: 2, Pages: 134--155, October 2020.
    DOI: \href{https://doi.org/10.3390/modelling1020009}{10.3390/modelling1020009}.

\bibitem{Sun2020}
    Congcong Sun, Vicenç Puig, Gabriela Cembrano.
    \textit{Real-Time Control of Urban Water Cycle under Cyber-Physical Systems Framework}.
    Water, MDPI AG, Volume: 12, Number: 2, Pages: 406, February 2020.
    DOI: \href{https://doi.org/10.3390/w12020406}{10.3390/w12020406}.

\bibitem{GarridoBaserba2020}
    Manel Garrido-Baserba, Lluís Corominas, Ulises Cortés, Diego Rosso, Manel Poch.
    \textit{The Fourth-Revolution in the Water Sector Encounters the Digital Revolution}.
    Environmental Science and Technology, American Chemical Society (ACS), Volume: 54, Number: 8, Pages: 4698--4705, March 2020.
    DOI: \href{https://doi.org/10.1021/acs.est.9b04251}{10.1021/acs.est.9b04251}.

\bibitem{Botin2021digital}
    Diego M. Botín-Sanabria, Jorge G. Lozoya-Reyes, Roberto C. Vargas-Maldonado, Karen L. Rodríguez-Hernández, Ricardo A. Ramírez-Mendoza, Mauricio A. Ramírez-Moreno, Jorge de J. Lozoya-Santos.
    \textit{Digital Twin for Urban Spaces: An Application}.
    Proceedings of the International Conference on Industrial Engineering and Operations Management, Monterrey, Mexico, Pages: 2880--2891, November 2021.

\bibitem{Udugama2021}
    Isuru A. Udugama, Pau C. Lopez, Carina L. Gargalo, Xueliang Li, Christoph Bayer, Krist V. Gernaey.
    \textit{Digital Twin in Biomanufacturing: Challenges and Opportunities Towards its Implementation}.
    Systems Microbiology and Biomanufacturing, Springer Science and Business Media LLC, Volume: 1, Number: 3, Pages: 257--274, March 2021.
    DOI: \href{https://doi.org/10.1007/s43393-021-00024-0}{10.1007/s43393-021-00024-0}.

\bibitem{Pedersen2021}
    Agnethe N. Pedersen, Morten Borup, Annette Brink-Kjær, Lasse E. Christiansen, Peter S. Mikkelsen.
    \textit{Living and Prototyping Digital Twins for Urban Water Systems: Towards Multi-Purpose Value Creation Using Models and Sensors}.
    Water, MDPI AG, Volume: 13, Number: 5, Pages: 592, February 2021.
    DOI: \href{https://doi.org/10.3390/w13050592}{10.3390/w13050592}.

\bibitem{Hietala2021}
    Heidi Hietala, Pekka M. Rossi, Elina Annanperä, Tero Päivärinta.
    \textit{Modes of Collaboration in Digital Transformation of Municipal Wastewater Management}.
    29th European Conference on Information Systems (ECIS 2021), Marrakech, Morocco (Virtual), Pages: 1470--1486, June 2021.

\bibitem{VanRooij2021}
    Frits van Rooij, Philip Scarf, Phuc Do.
    \textit{Planning the Restoration of Membranes in RO Desalination Using a Digital Twin}.
    Desalination, Elsevier BV, Volume: 519, Pages: 115214, December 2021.
    DOI: \href{https://doi.org/10.1016/j.desal.2021.115214}{10.1016/j.desal.2021.115214}.

\bibitem{Pesantez2022}
    Jorge E. Pesantez, Faisal Alghamdi, Shreya Sabu, G. Mahinthakumar, Emily Zechman Berglund.
    \textit{Using a Digital Twin to Explore Water Infrastructure Impacts During the COVID-19 Pandemic}.
    Sustainable Cities and Society, Elsevier BV, Volume: 77, Pages: 103520, February 2022.
    DOI: \href{https://doi.org/10.1016/j.scs.2021.103520}{10.1016/j.scs.2021.103520}.

\bibitem{Matheri2022}
    Anthony Njuguna Matheri, Belaid Mohamed, Freeman Ntuli, Esther Nabadda, Jane Catherine Ngila.
    \textit{Sustainable Circularity and Intelligent Data-Driven Operations and Control of the Wastewater Treatment Plant}.
    Physics and Chemistry of the Earth, Parts A/B/C, Elsevier BV, Volume: 126, Pages: 103152, June 2022.
    DOI: \href{https://doi.org/10.1016/j.pce.2022.103152}{10.1016/j.pce.2022.103152}.

\bibitem{Torfs2024}
    Elena Torfs, Niels Nicolaï, Saba Daneshgar, John B. Copp, Henri Haimi, David Ikumi, Bruce Johnson, Benedek B. Plosz, Spencer Snowling, Lloyd R. Townley, Borja Valverde-Pérez, Peter A. Vanrolleghem, Luca Vezzaro, Ingmar Nopens.
    \textit{The Transition of WRRF Models to Digital Twin Applications}.
    Modelling for Water Resource Recovery, IWA Publishing, Chapter: 6, Pages: 2840--2853, June 2024.
    DOI: \href{https://doi.org/10.2166/wst.2022.107}{10.2166/wst.2022.107}.

\bibitem{Pedersen2022}
    A. N. Pedersen, J. W. Pedersen, M. Borup, A. Brink-Kjær, L. E. Christiansen, P. S. Mikkelsen.
    \textit{Using Multi-Event Hydrologic and Hydraulic Signatures from Water Level Sensors to Diagnose Locations of Uncertainty in Integrated Urban Drainage Models Used in Living Digital Twins}.
    Water Science and Technology, IWA Publishing, Volume: 85, Number: 6, Pages: 1981--1998, February 2022.
    DOI: \href{https://doi.org/10.2166/wst.2022.059}{10.2166/wst.2022.059}.

\bibitem{Dodanwala2023}
    Tharindu C. Dodanwala, Rajeev Ruparathna.
    \textit{A Levels of Service (LOS) Digital Twin for Potable Water Infrastructure Systems}.
    Proceedings of the Canadian Society for Civil Engineering Annual Conference 2023, Springer Nature Switzerland AG, Pages: 15--37, 2023.

\bibitem{Ramos2023smart}
    Helena M. Ramos, Alban Kuriqi, Mohsen Besharat, Enrico Creaco, Elias Tasca, Oscar E. Coronado-Hernández, Rodolfo Pienika, Pedro Iglesias-Rey.
    \textit{Smart Water Grids and Digital Twin for the Management of System Efficiency in Water Distribution Networks}.
    Water, MDPI AG, Volume: 15, Number: 6, Pages: 1129, March 2023.
    DOI: \href{https://doi.org/10.3390/w15061129}{10.3390/w15061129}.

\bibitem{Menapace2024}
    Andrea Menapace, Ariele Zanfei, Manuel Herrera, Bruno Brentan.
    \textit{Graph Neural Networks for Sensor Placement: A Proof of Concept Towards a Digital Twin of Water Distribution Systems}.
    Water, MDPI AG, Volume: 16, Number: 13, Pages: 1835, June 2024.
    DOI: \href{https://doi.org/10.3390/w16131835}{10.3390/w16131835}.

\bibitem{Grievson022}
    Oliver Grievson, Timothy Holloway, Bruce Johnson.
    \textit{A Strategic Digital Transformation for the Water Industry}.
    IWA Publishing, ISBN: 9781789063400, September 2022.
    DOI: \href{https://doi.org/10.2166/9781789063400}{10.2166/9781789063400}.

\bibitem{Fu2022}
    Guangtao Fu, Yiwen Jin, Siao Sun, Zhiguo Yuan, David Butler.
    \textit{The Role of Deep Learning in Urban Water Management: A Critical Review}.
    Water Research, Elsevier BV, Volume: 223, Pages: 118973, September 2022.
    DOI: \href{https://doi.org/10.1016/j.watres.2022.118973}{10.1016/j.watres.2022.118973}.

\bibitem{Curl2019}
    Jason M. Curl, Tyler Nading, Kyle Hegger, Amer Barhoumi, Monika Smoczynski.
    \textit{Digital Twins: The Next Generation of Water Treatment Technology}.
    Journal AWWA, Wiley, Volume: 111, Number: 12, Pages: 44--50, December 2019.
    DOI: \href{https://doi.org/10.1002/awwa.1413}{10.1002/awwa.1413}.

\bibitem{Ramos2023}
    Helena M. Ramos, Alban Kuriqi, Oscar E. Coronado-Hernández, P. Amparo López-Jiménez, Modesto Pérez-Sánchez.
    \textit{Are Digital Twins Improving Urban-Water Systems Efficiency and Sustainable Development Goals?}.
    Urban Water Journal, Informa UK Limited, March 2023.
    DOI: \href{https://doi.org/10.1080/1573062x.2023.2180396}{10.1080/1573062x.2023.2180396}.

    Aditya P. Mathur, Nils Ole Tippenhauer.
    \textit{SWaT: A Water Treatment Testbed for Research and Training on ICS Security}.
    2016 International Workshop on Cyber-physical Systems for Smart Water Networks (CySWater), IEEE, April 2016.
    DOI: \href{https://doi.org/10.1109/cyswater.2016.7469060}{10.1109/cyswater.2016.7469060}.

\bibitem{Wei2022b}
    Yuying Wei, Adrian Wing-Keung Law, Chun Yang.
    \textit{Real-Time Data-Processing Framework with Model Updating for Digital Twins of Water Treatment Facilities}.
    Water, MDPI AG, Volume: 14, Number: 22, Pages: 3591, November 2022.
    DOI: \href{https://doi.org/10.3390/w14223591}{10.3390/w14223591}.

\bibitem{ambling}
    M.H. Homaei.
    \textit{Ambling}.
    \href{https://caucces.ambling.es/}{https://caucces.ambling.es/}, April 2024.

\bibitem{AEMET_OpenData}
    The State Meteorological Agency.
    \textit{AEMET OpenData}.
    \href{https://www.aemet.es/en/datos\_abiertos/AEMET\_OpenData}{https://www.aemet.es/en/datos\_abiertos/AEMET\_OpenData}, April 2024.
    Accessed: April 4, 2024.

\bibitem{Homaei2022}
    Mohammad Hossein Homaei, Andres Caro Lindo, Jose Carlos Sancho Núñez, Oscar Mogollón Gutiérrez, Javier Alonso Díaz.
    \textit{The Role of Artificial Intelligence in Digital Twin’s Cybersecurity}.
    XVII Reunión Española sobre Criptología y Seguridad de la Información (RECSI 2022), Editorial Universidad de Cantabria, ISBN: 9788419024145, September 2022.
    DOI: \href{https://doi.org/10.22429/euc2022.028}{10.22429/euc2022.028}.

\bibitem{Homaei2024}
    Mohammadhossein Homaei, Óscar Mogollón-Gutiérrez, José Carlos Sancho, Mar Ávila, Andrés Caro.
    \textit{A Review of Digital Twins and Their Application in Cybersecurity Based on Artificial Intelligence}.
    Artificial Intelligence Review, Volume: 57, Number: 8, Pages: 201, July 2024.
    DOI: \href{https://doi.org/10.1007/s10462-024-10805-3}{10.1007/s10462-024-10805-3}.

\bibitem{Guikema_2024}
    Seth Guikema, Roger Flage.
    \textit{Digital Twins as a Security Risk?}.
    Risk Analysis, Wiley, July 2024.
    DOI: \href{https://doi.org/10.1111/risa.15749}{10.1111/risa.15749}.

\bibitem{Niknam2023}
    Azar Niknam, Hasan Khademi Zare, Hassan Hosseininasab, Ali Mostafaeipour.
    \textit{Developing an LSTM Model to Forecast the Monthly Water Consumption According to the Effects of the Climatic Factors in Yazd, Iran}.
    Journal of Engineering Research, Elsevier BV, Volume: 11, Number: 1, Pages: 100028, March 2023.
    DOI: \href{https://doi.org/10.1016/j.jer.2023.100028}{10.1016/j.jer.2023.100028}.



\end{thebibliography}


\newpage
\appendix

\section{Lemma and Proof of Pearson’s Correlation Coefficient}
\label{sec:LemmaPearson}

\textbf{Lemma:} The Pearson correlation coefficient \( R \) is a measure of the linear relationship between two variables and is invariant under changes in the location and scale of the variables.

\textbf{Proof:} 
Let \( X' \) and \( Y' \) be two new variables derived from \( X \) and \( Y \) by linear transformations:

\begin{equation} \label{eq:linear_transformations}
X' = \alpha X + \beta \quad \text{and} \quad Y' = \gamma Y + \delta
\end{equation}

where \( \alpha \) and \( \gamma \) are non-zero constants, and \( \beta \) and \( \delta \) are constants.

The covariance of \( X' \) and \( Y' \) is given by:

\begin{equation} \label{eq:covariance_transformed}
\text{Cov}(X', Y') = \text{Cov}(\alpha X + \beta, \gamma Y + \delta) = \alpha \gamma \cdot \text{Cov}(X, Y)
\end{equation}

The standard deviations of \( X' \) and \( Y' \) are:

\begin{equation} \label{eq:std_dev_transformed}
\sigma_{X'} = |\alpha|\sigma_X \quad \text{and} \quad \sigma_{Y'} = |\gamma|\sigma_Y
\end{equation}

The Pearson correlation coefficient for \( X' \) and \( Y' \) is:

\begin{equation} \label{eq:pearson_invariance}
R' = \frac{\text{Cov}(X', Y')}{\sigma_{X'} \sigma_{Y'}} = \frac{\alpha \gamma \cdot \text{Cov}(X, Y)}{|\alpha| \sigma_X |\gamma| \sigma_Y} = \frac{\text{Cov}(X, Y)}{\sigma_X \sigma_Y} = R
\end{equation}

Thus, the Pearson correlation coefficient \( R \) remains unchanged under linear transformations, proving its invariance under changes in location and scale.

\section{Explanation of Non-Overlap Constraints}
\label{appendix:non_overlap_explanation}

\textbf{Explanation of Constraints \eqref{eq:non_overlap_corrected_1} and \eqref{eq:non_overlap_corrected_2}:}

\begin{itemize}
    \item \textbf{When} \( x_{ijkl} = 1 \):
    \begin{itemize}
        \item Constraint \eqref{eq:non_overlap_corrected_1} simplifies to:
        \begin{equation} \label{eq:non_overlap_case1_constraint1}
        C_{ik} \leq s_{jl}
        \end{equation}
        This ensures that segment \( k \) of task \( i \) finishes before segment \( l \) of task \( j \) starts.
        \item Constraint \eqref{eq:non_overlap_corrected_2} becomes:
        \begin{equation} \label{eq:non_overlap_case1_constraint2}
        C_{jl} \leq s_{ik} + M \times 1
        \end{equation}
        Since \( M \) is a large number, this constraint does not restrict the scheduling and is effectively redundant.
    \end{itemize}
    \item \textbf{When} \( x_{ijkl} = 0 \):
    \begin{itemize}
        \item Constraint \eqref{eq:non_overlap_corrected_1} becomes:
        \begin{equation} \label{eq:non_overlap_case2_constraint1}
        C_{ik} \leq s_{jl} + M \times 1
        \end{equation}
        Again, this constraint is redundant due to the large \( M \).
        \item Constraint \eqref{eq:non_overlap_corrected_2} simplifies to:
        \begin{equation} \label{eq:non_overlap_case2_constraint2}
        C_{jl} \leq s_{ik}
        \end{equation}
        This ensures that segment \( l \) of task \( j \) finishes before segment \( k \) of task \( i \) starts.
    \end{itemize}
\end{itemize}

These constraints guarantee that for any two segments, either one must finish before the other starts, thereby preventing any overlap on the single machine.

\textbf{Setting the Value of \( M \):}

The constant \( M \) must be chosen carefully. It should be a value larger than the maximum possible difference between any task segment's start and completion times within the scheduling horizon. This ensures that when a constraint is intended to be inactive (due to the value of \( x_{ijkl} \)), it does not inadvertently impose any restrictions on the scheduling variables.

\end{document}